\documentclass[fleqn,usenatbib]{mnras}
\usepackage{newtxtext,newtxmath}
\usepackage[T1]{fontenc}
\DeclareRobustCommand{\VAN}[3]{#2}
\let\VANthebibliography\thebibliography
\def\thebibliography{\DeclareRobustCommand{\VAN}[3]{##3}\VANthebibliography}
\usepackage{graphicx}	
\usepackage{amsmath}	
\usepackage{amssymb}	

\title[Spatially resolved galaxy interactions]{Spatially resolved star formation and fuelling in galaxy interactions}

\author[J. Moreno et al.]
{\parbox{17.5cm}{
Jorge Moreno$^{1}$\thanks{E-mail: jorge.moreno@pomona.edu}, Paul Torrey$^{2}$, Sara~L. Ellison$^{3}$, David~R. Patton$^{4}$, Connor Bottrell$^{3}$, Asa~F.~L. Bluck$^{5,6}$, Maan~H. Hani$^{3,\dagger}$,  Christopher~C. Hayward$^{7}$, James~S. Bullock$^{8}$, Philip~F. Hopkins$^{9}$ and Lars Hernquist$^{10}$ }\vspace{0.3cm}\\
$^{1}$Department of Physics and Astronomy, Pomona College, Claremont, CA 91711, USA\\
$^{2}$Department of Astronomy, University of Florida, 211 Bryant Space Sciences Center, Gainesville, FL, USA\\
$^{3}$Department of Physics \& Astronomy, University of Victoria, Finnerty Road, Victoria, British Columbia, V8P 1A1, Canada\\
$^{4}$Department of Physics \& Astronomy, Trent University, 1600 West Bank Drive, Peterborough, Ontario, K9L 0G2, Canada\\
$^{5}$Kavli Institute for Cosmology \& Cavendish Astrophysics, University of Cambridge, Madingley Road, Cambridge, CB3 0HA, UK 14\\
$^{6}$Hughes Hall College, University of Cambridge, Wollaston Road, CB1 2EW, UK\\
$^{7}$Center for Computational Astrophysics, Flatiron Institute, 162 Fifth Avenue, New York, NY 10010, USA\\ 
$^{8}$Department of Physics and Astronomy, University of California, Irvine, CA 92697 USA\\
$^{9}$TAPIR, Mailcode 350-17, California Institute of Technology, Pasadena, CA 91125, USA\\
$^{10}$Harvard-Smithsonian Center for Astrophysics, 60 Garden Street, Cambridge, MA, 02138, USA\\ 
$\dagger$Vanier Fellow\\
}

\date{Accepted XXX. Received YYY; in original form ZZZ}

\pubyear{2020}

\begin{document}
\label{firstpage}

 \maketitle

\begin{abstract}
We investigate the spatial structure and evolution of star formation and the interstellar medium (ISM) in interacting galaxies. We use an extensive suite of parsec-scale galaxy merger simulations (stellar mass ratio $=$ 2.5:1), which employs the "Feedback In Realistic Environments-2" model (\textsc{fire-2}). This framework resolves star formation, feedback processes, and the multi-phase structure of the ISM. We focus on the galaxy-pair stages of interaction. We find that close encounters substantially augment cool (HI) and cold-dense (H$_{2}$) gas budgets, elevating the formation of new stars as a result. This enhancement is centrally-concentrated for the secondary galaxy, and more radially extended for the primary. This behaviour is weakly dependent on orbital geometry. We also find that galaxies with elevated global star formation rate (SFR) experience intense nuclear SFR enhancement, driven by high levels of either star formation efficiency (SFE) or available cold-dense gas fuel. Galaxies with suppressed global SFR also contain a nuclear cold-dense gas reservoir, but low SFE levels diminish SFR in the central region. Concretely, in the majority of cases, SFR-enhancement in the central kiloparsec is fuel-driven (55\% for the secondary, 71\% for the primary) -- whilst central SFR-suppression is efficiency-driven (91\% for the secondary, 97\% for the primary). Our numerical predictions underscore the need of substantially larger, and/or merger-dedicated, spatially-resolved galaxy surveys -- capable of examining vast and diverse samples of interacting systems -- coupled with multi-wavelength campaigns aimed to capture their internal ISM structure.
\end{abstract}


\section{Introduction}
\label{sec:intro}

It has been eighty years since the publication of very first observational and `numerical' investigations on the nature of galaxy encounters \citep{Holmberg1940,Holmberg1941}. Decades later, the emergence of computers allowed researchers to conduct the first numerical experiments of idealised (non-cosmological) galaxy merging systems \citep{TT72,Hernquist1989,BH91,BH96,MH96} -- which supplied a theoretical framework to explain tidally distorted galaxies \citep{Arp1966,LarsonTinsley1978}, and a possible connection between galaxy mergers, starburts, and quasars \citep{Sanders1988,SandersMirabel1996,CanalizoStockton2001}. 

Contemporaneously, galaxy mergers were recognized as naturally occurring events within the hierarchical $\Lambda$ Cold Dark Matter ($\Lambda$CDM) paradigm \citep{WhiteRees1978,Blumenthal1984,White1991} -- and now form a crucial ingredient in semi-analytic models (SAMs) of galaxy formation \citep{Cole2000,Bower2006,Croton2006,DeLucia2007,Henriques2011,Benson2012,Guo2012,Lagos2018,Lagos2019}. Often, these SAMs rely on idealised galaxy merger simulations for guidance. For instance, \cite{Hopkins2008} and \cite{Somerville2008} directly implement results from idealised galaxy merger simulations by \cite{Hopkins2005} in their cosmological recipes. 

Unfortunately, the great majority of SAMs entirely ignore the early stages of interaction -- i.e., when the merging galaxies can still be identified as two distinct units \citep[but see][for an exception]{Menci2004}. This is despite the fact that a vast number of observations reveal their definitive importance.  Concretely, in the local Universe, interaction-induced star formation is enhanced in galaxies with close companions \citep{Patton1997,Barton2000,Lambas2003,Ellison2008,Scudder2012,Robotham2014}. Close galaxy encounters also diminish nuclear metallicity \citep{RupkeGradients2010,Rich2012,Scudder2012}, augment molecular gas content \citep{Violino2018}, mould the circumgalactic medium \citep{Hani2018,Smith2018,Smith2019}, and ignite active galactic nuclei \citep[AGN,][]{Ellison2011,Treister2012,Sabater2013,Satyapal2014,Ellison2019}. Idealised simulations confirm these effects during the pre-merger `galaxy-pair' period: including enhanced star formation \citep{DiMatteo2007,DiMatteo2008,Moreno2015}, decrements in nuclear metallicity  \citep{Torrey2012}, alterations in the structure of interstellar medium \citep[ISM,][]{Moreno2019}, and triggered AGN \citep{TDiMatteo2005,Callegari2009,Capelo2015,Capelo2017}. 

Unlike the dramatic, albeit brief, turmoil experienced by merging galaxies at coalescence, the effects sparked by the influence of a close neighbour during the early stages of interaction tend to be gentler and of longer duration \citep{Moreno2015}. Observations by \cite{Patton2013} suggest that interaction-driven effects extend out to $\sim$150 kpc in projected separation. \cite{Patton2020} confirm this effect in cosmological simulations, and demonstrate that close encounters affect galaxy pairs out to separations of $\sim$280 kpc in 3D space. Each interaction and fly-by \citep{Moreno2012,Sinha2012,LHuillier2015,An2019} is capable of inciting bar formation \citep{Lokas2016,Lokas2019,Pettitt2018,Cavanagh2020} and promoting bulge mass growth \citep{Just2010,Bekki2011}. But more importantly, the cumulative effect of multiple -- frequently occurring and long-lived -- galaxy encounters may ultimately stimulate the transformation of spirals into lenticulars in dense environments \citep{Moore1996,BoselliGavazzi2006,Cappellari2013,Joshi2020}. In sum, galaxy pairs offer a unique and powerful window to understand how close encounters affect both global properties and the internal structure of galaxies.

Coincidentally, recent years have also witnessed the emergence of integral-field unit (IFU) surveys targeting the local Universe -- such as the Calar Alto Legacy Integral Field Area Survey \citep[CALIFA,][]{Sanchez2012}, \textcolor{black}{the Sydney-AAO Multi-object Integral field spectrograph Galaxy Survey} \citep[SAMI,][]{Croom2012}, and the Mapping Nearby Galaxies at APO Survey \citep[MaNGA,][]{Bundy2015} -- which open a new avenue for studying the connection between galactic structure and interaction history. These spatially-resolved campaigns allow us to go beyond asking only about global properties -- and permit us to analyse the spatial extent of SFR enhancements \citep[along with the flattening of metallicity gradients,][]{BarreraBallesteros2015sfr,Pan2019sfr,Thorp2019}, plus kinematic signatures \citep{BarreraBallesteros2015kin,Hung2016,Bloom2017,Li2019,Feng2020} in interacting galaxies. Coupled with interferometric follow-up observations focused on H$_2$ content and its structure -- such as EDGE-CALIFA\footnote{EDGE stands for the Extragalactic Database for Galaxy Evolution Survey.} \citep{Bolatto2017} and ALMaQUEST\footnote{ALMaQUEST stands for ALMA-MaNGA QUEnching and STar formation Survey, whilst ALMA stands for Atacama Large Millimeter/submillimeter Array.} \citep{Lin2019}, we now have the potential to clearly define the spatial extent in which galaxy interactions affect the ISM, and how this process fuels star formation.

On the numerical side, it is understandable to ask if idealised (non-cosmological) galaxy merger simulations, like the ones we present in this paper, are the optimal tool for these kind of studies. Indeed, cosmological simulations \citep[e.g.,][]{Perez2006,Perez2011,Bustamante2018,Blumenthal2020,Patton2020,Hani2020} and zoom-in simulations \citep[e.g.,][]{Sparre2016} also provide a viable path. Furthermore, such simulations naturally provide the diversity and cosmological context experienced by merging galaxies in the real Universe \citep{Martig2008,Moreno2013}. However, unlike cosmologically-selected galaxy pairs, idealised simulations offer unparalleled spatial and temporal resolution. Moreover, this non-cosmological framework renders the user control to conduct numerical experiments where specific initial orbital conditions can be designed to answer specific questions -- e.g., the effect of spin-orbit inclination \textcolor{black}{and} impact parameter, to name a few. In principle, one can also address these questions with cosmological simulations. Unfortunately, when those simulations include hydrodynamics and explicit stellar feedback, their box sizes tends to be small -- which may severely limit the diversity in merging orbits. Maximising high resolution and sizable volumes remains a challenge for such large-scale simulations -- although the use of `genetically-modified' zoom-in technology offers promise to overcome this limitation \citep{Noth2016,Rey2018,Stopyra2020}.

Similarly, in the idealised galaxy-merger approach, expanding diversity in orbital geometries also conspires against increments in resolution. Merger libraries capable of broadly exploring orbital parameter space \citep[e.g.,][]{DiMatteo2008,Moreno2015} are often forced to employ low-resolution schemes to optimise computational resources. Increases in resolution must also be accompanied with physically-motivated sub-grid recipes capable of capturing the multi-phase structure of the interstellar medium at those scales. Conversely, when high-resolution programmes (with improved physics prescriptions) attempt to simulate galaxy mergers at the parsec scale, they do so at the expense of {\it not} being able to explore orbital parameter space in great detail \citep{Renaud2009,Kim2009,Teyssier2010,Karl2010,Karl2011,Karl2013,Hopkins2013mergers,Renaud2014,Renaud2015,Renaud2019,Renaud2020}.

In this paper, we employ an extensive suite of parsec-scale galaxy merger simulations \citep{Moreno2019} based on the `Feedback In Realistic Environment-2' (\textsc{fire-2}) model \citep{FIRE2}. This framework captures the multi-phase structure of the interstellar medium and resolves the physics of relevant feedback processes that regulate star formation. Our suite consists of 24 orbital configurations, making it {\it the largest library of galaxy merger simulations at the parsec scale to date.} This uniquely positions us to conduct spatially-resolved studies of star-formation and the evolution of the ISM in interacting galaxies from a numerical perspective. This work expands on \cite{Moreno2019}, who only addresses interaction-induced effects on the ISM for the entire two-galaxy system -- and \cite{Moreno2015}, who analyse the spatial extent of interaction-induced SFR using an older model.
With this in mind, this paper addresses the following questions:
\begin{enumerate}
    \item[1.] How are the new-stellar and ISM budgets -- as well as the instantaneous star formation rate (SFR) and efficiency (SFE, equation~\ref{eqn:sfe}) -- in a galaxy affected by the presence of a companion? 
    \item[2.] How do these quantities evolve globally, within the central region, and in the rest of the galaxy?
    \item[3.] How do close galaxy encounters affect the radial structure of the above baryonic components?
    \item[4.] How does the radial structure of galaxies evolve in time during the interaction?
    \item[5.] How does our choice of orbital merging geometry affect the radial structure in these baryonic components?
    \item[6.] Is there a connection between global SFR enhancement and the radial structure of SFR, SFE, and fuel availability?
    \item[7.] What drives SFR enhancement (or suppression) in the central kiloparsec, SFE or fuel availability?
    \item[8.] Do the primary and secondary galaxies exhibit different behaviour during the interaction?
\end{enumerate}

This manuscript is organised as follows. Section~\ref{sec:simulations_and_terminology} introduces our simulations and relevant terminology (\textbf{\textit{boldface italics}}). Sections~\ref{subsec:time_evolution}, \ref{subsec:averages}, and \ref{subsec:profiles} respectively focus on time evolution, sample-wide averages, and radial structure. We investigate the role of orbital merging geometry in Section~\ref{subsec:profiles_subsuites}. In Section \ref{subsec:driving_global_sfr} we address connections between global SFR deviations and radial structure, and in Section~\ref{subsec:fuel} we focus on what factors drive SFR in the central kiloparsec. Section~\ref{sec:summary} summarises our findings.

\section{Simulations and terminology}
\label{sec:simulations_and_terminology} 

Our galaxy merger simulations employ the `Feedback In Realistic Environments-2' (\textsc{fire-2}) physics model. See \cite{FIRE,FIRE2} for details -- we only provide a brief summary here. \textbf{\textit{Star formation}} is constrained to self-gravitating, self-shielding \citep{Krumholz2011} gas denser than 1000 cm$^{-3}$ \citep{Hopkins2013}. \textcolor{black}{Once these conditions are met, gas is converted into stars at 100\% efficiency per local dynamical time.} We incorporate free-free, photo-ionisation/recombination, Compton, photoelectric, dust-collisional, cosmic ray, molecular, metal-line and fine-structure processes in our treatment of radiative heating and cooling. Our \textbf{\textit{feedback model}} includes momentum flux from radiation pressure; energy, momentum, mass and metal injection from Type Ia and II SNe, plus mass loss from OB and AGB stars. We use \textsc{starburst99} \citep{Leitherer1999} to tabulate stellar masses, ages, metallicities, feedback event rates, luminosities, energies and mass-loss rates. \textcolor{black}{Our simulations employ the meshless finite mass (MFM) mode of the GIZMO hydro solver} \citep{gizmo2}. We do not incorporate AGN feedback because (1) we wish to focus on the role of stellar feedback alone, and (2) the coupling of AGN fueling and feedback with the surrounding multi-phase ISM at the scales probed in this paper is not yet fully understood \citep[but see, e.g.,][for recent work exploring this question]{Hopkins2016,Angles2017,Angles2020}.

\begin{table}
  \begin{center}
    \begin{tabular}{l|l|l|} 
        \hline

        ISM regime &  Temperature-density demarcations \\
        \hline \hline
        cold-dense & $(T<300\,{\rm K}, n>10 \,{\rm cm}^{-3})$  \\
        \hline
        cool & $(T<8000\,{\rm K}, 0.1 \,{\rm cm}^{-3}<n<10 \,{\rm cm}^{-3})$ \\
           & ${\rm \&} \,\, (300\,{\rm K}<T< 8000\,{\rm K},n<0.1\, {\rm cm}^{-3})$  \\
        \hline

    \end{tabular}
  \end{center}
\caption{Temperature-density demarcations: the cold-dense and cool ISM regimes, adopted to approximately represent H$_{2}$ and HI gas, respectively. See \citet{Moreno2019} for details.}
\label{table:phases}
\end{table}

\begin{figure*}
\centerline{\vbox{
\vspace{0in}
\hbox{
\includegraphics[width=7.1in]{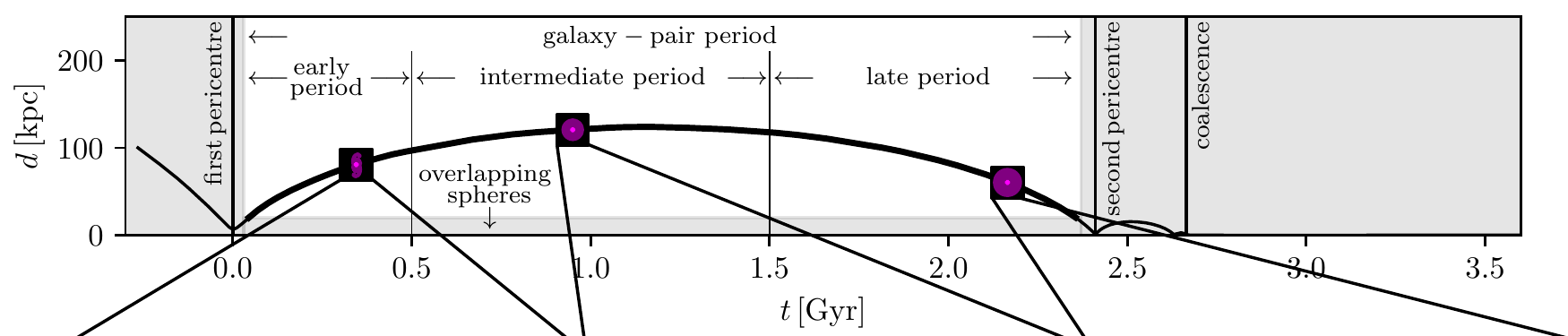}
}
\vspace{-.0in}
\hbox{
\includegraphics[width=2.605in]{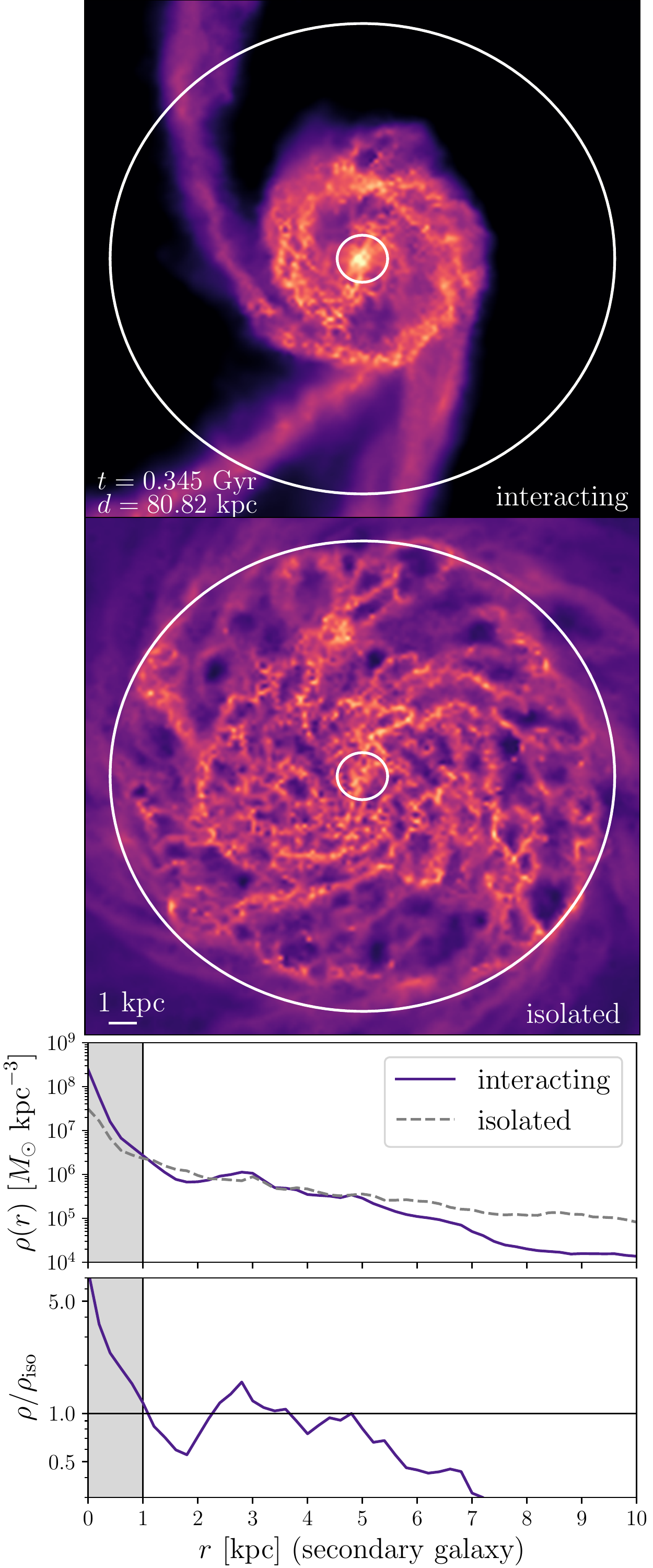}
\includegraphics[width=2.24in]{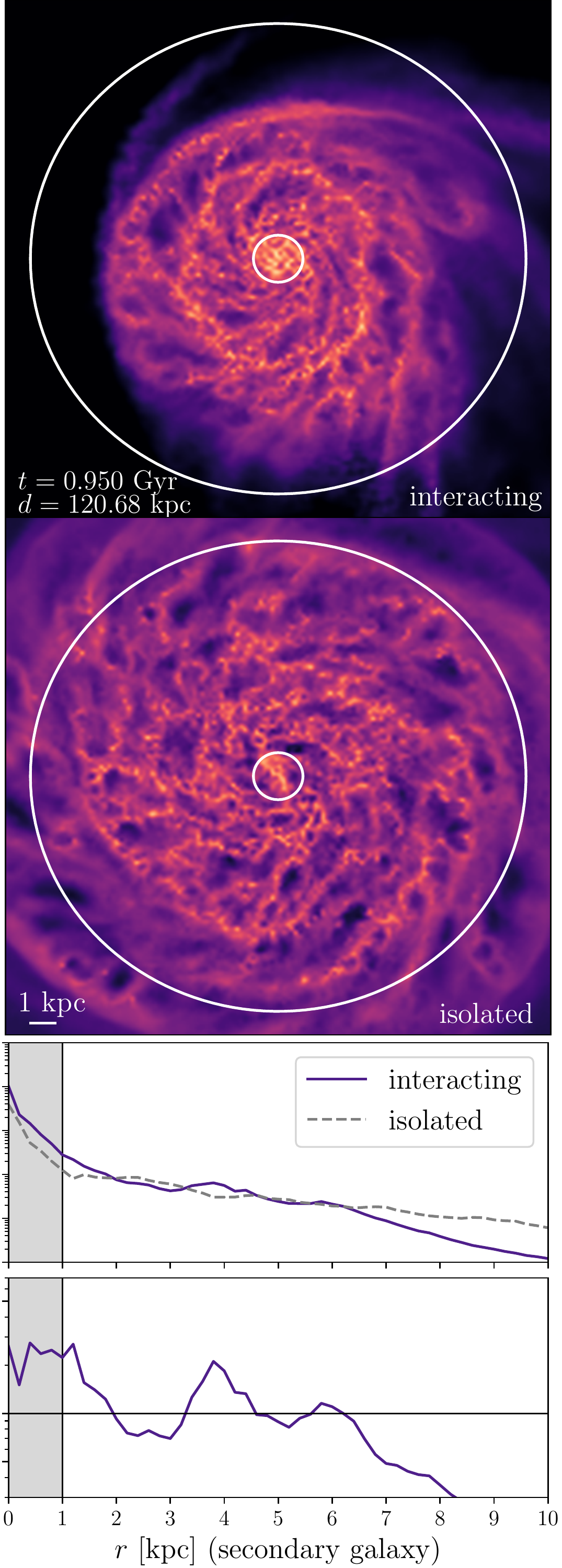}
\includegraphics[width=2.24in]{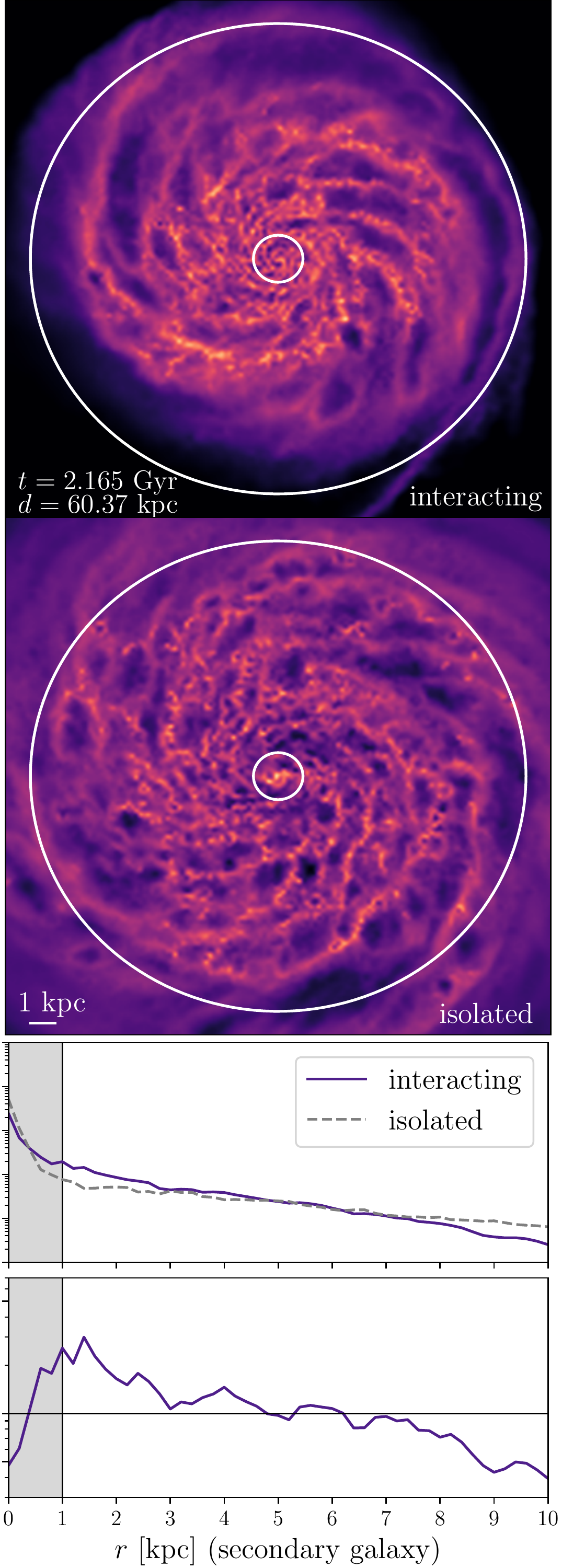}
}
\vspace{-.0in}
}}
\caption{Terminology and three illustrative examples. {\it Top panel:} Galaxy-galaxy distance versus time (fiducial run). The thick vertical lines from left to right denote first and second pericentric passages, plus coalescence. Time is shifted to zero at first passage. \textcolor{black}{The thick portion of the curve and the area not covered by gray boxes denote the galaxy-pair period, defined to be between first and second passages, with separations of at least 20 kpc.} The thin vertical lines indicate the early, intermediate, and late periods, chosen arbitrarily to describe the evolution of the interaction. The solid purple symbols depict three specific times, described further in subsequent panels (increasing in time from left-to-right). \textcolor{black}{{\it Second (third) row panels:} Face-on surface density maps of the entire gas budget -- i.e., all ISM phases -- for the interacting (isolated) galaxy at three illustrative times, indicated by the three vertical rows.} The colour scale is logarithmic (mass-weighted) and the same for both rows (all six images). We only show the secondary galaxy and its isolated counterpart. The white circles indicate the centre (0$-$1 kpc) and the outskirts (1$-$10 kpc). The galaxy-pair period excludes times when the larger circles belonging to each galaxy overlap. The keys indicate time and galaxy-galaxy distance after first pericentric passage, plus the spatial scale. See supplementary materials for videos associated with these images. {\it Fourth (fifth) row panels:} 3D radial gas mass density profiles for the interacting (purple) and isolated (dashed-gray) galaxy, plus their ratio (purple). The vertical line at 1 kpc and the gray box indicate the central region. The horizontal line indicates unity. 
}
\label{fig:terminology}
\end{figure*}

Our \textbf{\textit{galaxy merger suite}} consists of 24 major mergers (stellar mass ratio $=$ 2.5:1) with three \textbf{\textit{spin-orbit orientations}}: near-prograde, near-polar, and near-retrograde \citep{Moreno2015}. Each orientation spans the following separations at first pericentric passage: $\sim$7 kpc (three orbits), $\sim$16 kpc (three orbits), and $\sim$27 kpc (two orbits). Initially, the secondary galaxy has the following properties: stellar mass $=$ 1.2 $\times 10^{10}$ M$_{\odot}$, bulge mass $=$ 7.0 $\times 10^{9}$ M$_{\odot}$, and gas mass $=$ 7.0 $\times 10^{10}$ M$_{\odot}$ -- and the primary has stellar mass $=$ 3.0 $\times 10^{10}$ M$_{\odot}$, bulge mass $=$ 2.5 $\times 10^{10}$ M$_{\odot}$, and gas mass $=$ 8.0 $\times 10^{10}$ M$_{\odot}$. We follow \cite{Mendel2014} and \cite{Saintonge2016} for our bulge and gas mass choices. For comparison, we also simulate these two galaxies in isolation. \textcolor{black}{Time outputs are stored at 5 Myr resolution. We adopt gas and stellar particle masses of $1.4 \times 10^{4} M_{\odot}$ and $1.9 \times 10^{3} M_{\odot}$, respectively. We achieve a maximum spatial resolution of 1.1 parsec and a maximum gas-density resolution of $5.8 \times 10^{5}$ cm$^{-3}$. With these choices, $\sim$0.15\% of the gas exceeds the 1000 cm$^{-3}$ density threshold required (but not necessarily sufficient) for star formation.} See Tables 1 and 2 of \cite{Moreno2019} for more details on our adopted initial conditions.

We characterise the interstellar medium in terms of four density-temperature demarcations: hot, warm, cool, and cold-dense regimes -- meant to represent gas above 1 million Kelvin, warm-ionised gas, HI gas, and H$_2$ gas \citep{Moreno2019}. In this work, we only discuss the \textbf{\textit{cool}} and \textbf{\textit{cold-dense}} gas components (Table~\ref{table:phases}) because these phases are more closely connected to star formation fuelling. In particular, we adopt the temperature and density thresholds of \cite{Orr2018} for our cold-dense gas regime because ISM at densities above $n>10 \, {\rm cm}^{-3}$ (at solar luminosity) captures the majority of the H$_{2}$ gas \citep[][]{Semenov2017}.

\textcolor{black}{One can write the instantaneous star formation rate as
\begin{equation}
\label{eqn:sfe}
{\rm SFR} = \frac{\rm SFR}{M_{\rm cold-dense}} \times M_{\rm cold-dense} \equiv  {\rm SFE} \times M_{\rm cold-dense},
\end{equation}
where SFE denotes the \textbf{\textit{star formation efficiency}}.} 
Writing SFR in this particular format facilitates our discussion of how these three quantities are interconnected (Sections~\ref{subsec:time_evolution}, \ref{subsec:averages}, \ref{subsec:driving_global_sfr}, and \ref{subsec:fuel}). Indeed, the ability to probe variations in SFE is one of the advantages of using simulations like \textsc{fire}. We note this exercise is not entirely appropriate for older simulations that tune their star-formation recipes to the observed KS \citep{Kennicutt1998} law \citep[e.g.,][]{SH03}. However,  simulations like ours, capable of capturing the multi-phase structure of the ISM -- and for which the KS naturally emerges as an output \citep{Orr2018} -- provide an appropriate framework to study variations in star formation efficiency.

\begin{figure*}
\centerline{\vbox{
\hbox{
\includegraphics[width=3.5in]{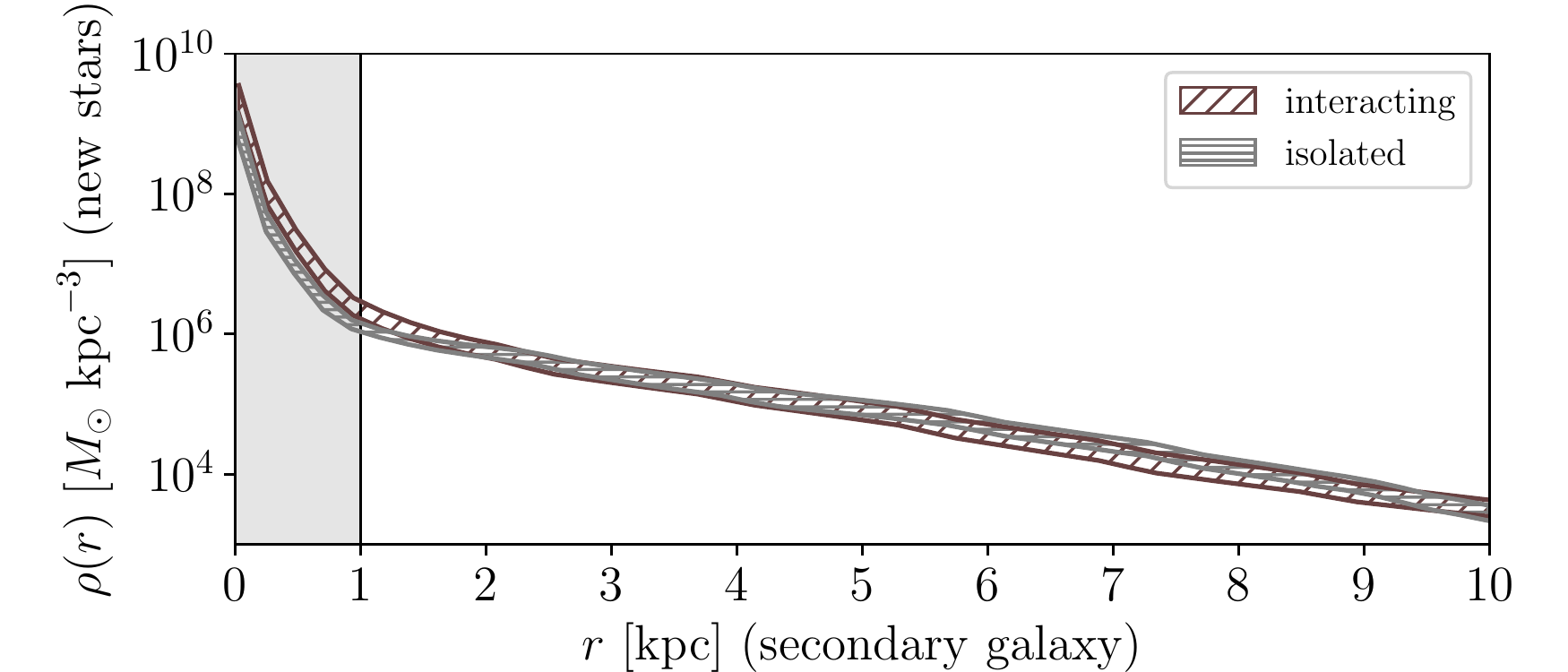}
\includegraphics[width=3.5in]{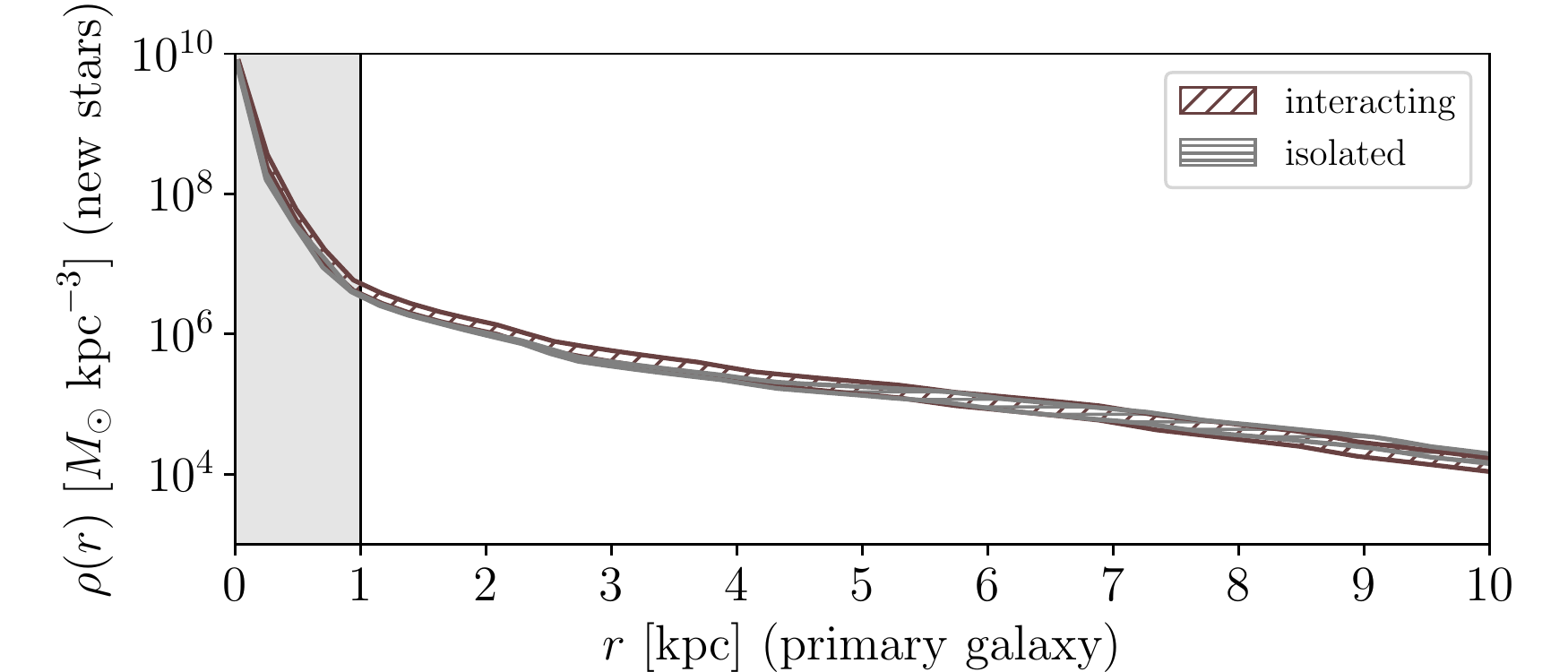}
}
\vspace{-.02in}
\hbox{
\includegraphics[width=3.5in]{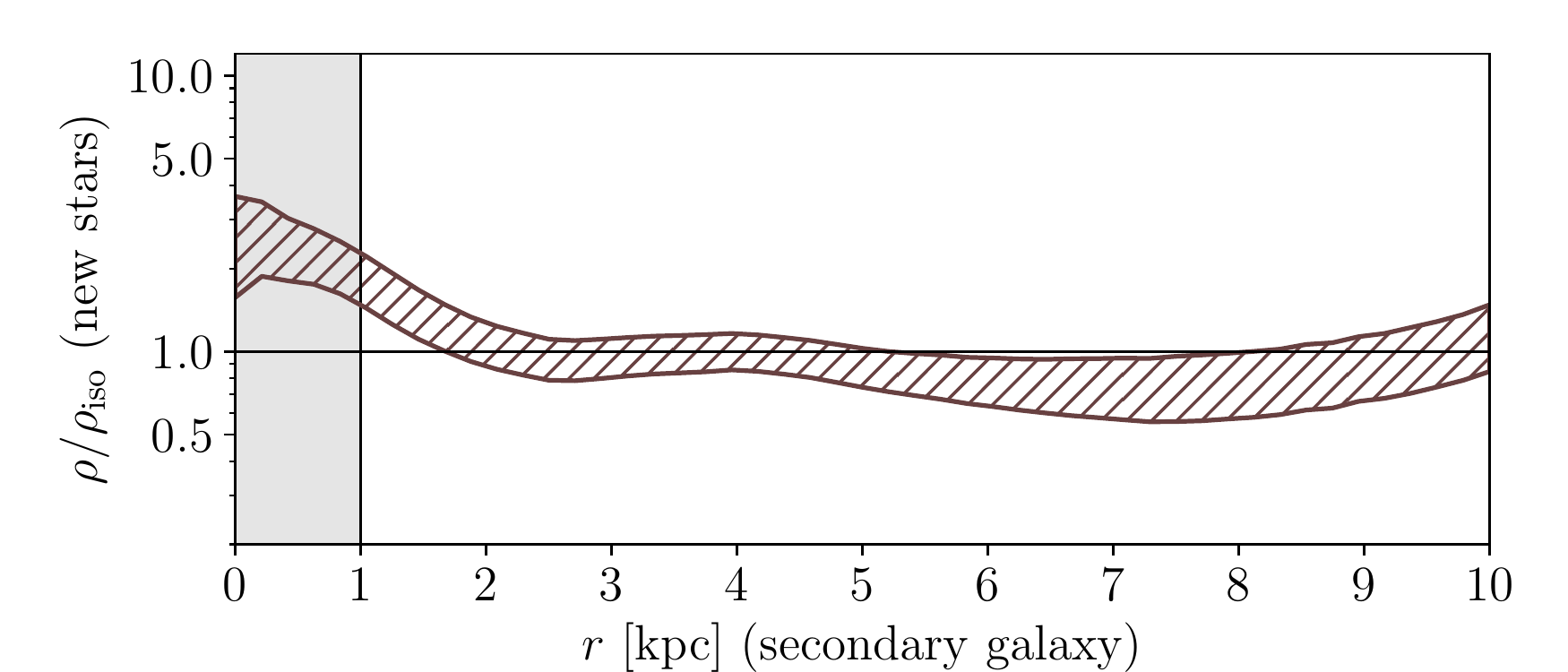}
\includegraphics[width=3.5in]{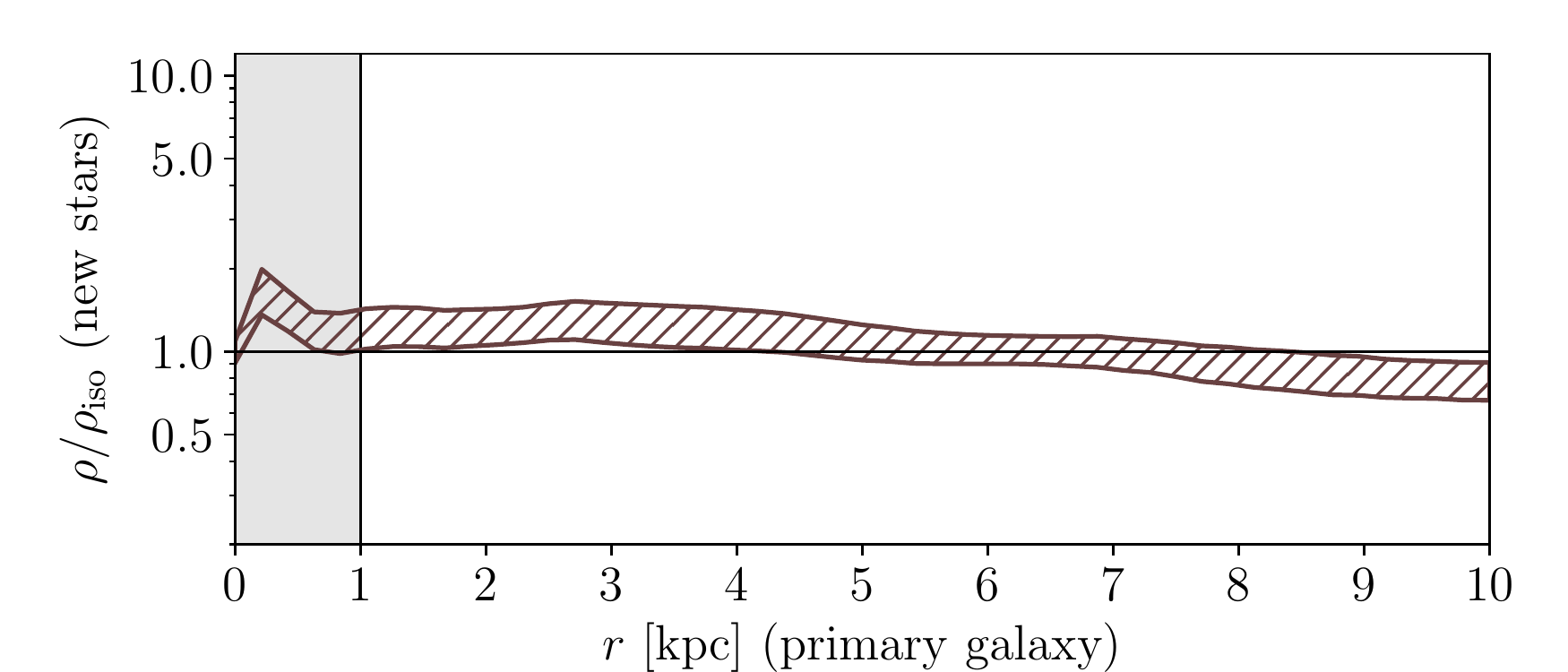}
}
\vspace{-.01in}
}}
\caption{Definitions: average density profile and average profile ratio (mass in new stars, entire merger suite, galaxy-pair periods only). {\it Left (right) panels}: secondary (primary) galaxy. {\it Top panels}: The diagonally-hatched brown band represents the average density profile in new stellar mass versus 3D radial distance for the interacting galaxy -- calculated by averaging across every configuration in our merger suite within their respective galaxy-pair periods. The gray horizontally-hatched band is the result of the same exercise for its isolated counterpart. Band thickness represents one standard deviation.  {\it Bottom panels}: The brown hatched band displays the average profile ratio, which is the result of averaging the ratios of individual density profiles for the interacting galaxy and its isolated counterpart. The vertical black line and gray box highlights the central region ($r<1$ kpc). The horizontal line indicates unity. 
}
\label{fig:profile_definitions}
\end{figure*}

To gain insight, we devote special attention to a specific \textbf{\textit{fiducial run}}: a nearly prograde configuration with small impact parameter ($\sim$7 kpc), selected to maximise the effects of the encounter \citep{Moreno2019}. Figure~\ref{fig:terminology} illustrates this run at three representative times. The top panel shows galaxy-galaxy separation versus time. For each galaxy, we use the location of the central supermassive black hole (SMBH) as a proxy for galactic centre. \textcolor{black}{We note that, in our simulations, this SMBH does not play a dynamical role, but only serves to trace the potential minimum.} The thick vertical lines (from left to right) indicate first and second pericentric passages, plus coalescence. Throughout this paper, time is shifted to zero at first pericentric passage. This work focuses on the \textbf{\textit{galaxy-pair period}}, corresponding to times between first and second pericentric passage. The gray boxes and thin portions of the galaxy-separation curve indicate times {\it outside} the galaxy-pair period. To avoid contamination from the companion galaxy, we {\it exclude} times at which two 10-kpc spheres surrounding each galaxy overlap.  Without this extra condition, one cannot disentagle dynamical effects driven by a recent encounter from the simple presence of new material `belonging' to (i.e., being closer to the centre of) the companion galaxy. The \textcolor{black}{thin horizontal gray box} and the small extra gray spaces right and left of the first and second thick vertical lines, respectively, highlight this extra condition.

To describe how the spatial extent of various baryonic components evolves in time, we split the galaxy-pair period into three portions: the \textbf{\textit{early}} ($t<0.5$ Gyr), \textbf{\textit{intermediate}} ($0.5<t<1.5$ Gyr), and \textbf{\textit{late}} ($t>1.5$ Gyr) time periods. The two thin vertical lines at $t=0.5$ and 1.5 Gyr separate these three periods (Figure~\ref{fig:terminology}, top panel). We keep these lines in figures displaying time evolution throughout the paper to facilitate cross-reference between Sections~\ref{subsec:time_evolution} and \ref{subsec:profiles}. There is no fundamental reason behind these three specific choices. One possibility is to split the galaxy-pair period according to where the two galaxies are located in their merging orbit \citep{Privon2014,Pan2019sfr} -- i.e., if they are close to first pericentric passage, apocentre, etc. However, this would require replacing our universal demarcations (at 0.5 and 1.5 Gyr) with orbit-dependent choices, because the duration of the galaxy-pair period varies from merger to merger. We elect to avoid this approach because the large-scale duration of the interaction does not necessarily dictate the timescales governing small-scale hydrodynamics and feedback-regulated baryonic physics. In the absence of a rigourous option, we visually inspect the gas component in our fiducial secondary galaxy to guide our prima facie choices\footnote{\textcolor{black}{For videos of our galaxy merger simulations, see the online supplementary materials and please visit \url{https://research.pomona.edu/galaxymergers/videos/}.}}.  We focus on the gaseous component because interaction-induced disturbances are more extended and visually evident \citep{Bournaud2004}. We inspect the secondary galaxy in detail here (and typically describe it first throughout the paper -- i.e., by placing information pertaining to the primary galaxy on the right or bottom panels in figures) because its shallower gravitational potential makes it more susceptible to the effects of the encounter \citep{BH96,DiMatteo2007,Moreno2015}.

Our visual inspection reveals the following sequence of events. (1) Soon after its first pericentric passage, the gaseous component of the galaxy exhibits extended tidal tails, whilst its disc shrinks and develops a prominent central concentration. (2) External material originally launched into the bridge and tidal tails is re-accreted onto the outskirts of the disc, which begins to recover in radial extent. (3) Lastly, disturbances produced by the encounter fade away and the central concentration becomes diluted. See e.g., \cite{Blumenthal2018} for a more rigourous description of this process. Our demarcations at 0.5 and 1.5 Gyr approximately mark the timescales at which these three steps unfold. From left-to-right, the second-row panels in Figure~\ref{fig:terminology} display face-on surface-density maps of the gas component (all ISM phases) at specific representative times selected from each of these three periods. Third-row images show similar maps for its isolated counterpart. All six images share the same logarithmic mass-weighted colour scale. To facilitate our analysis \citep[as in][]{Moreno2015}, the white circles split each of our galaxies into the following spherically-symmetric 3D regions: the \textbf{\textit{centre}} ($r<1$ kpc), the \textbf{\textit{outskirts}} ($1<r<10$ kpc), and the \textbf{\textit{entire galaxy}} ($r<10$ kpc) -- see also \cite{Patton2013}, who use the same definition for their central region. 

To quantify the spatial extent of the baryonic content in our simulated galaxies, we employ 3D radial density profiles encompassing distances between 0 and 10 kpc from the centre of each galaxy. The fourth-row panels of Figure~\ref{fig:terminology} show mass density profiles corresponding to the second- and third-row images directly above them. The vertical line and gray box highlight the centre ($r<1$ kpc). The solid-purple and dashed-gray lines represent the profile corresponding to the interacting and isolated galaxy, respectively. To disentangle the effect of the interactions from secular effects, we calculate the ratio of the two. We show this in the fifth-row panels. The horizontal line represents unity. Values above this line indicate enhancement or excess, whilst those below indicate suppression or deficit. We informally use the word \textbf{\textit{enhancement}} to encompass these terms (e.g., sub-unity `enhancement' means deficit). One must be careful whilst interpreting the meaning of a profile ratio. For instance, from left to right, the profile ratios in the fifth-row panels plummet at large radii. This does not mean that there is no gas there, but rather, that the gas mass content in the interacting galaxy is orders of magnitude below that of its isolated counterpart (purple versus gray lines in the fourth-row panels, and regions immediately inside the larger white circles in the second-row versus third-row images). Similarly, the profile ratio in the centre evolves from strongly enhanced to suppressed. The central deficit at late times does not necessarily mean that there is a `hole' in the inner gas distribution -- rather, it generally means that the original central peak is diminished relative to the central concentration in the secularly-evolving isolated galaxy. Although, occasionally, we do witness brief gas evacuation episodes of the central 100-parsec region, which is then quickly replenished by new gas from the surroundings \citep{Torrey2017}.

It is impractical to analyse every individual radial density profile and profile ratio for 24 galaxy mergers at 5 Myr time resolution, which corresponds to approximately twenty thousand individual profiles per baryonic component in the galaxy-pair period alone! Rather, we employ \textbf{\textit{average density profiles}} and \textbf{\textit{average profile ratios}}, which are the result of averaging profiles (or profile ratios) over multiple times and orbital configurations in our merger suite. For the profile ratios, we match the interacting and isolated galaxies to be situated at the {\it same time} after the start of the simulation. Figure~\ref{fig:profile_definitions} illustrates this. We display the mass in new stars here (rather than gas, as in Figure~\ref{fig:terminology}) because explaining the effectiveness of interaction-induced star formation is one of the central goals of this paper (Section~\ref{subsec:profiles_subsuites}). The secondary galaxy is placed on the left-hand panels, whilst the primary on the right-hand panels. The diagonally-hatched brown bands in the top panels show the average density profiles for galaxies with a companion, whilst the horizontally-hatched gray bands correspond to their isolated counterparts. Band thickness represents one standard deviation. The vertical line and gray box highlight the centre ($r<1$ kpc). Similarly, the bottom panels show the result of averaging profile ratios. The horizontal line indicates unity. Note that the average profile ratio (hatched band, bottom panels) is {\it not} merely the ratio of the average density profiles (brown band divided by gray band, top panels). This is because the average of the ratios is not necessarily equal to the ratio of the averages.

Whilst it is true that the results in Figure~\ref{fig:profile_definitions} offer interesting average trends, teasing out the interaction-induced effects that govern the spatial distribution and evolution of new stars in galaxies is not trivial and depends on many factors. These include (1) the time of observation after the first encounter; (2) the orbital geometry of the encounter; (3) the amount of fuel available; and (4) the star formation efficiency -- to name a few. To address the first point, in Section~\ref{subsec:profiles} we make use of the early, intermediate, and late periods defined above. For the second point, we group our 24 mergers into three subsuites:
\begin{itemize}
    \item The \textbf{\textit{typical subsuite}} (66.7\%, 16/24 mergers, fiducal run included): \{near-prograde\} $\cup$ \{near-polar with first passage at $\sim$16 and $\sim$27 kpc\} $\cup$ \{near-retrograde with first passage at $\sim$7 kpc\}. 
    \item The \textbf{\textit{vigorous subsuite}} (12.5\%, 3/24 mergers): \{near-polar with first pericentric passage at $\sim$7 kpc\}.
    \item The \textbf{\textit{gentle subsuite}} (20.1\%, 5/24 mergers): \{near-retrograde with first passage at $\sim$16 and $\sim$27 kpc\}.
\end{itemize}
We explain these subsuite-naming conventions in Section~\ref{subsec:profiles_subsuites}. To address the third and fourth factors mentioned above, we also split the merger suite by {\it global} SFR enhancement into three populations, or \textbf{\textit{star-forming (SF) types}}:
\begin{itemize}
    \item \textbf{\textit{Enhanced star-formers}}: $\log$ SFR$/$SFR$_{\rm iso} > +0.3$,
    \item \textbf{\textit{Regular star-formers}}: $-0.3 < \log$ SFR$/$SFR$_{\rm iso} < 0.3$, 
  \item \textbf{\textit{Suppressed star-formers}}: $\log$ SFR$/$SFR$_{\rm iso} < -0.3$,
\end{itemize}
where SFR$/$SFR$_{\rm iso}$ denotes {\it global} SFR enhancement. We note that any particular run can `visit' all three SF-type regimes at different times during the interaction. We justify our global SFR cuts in Section~\ref{subsec:driving_global_sfr}, where we investigate the connection between location relative to the global SFR$/$SFR$_{\rm iso}=1$ line and the radial structure of SFR, SFE, and available fuel. 
Lastly, in Section~\ref{subsec:fuel} we define
\begin{itemize}
    \item \textcolor{black}{\textbf{\textit{Efficiency-driven enhanced star formation:}}}\\   \textcolor{black}{${\rm \,\,\,\,\,\,\,\,\,\,\,\, SFE}/{{\rm SFE}_{\rm iso}} > {M_{\rm cold-dense}}/{M_{\rm cold-dense, \, iso}}$,}
    \item  \textcolor{black}{\textbf{\textit{Efficiency-driven suppressed star formation:}}}\\   \textcolor{black}{${\rm \,\,\,\,\,\,\,\,\,\,\,\, SFE}/{{\rm SFE}_{\rm iso}} < {M_{\rm cold-dense}}/{M_{\rm cold-dense, \, iso}}$,}
    \item  \textcolor{black}{\textbf{\textit{Fuel-driven enhanced star formation:}}}\\  \textcolor{black}{${\rm \,\,\,\,\,\,\,\,\,\,\,\, SFE}/{{\rm SFE}_{\rm iso}} < {M_{\rm cold-dense}}/{M_{\rm cold-dense, \, iso}}$,}
    \item  \textcolor{black}{\textbf{\textit{Fuel-driven suppressed star formation:}}}\\  \textcolor{black}{${\rm \,\,\,\,\,\,\,\,\,\,\,\, SFE}/{{\rm SFE}_{\rm iso}} > {M_{\rm cold-dense}}/{M_{\rm cold-dense, \, iso}}$,}

\end{itemize}
to evaluate the relative contribution of SFE and available cold-dense fuel in driving SFR in the central kiloparsec.

\begin{figure*}
\centerline{\vbox{
\hbox{
\includegraphics[width=3.5in]{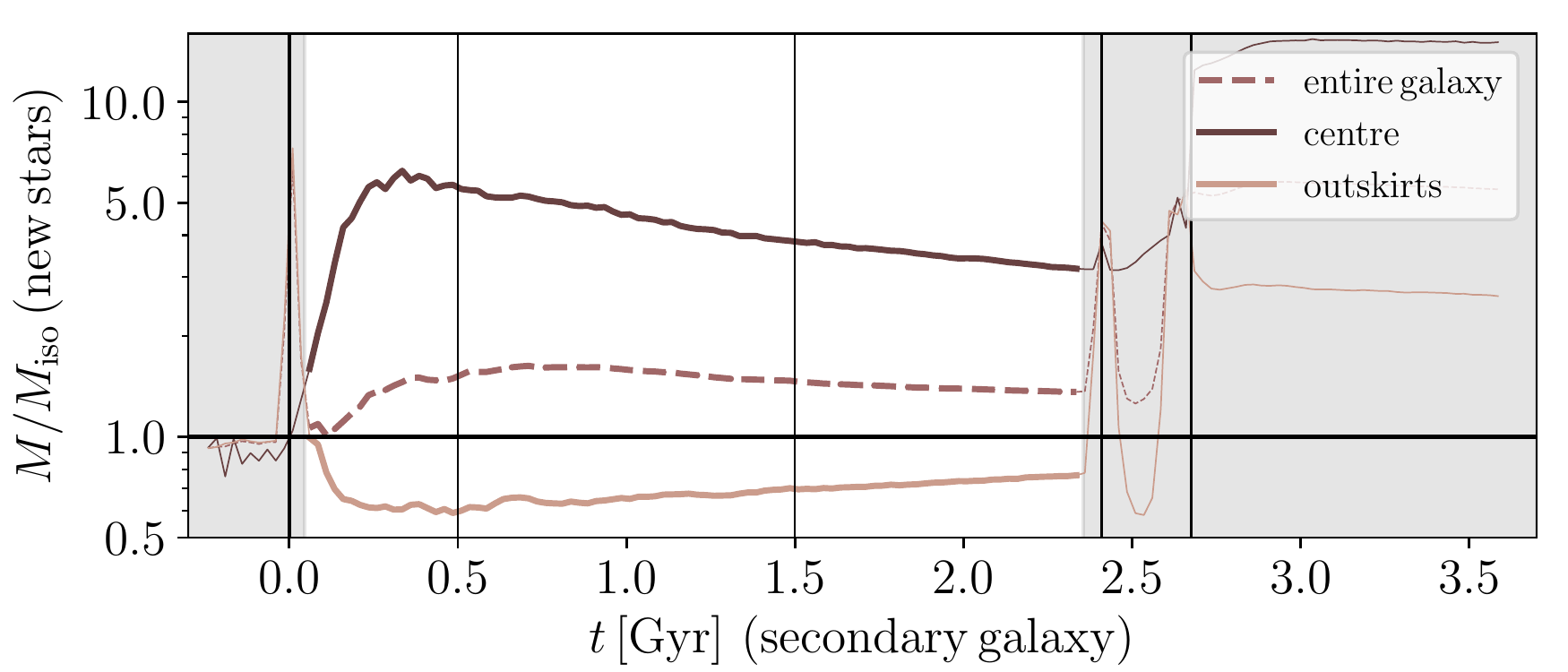}
\includegraphics[width=3.5in]{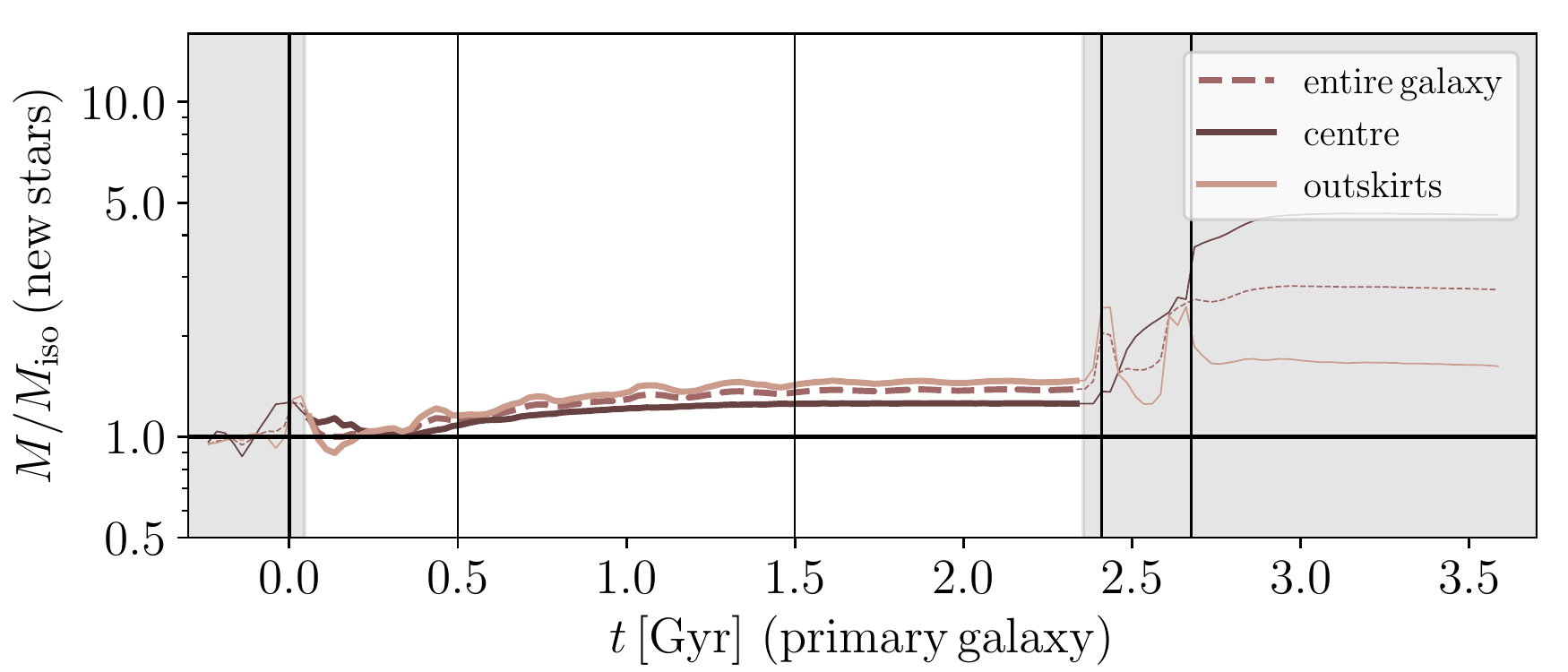}
}
\vspace{-.033in}
\hbox{
\includegraphics[width=3.5in]{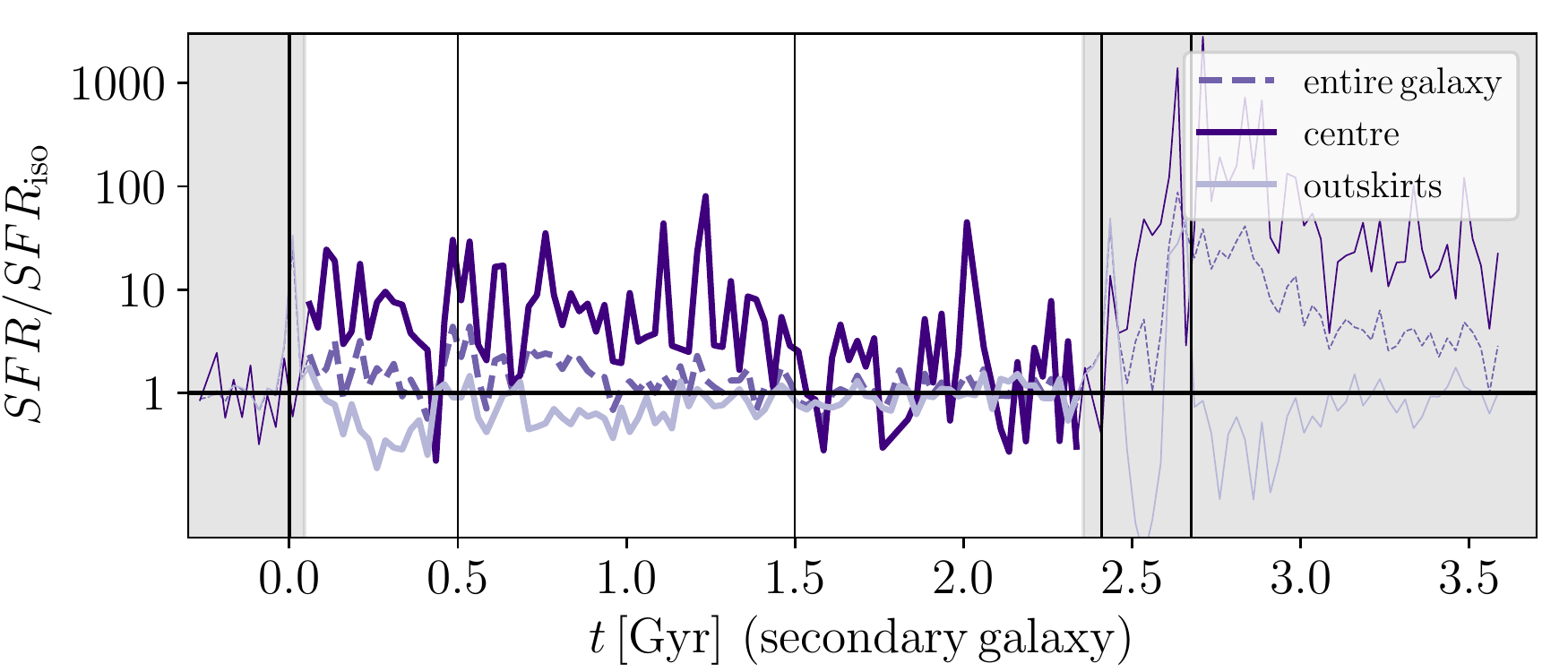}
\includegraphics[width=3.5in]{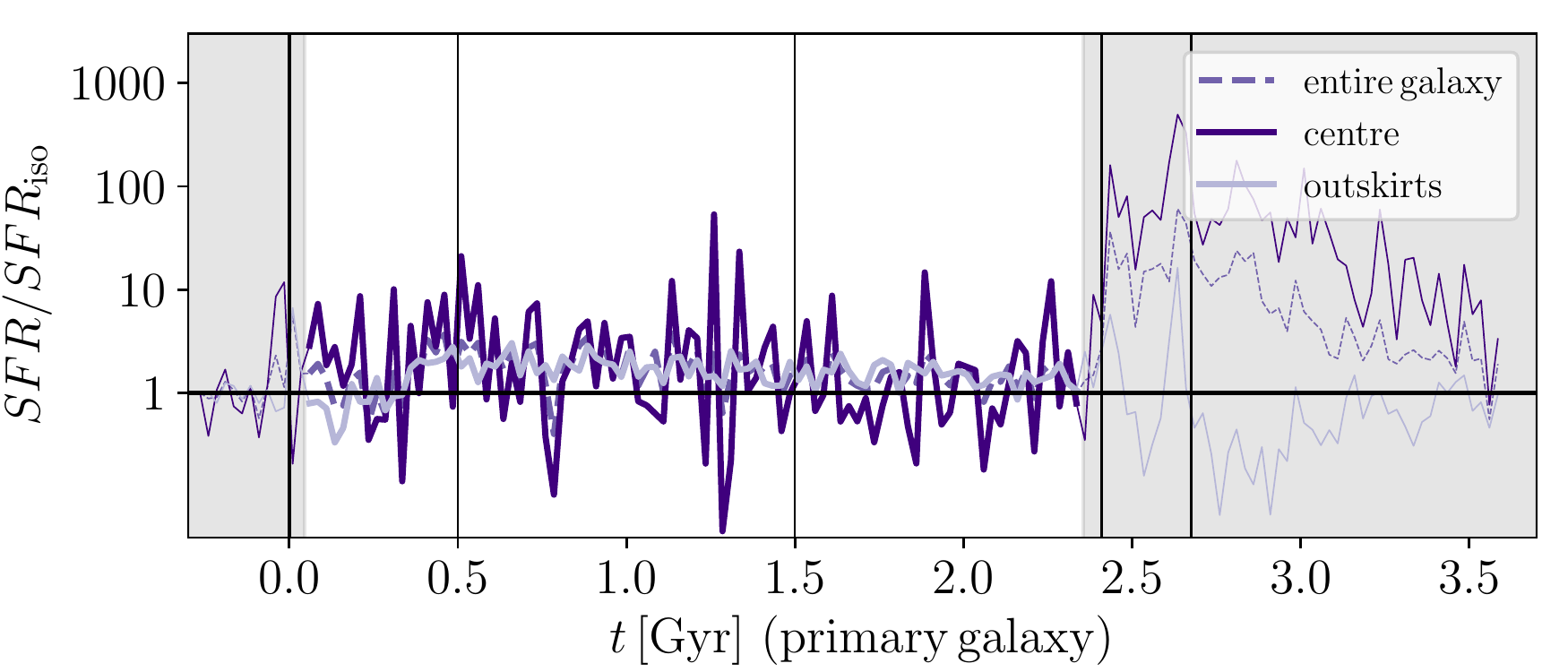}
}
\vspace{-.033in}
\hbox{
\includegraphics[width=3.5in]{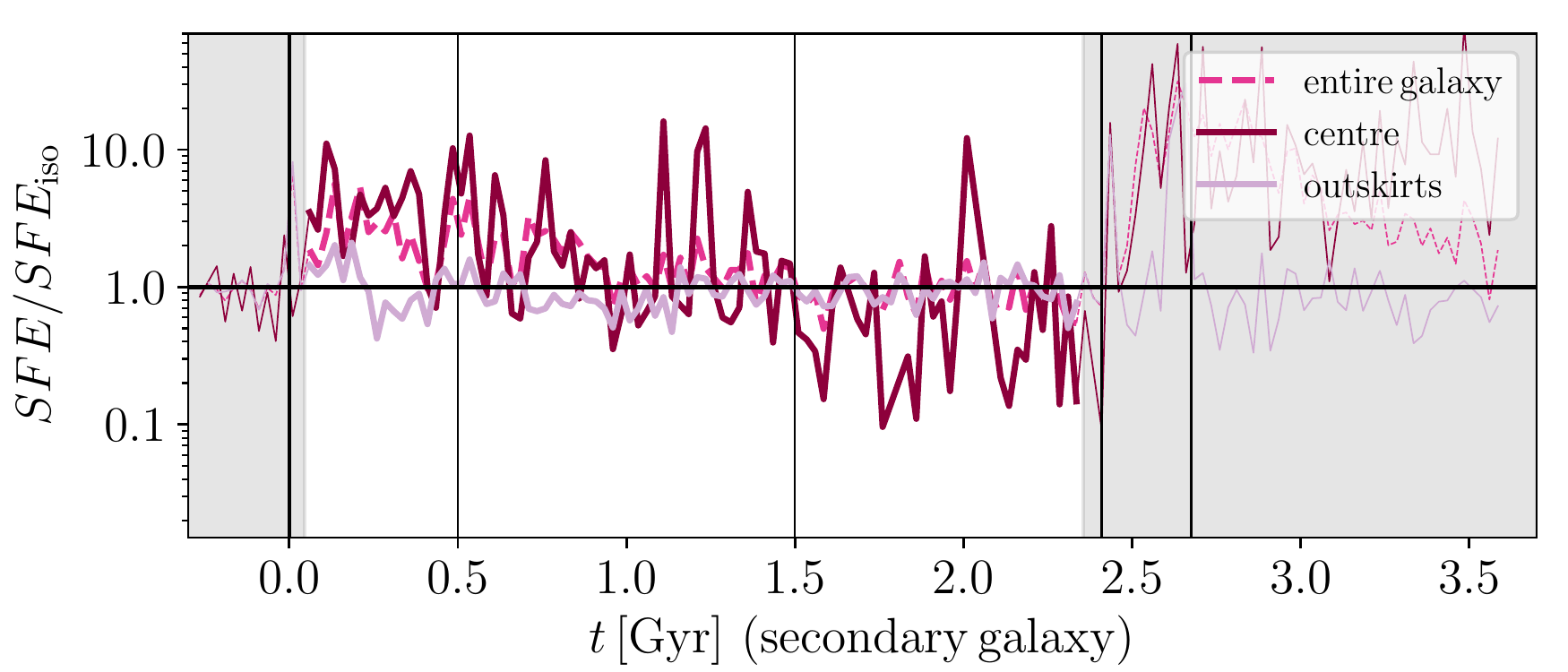}
\includegraphics[width=3.5in]{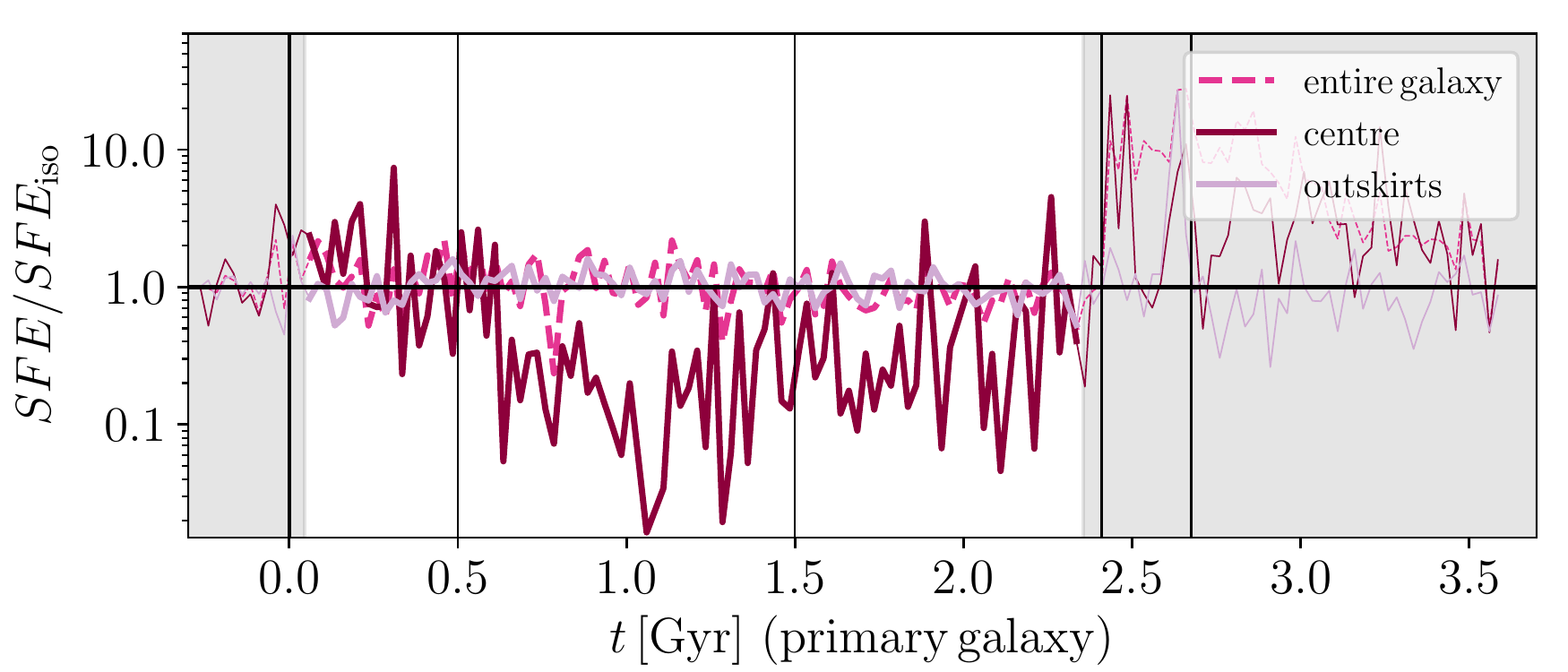}
}
\vspace{-.033in}
\hbox{
\includegraphics[width=3.5in]{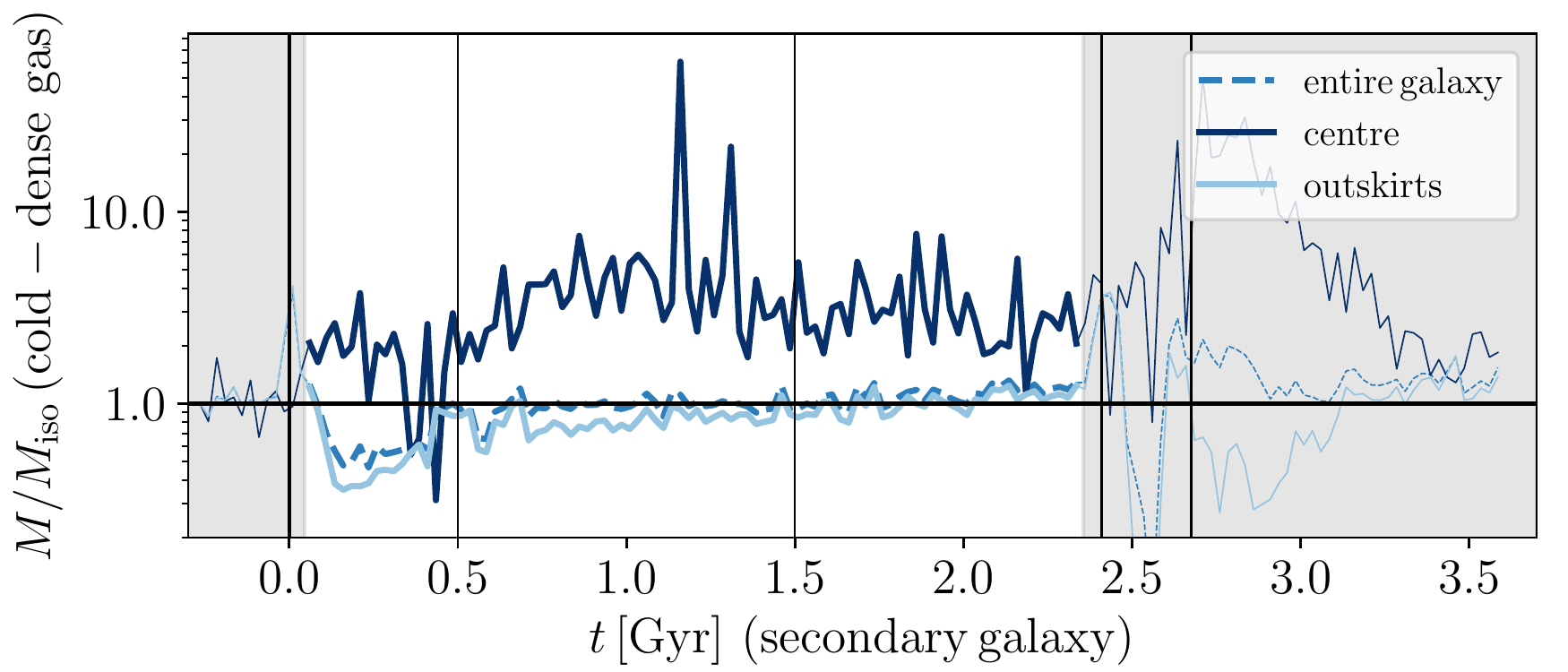}
\includegraphics[width=3.5in]{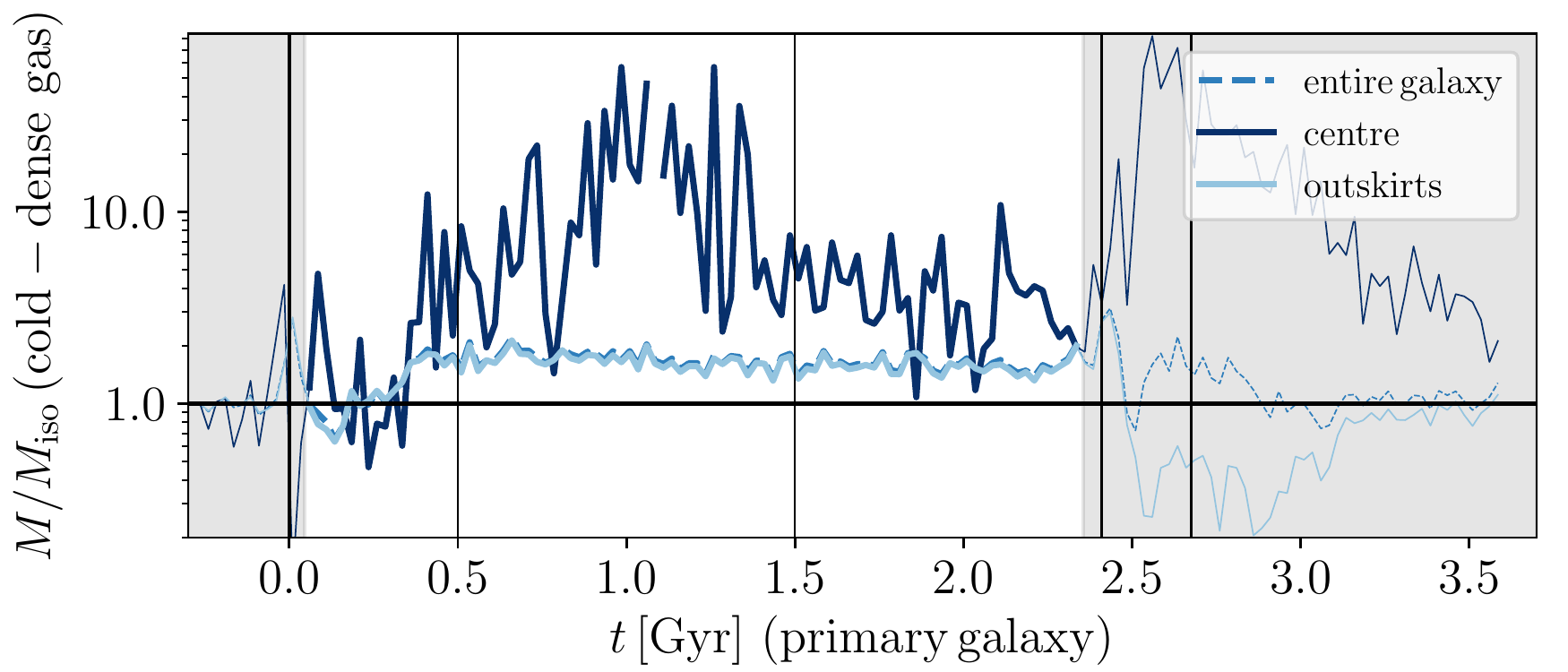}
}
\vspace{-.03in}
\hbox{
\includegraphics[width=3.5in]{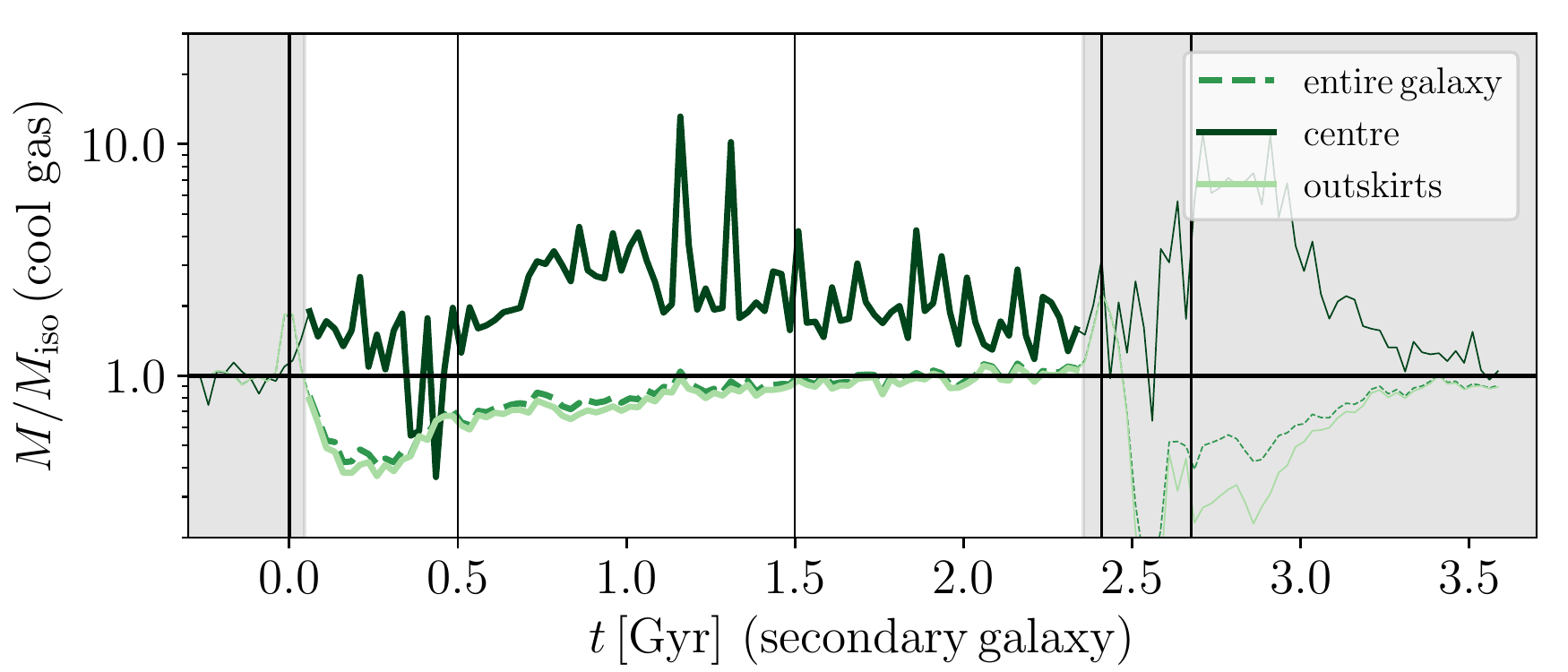}
\includegraphics[width=3.5in]{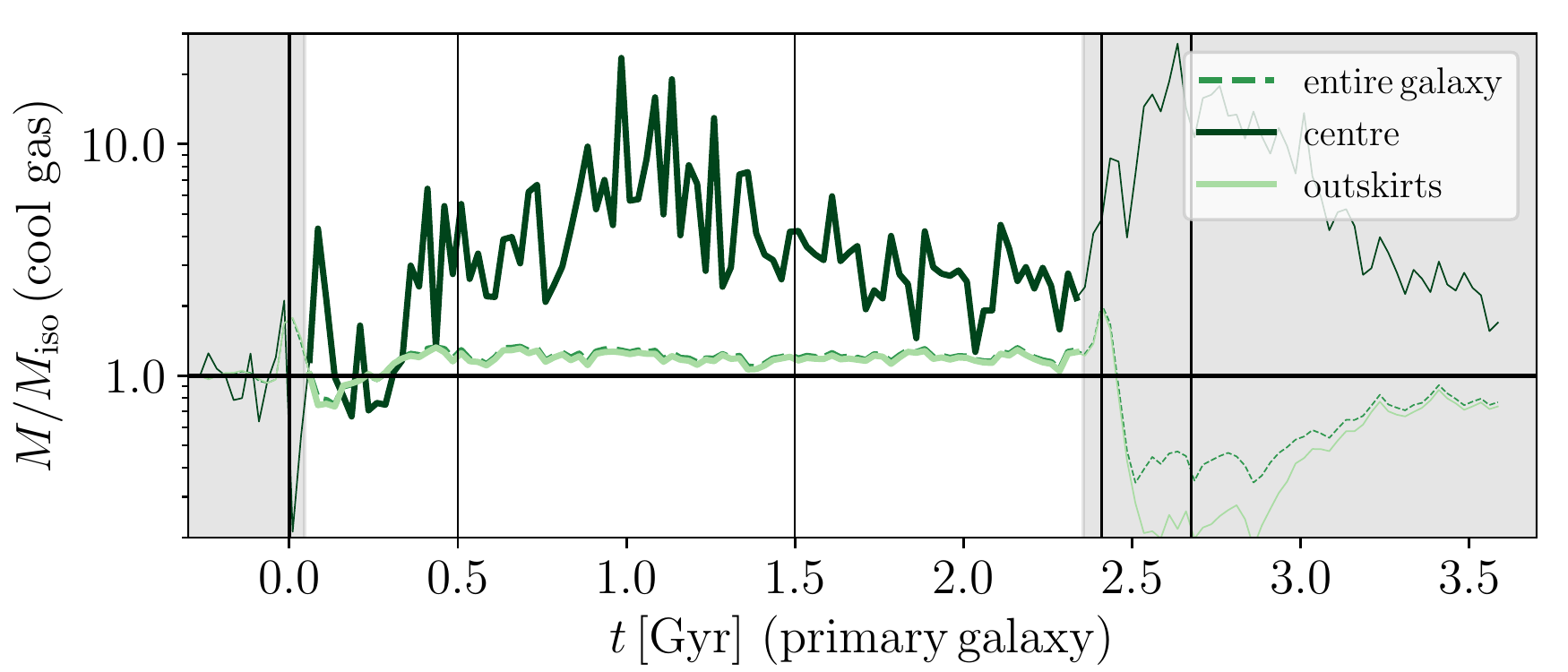}
}
\vspace{-.02in}
}}
\caption{Time evolution of the fiducial run. {\it Left (right) panels}: secondary (primary) galaxy. {\it Top-to-bottom panels}: new stellar mass (brown), SFR (purple), SFE (pink), cold-dense gas mass (blue), and cool gas mass (green) enhancement. The vertical scales are different for each row. The dark solid lines represent the centre (0$-$1 kpc), the light solid line represent the outskirts (1$-$10 kpc), and the medium dashed lines represent the entire galaxy (0$-$10 kpc). Time is shifted to zero at first pericentric passage. The thick vertical lines from left to right indicate first and second pericentric passages, plus coalescence. The gray boxes and thin portions of the coloured curves represent times outside the galaxy-pair period. The thin vertical lines split the galaxy-pair period into the early (0$-$0.5 Gyr), intermediate (0.5$-$1 Gyr), and late periods ($>$1 Gyr). The horizontal line indicates unity. 
}
\label{fig:time_evolution_fiducial}
\end{figure*}

\begin{figure*}
\centerline{\vbox{
\hbox{
\includegraphics[width=3.5in]{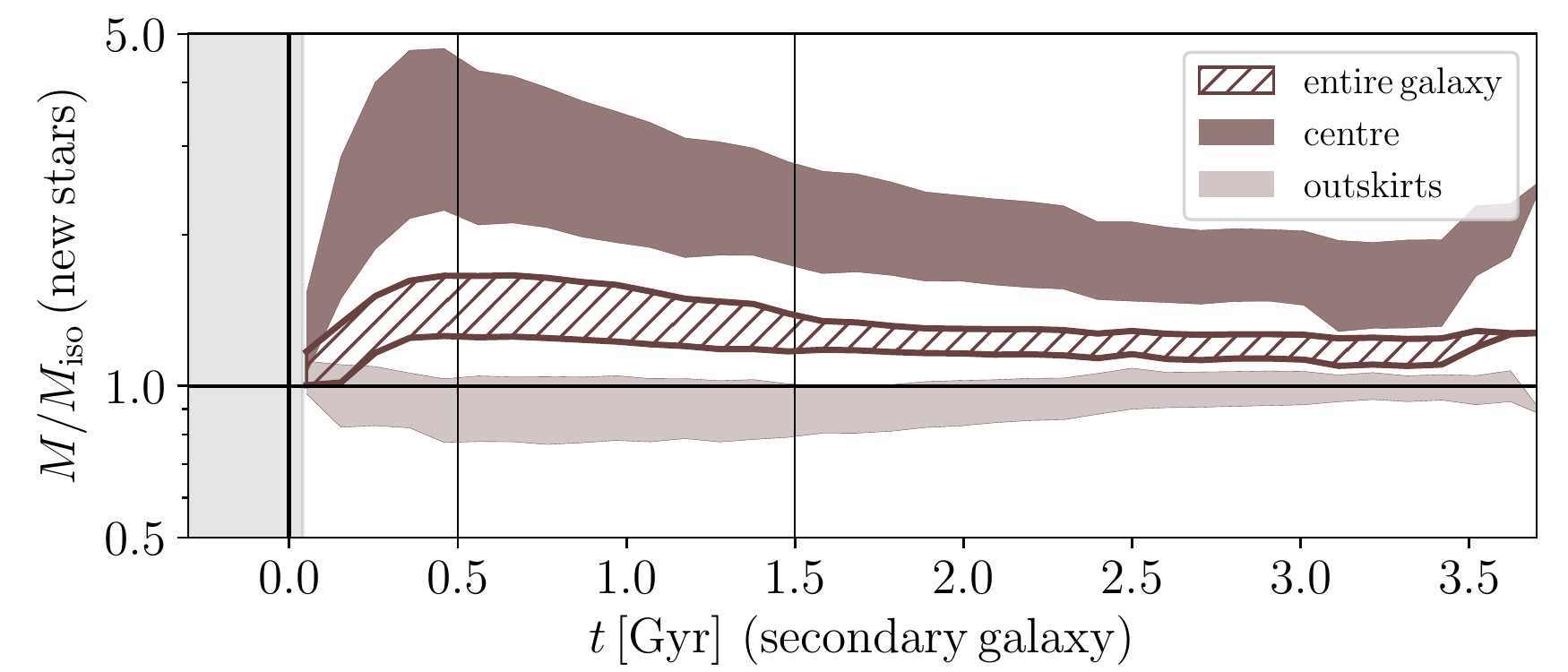}
\includegraphics[width=3.5in]{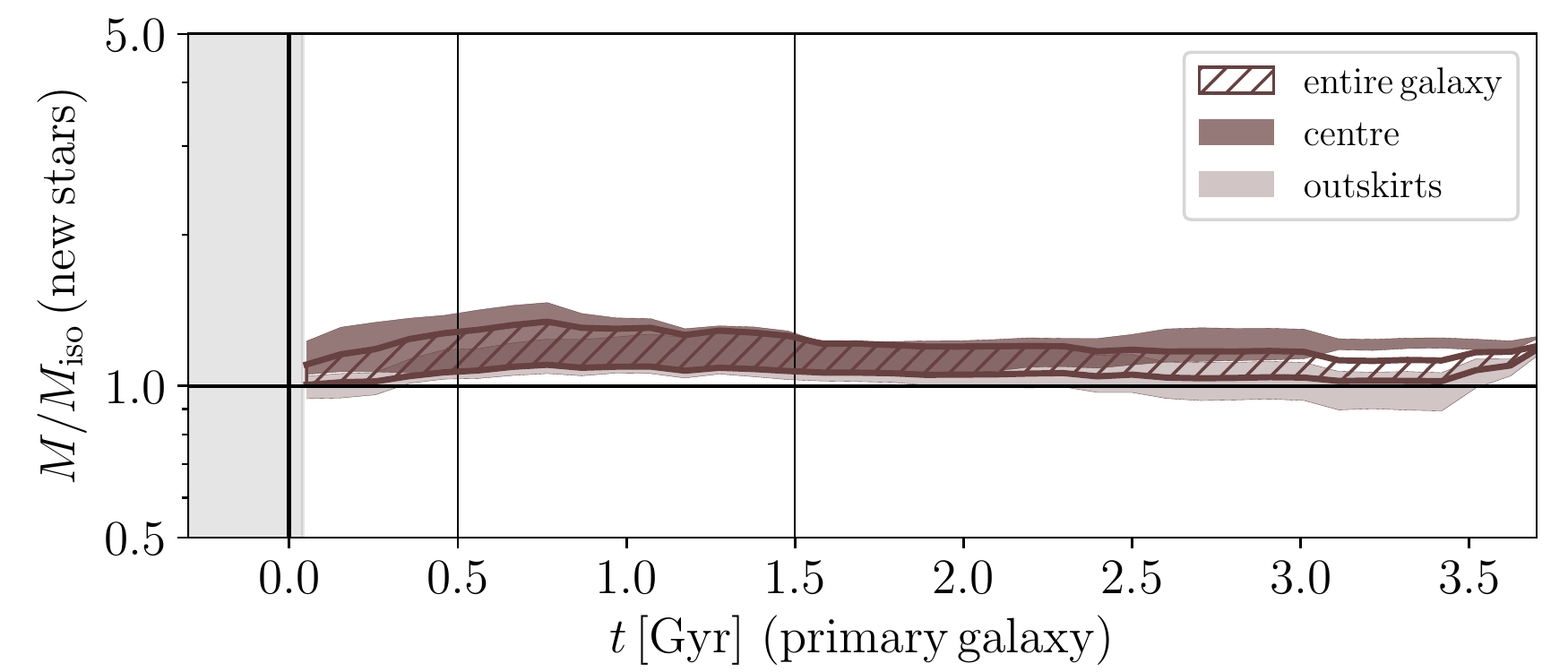}
}
\vspace{-.033in}
\hbox{
\includegraphics[width=3.5in]{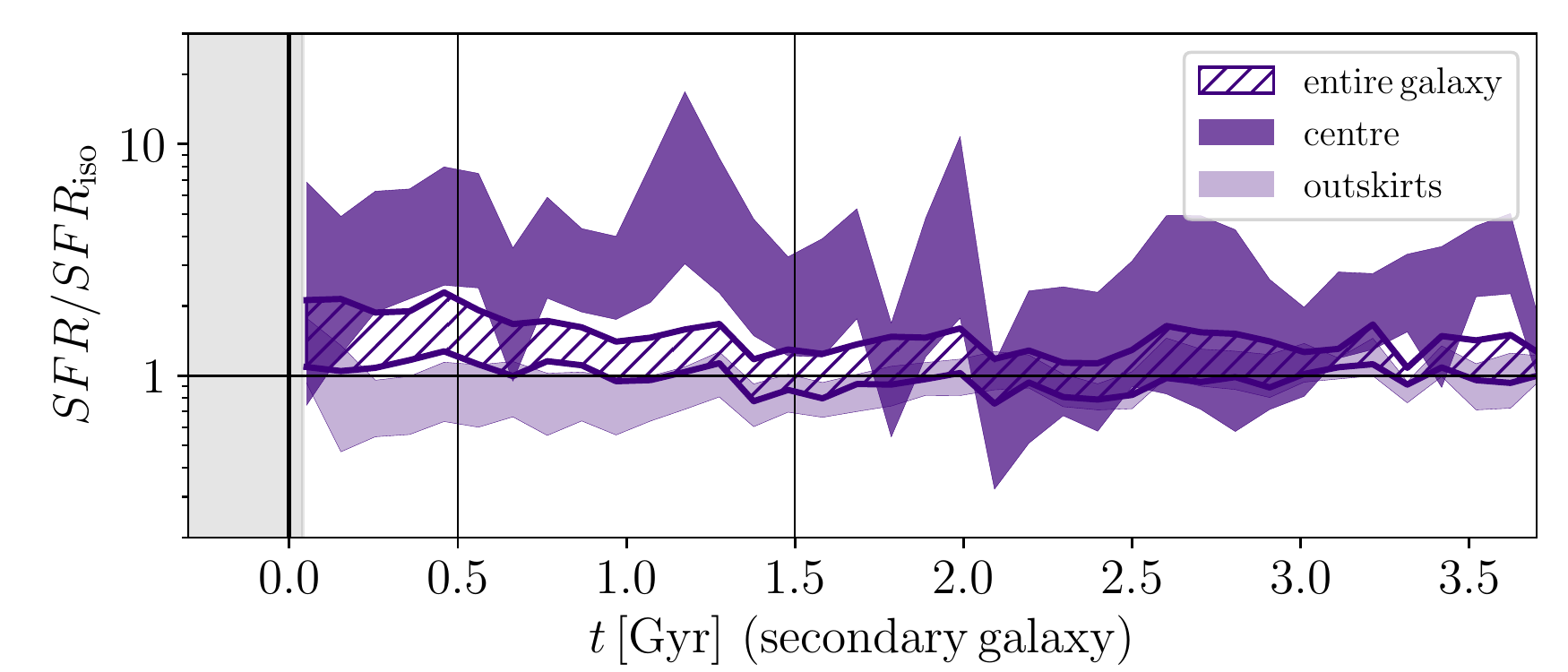}
\includegraphics[width=3.5in]{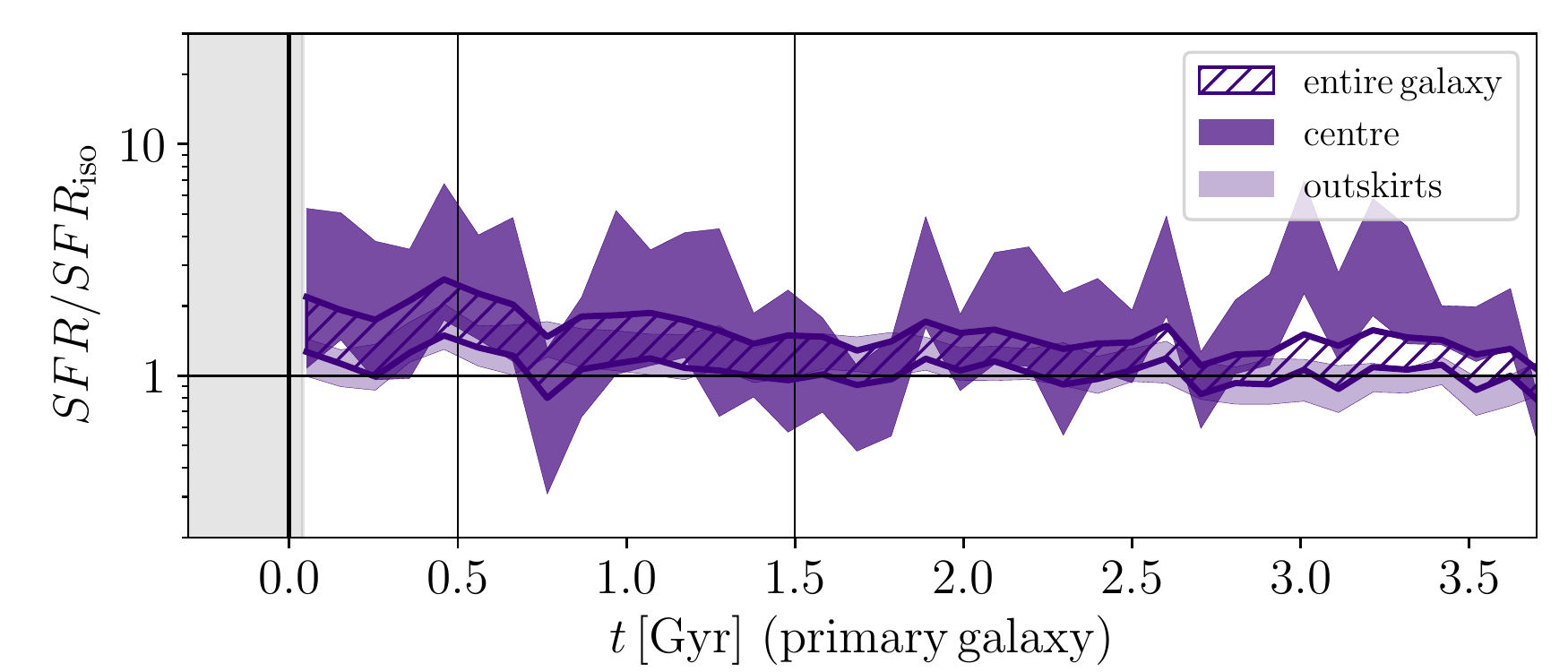}
}
\vspace{-.033in}
\hbox{
\includegraphics[width=3.5in]{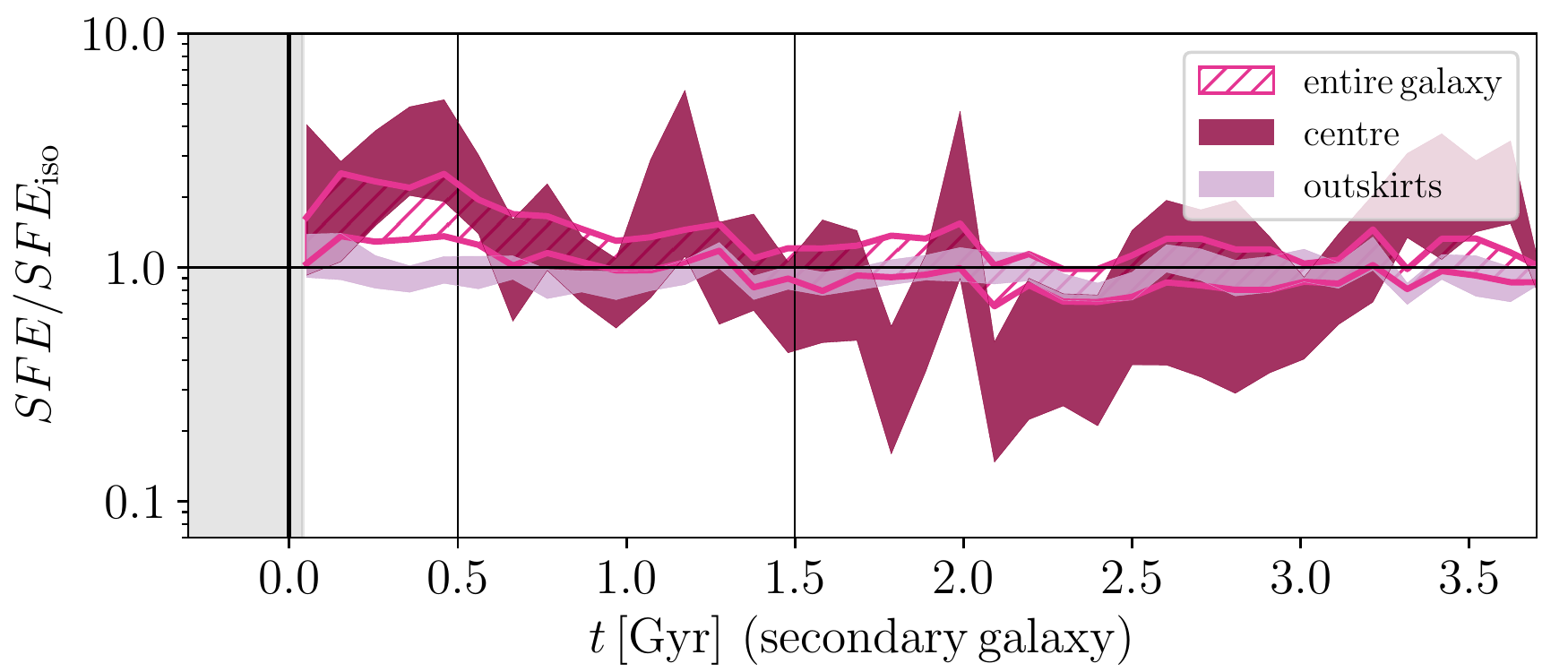}
\includegraphics[width=3.5in]{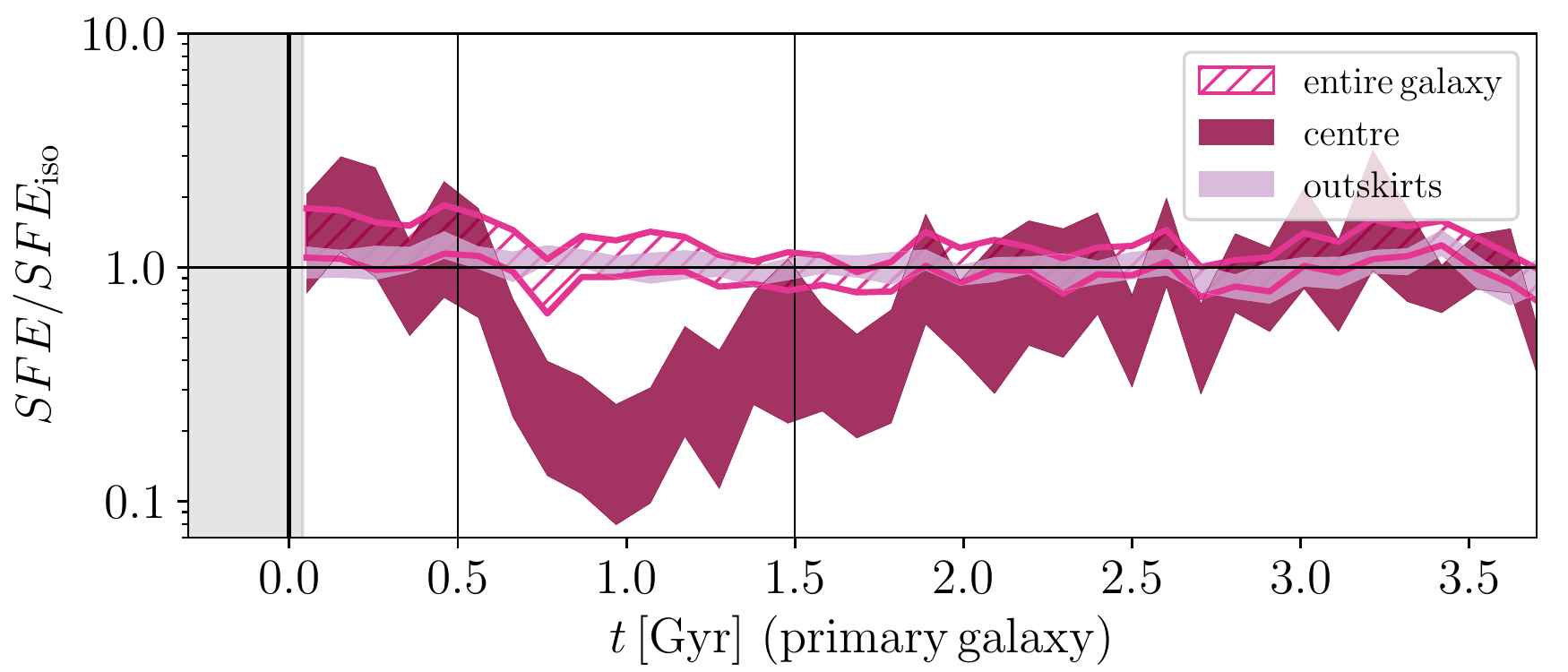}
}
\vspace{-.033in}
\hbox{
\includegraphics[width=3.5in]{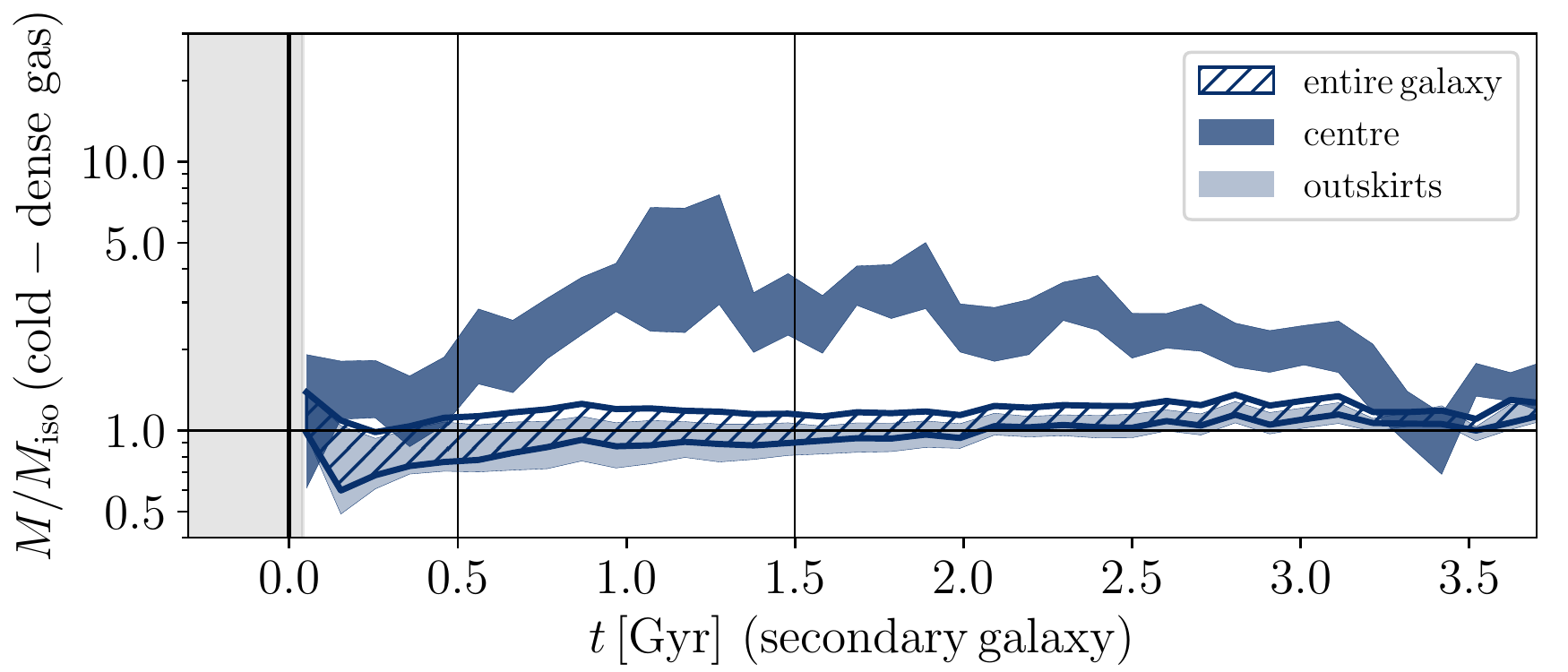}
\includegraphics[width=3.5in]{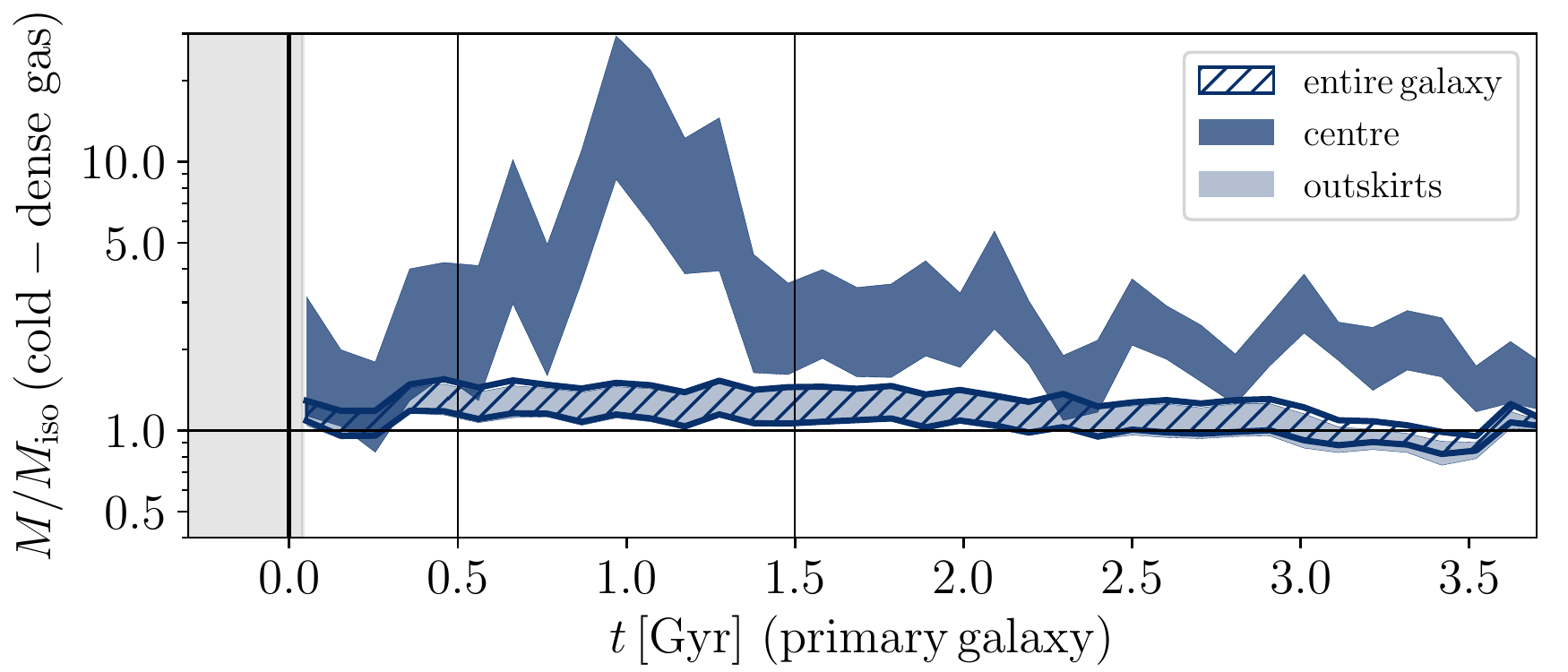}
}
\vspace{-.033in}
\hbox{
\includegraphics[width=3.5in]{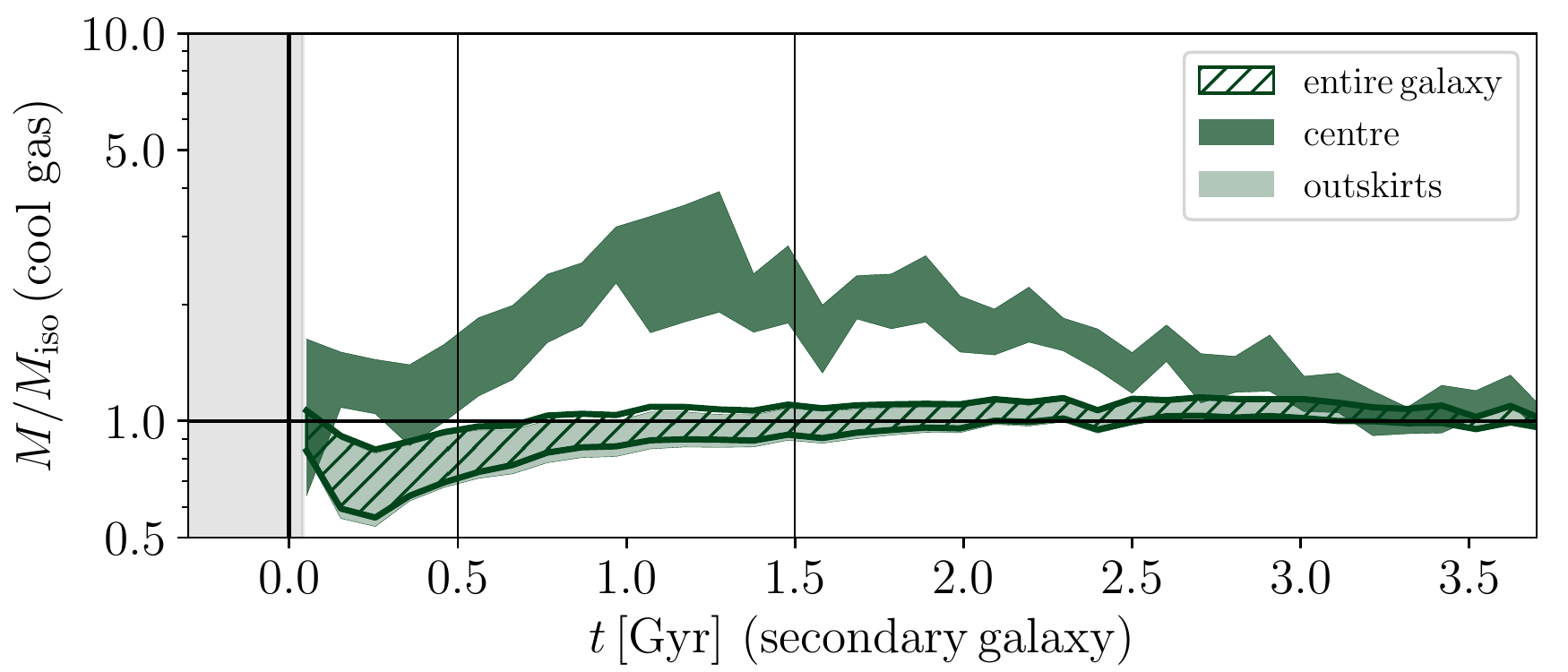}
\includegraphics[width=3.5in]{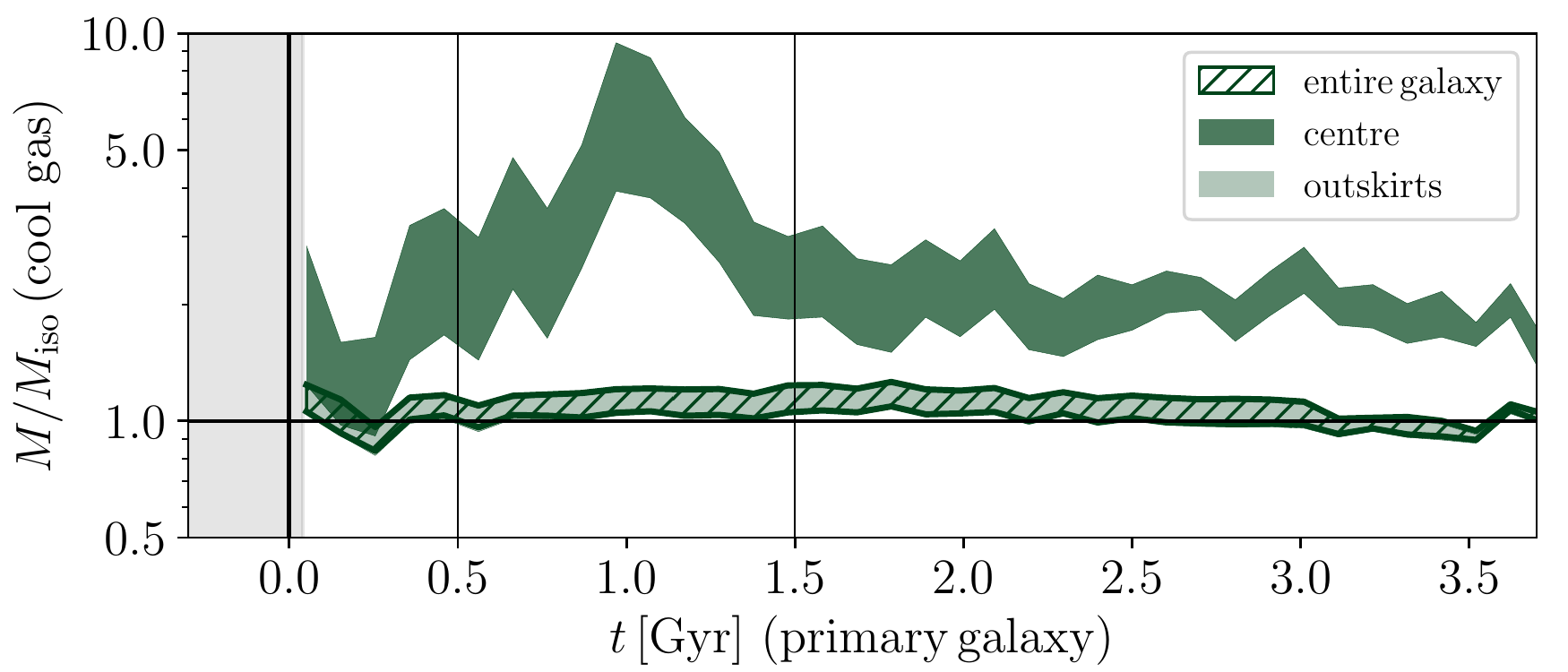}
}
\vspace{-.1in}
}}
\caption{Time evolution for the entire merger suite. {\it Left (right) panels}: secondary (primary) galaxy. {\it Top-to-bottom panels}: new stars (brown), SFR (purple), SFE (pink), cold-dense gas (blue), and cool gas (green). The vertical scales are different for each row. The dark coloured solid bands represent the centre (0$-$1 kpc), the light solid bands represent the outskirts (1$-$10 kpc), and the hatched bands represent the entire galaxy (0$-$10 kpc). Band thickness corresponds to one standard deviation. Time is shifted to zero at first pericentric passage (thick vertical line). The gray box represents times outside the galaxy-pair period, defined to be between first and second pericentric passage, with separation greater than 20 kpc. We do not include lines corresponding to second pericentric passage and coalescence, nor the gray box indicating times after the galaxy-pair period, because these vary from merger to merger in the suite. The thin vertical lines split the galaxy-pair period into the early (0$-$0.5 Gyr), intermediate (0.5$-$1 Gyr), and late periods ($>$1 Gyr). The horizontal line indicates unity.  
}
\label{fig:time_evolution_suite}
\end{figure*}

\section{Results \& Discussion}
\label{sec:results_and_discussion}

\subsection{Time evolution}
\label{subsec:time_evolution}

Before discussing the spatial extent of the new stars and the cold-dense$/$cool ISM phases, Figure~\ref{fig:time_evolution_fiducial} addresses the time evolution of our fiducial run in three regions: globally (within 10 kpc, medium-coloured dashed lines), in the centre (within 1 kpc, dark solid lines), and in the outskirts (between 1$-$10 kpc, light solid lines). We focus on the galaxy-pair period (thick portions of the coloured lines that are not masked by gray boxes). The secondary galaxy (left panels) exhibits the following behaviour. The mass in new stars in the centre (dark brown) is enhanced suddenly and to high levels within the first $\sim$0.3 Gyr after the first encounter, and decreases at later times as stellar-mass growth in the interacting galaxy slows down relative to the isolated one. This growth is driven by the generally strongly-enhanced SFR (dark purple) during the galaxy-pair period. We note that the connection between new-stellar mass enhancement and instantaneous SFR enhancement is {\it not} trivial:
\begin{equation}
\label{eqn:mnew_vs_sfr}
\frac{M_{\rm new \,\, stars}(t)}{M_{\rm new \,\, stars, \,\, iso}(t)} \equiv \frac{\int^{t}_0 {\rm SFR}(t') {\rm \, d}t'}{\int^{t}_0 {\rm SFR}_{\rm iso}(t') {\rm \, d}t'} \neq \int^{t}_0 \frac{{\rm SFR}(t')}{{\rm SFR_{\rm iso}}(t')}{\rm \, d}t'.
\end{equation}
Changes in nuclear SFR are the result of the combination of changes in SFE (dark pink) and available cold-dense gas fuel (dark blue). The centre experiences strong enhancement in both cold-dense and cool (dark green) gas. SFE in this region is generally enhanced, especially at early times, but reaches sub-unity levels at later times. The outskirts exhibit a deficit in new stellar mass (light brown), caused by suppressed SFR (light purple) during the first $\sim$1.2 Gyr. SFR recovers eventually, but this is not enough for the mass in new stars to catch up with the isolated galaxy. Generally, this SFR deficit is caused by suppression in both SFE (light pink) and cold-dense gas mass (light blue). The mass content in both cold-dense gas and cool gas (light green) is suppressed for most of the galaxy-pair period, with only a mild recovery at late times.

Unlike the secondary galaxy, the primary (right panels) exhibits new-stellar mass and SFR enhancement in both the centre and the outskirts. This enhancement is weak in the outskirts, and weaker in the centre. Even though the centre of this galaxy experiences stronger levels of mass enhancement in cold-dense gas (and cool-gas), compared to the centre of the secondary galaxy -- especially at intermediate times -- SFE is strongly suppressed in that region. Section~\ref{subsec:fuel} explores the connection between SFR, SFE and cold-dense gas mass in more detail. In the outskirts, enhanced \textcolor{black}{SFR} is driven by an elevated presence of cold-dense gas, whilst SFE remains close to unity. In other words, both galaxies experience significant increases in cool$/$cold-dense gas, but SFR does not always increase accordingly.

Figure~\ref{fig:time_evolution_suite} tells a similar story for the entire merger suite. The coloured bands are the result of taking the average of quantities described by the coloured lines in Figure~\ref{fig:time_evolution_fiducial} over our 24 mergers within their respective galaxy-pair periods. Band thickness represents one standard deviation. Note that these panels do not include lines corresponding to second pericentric passage and coalescence. This is because the specific timing of these events varies from merger to merger in the suite. In other words, the late-time regime in this Figure is likely to be dominated by contributions from long-lived interactions. See Figure~4 of \cite{Moreno2019} for the diversity in duration and separation extend experienced by the mergers in our suite. In comparing Figures~\ref{fig:time_evolution_fiducial} and \ref{fig:time_evolution_suite}, one notable difference is that, although the fiducial run experiences suppression of star formation between 1$-$10 kpc, this effect is not statistically significant when we average over a diverse set of interacting orbits.

Our SFRs are bursty, which is common in our \textsc{fire} simulations \citep{Sparre2017,Orr2017bursty,Emami2019,Flores2020}. This is strikingly different to predictions by older effective equation-of-state (EOS) models with star-formation recipes tuned to match either the \cite{Kennicutt1998} law \citep{Cox2006,Hayward2014arepo,Moreno2015} or a narrow range of SFRs in their isolated galaxies \citep{DiMatteo2007,DiMatteo2008}. In such models, SFR is gentle (non-bursty) and, after first passage, global SFR is initially enhanced, and becomes heavily suppressed at late times. We do not witness this behaviour in our simulations: at most, global SFR enhancement declines to unity at late times. Making more detailed comparisons is beyond the scope of this paper. We point the interested reader to \cite{Hopkins2013mergers}, who conduct such a comparison between EOS models and a (pre-\textsc{fire}) physics model similar to ours with resolved star formation and feedback.

\subsection{Average sample-wide enhancements}
\label{subsec:averages}

\begin{figure}
\centerline{\vbox{
\hbox{
\includegraphics[width=3.5in]{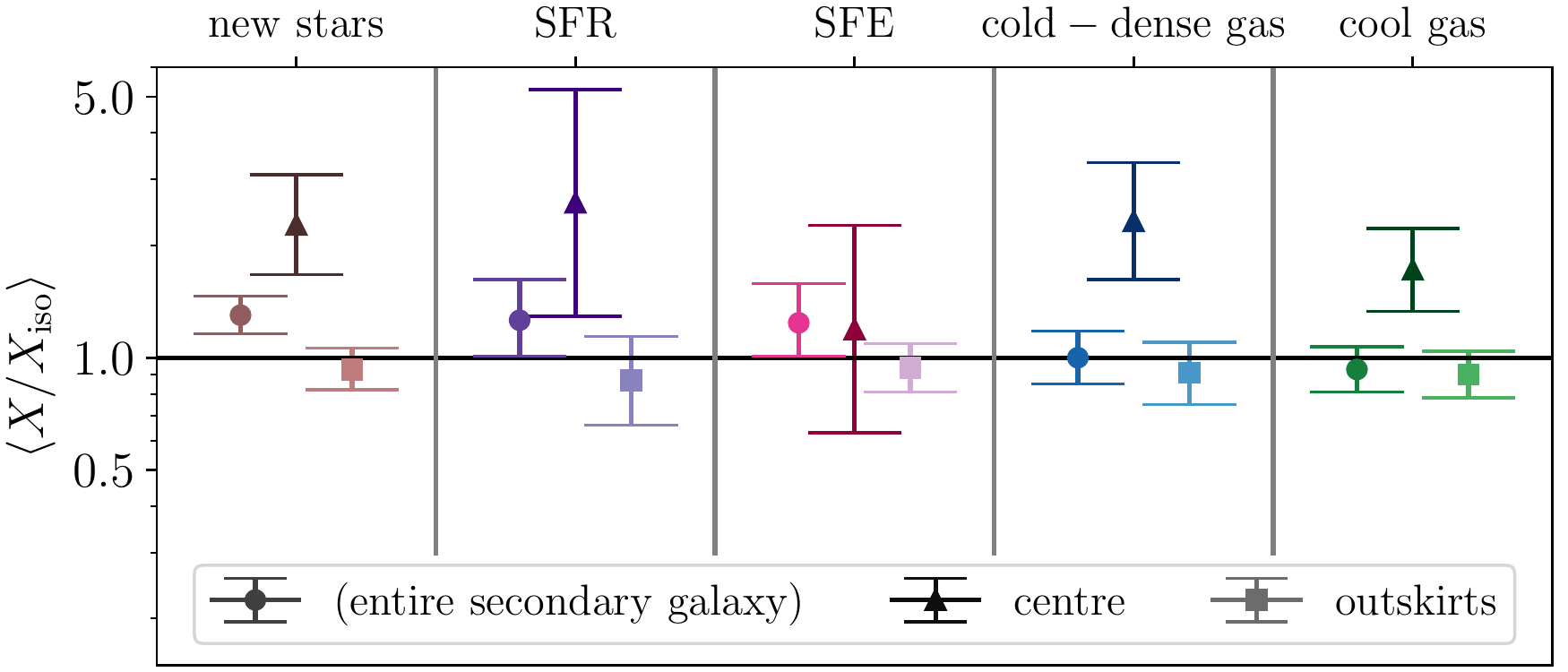}
}
\vspace{-.01in}
\hbox{
\includegraphics[width=3.5in]{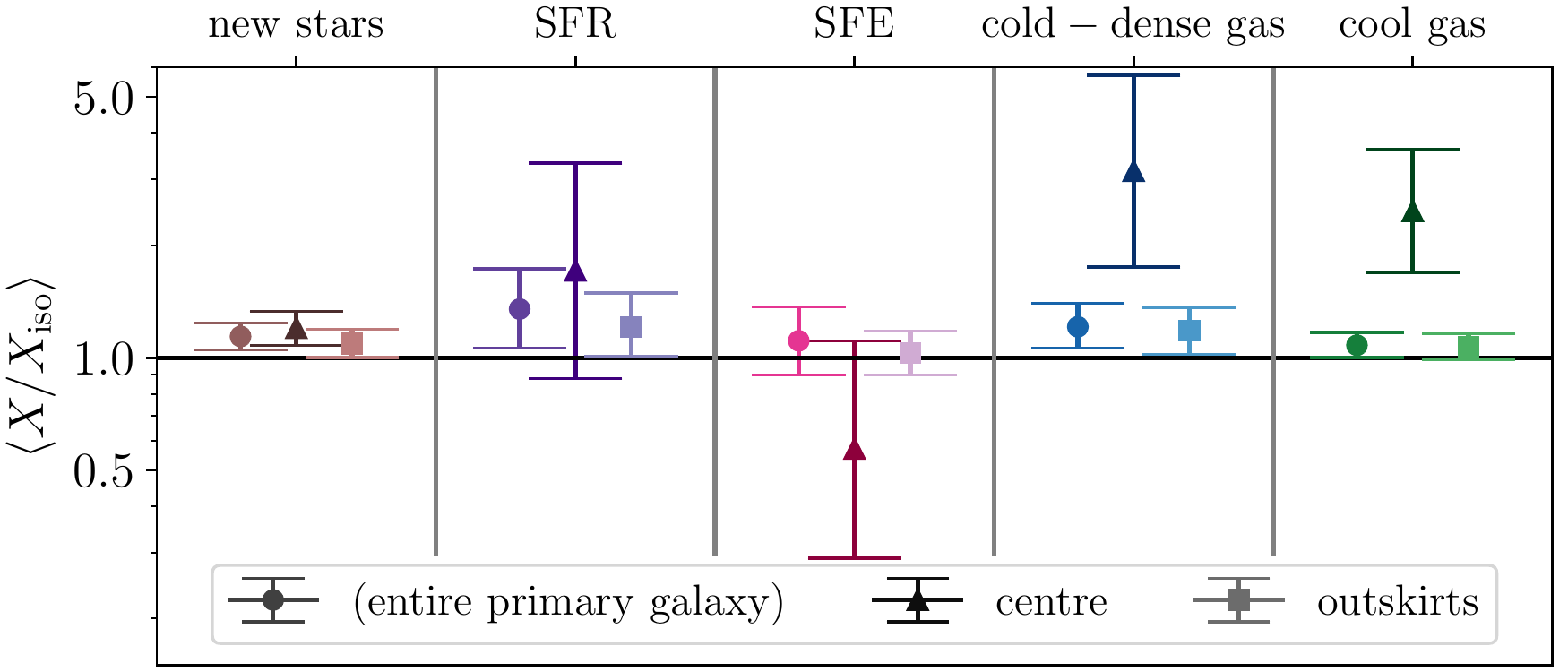}
}
\vspace{-.1in}
}}
\caption{Average sample-wide enhancements across time (within the galaxy-pair period), radial distance (within indicated region), and runs in the merger suite. {\it Top (bottom) panel}: Secondary (primary) galaxy. Quantity $X$ represents mass, SFR, or SFE. {\it Left-to-right} symbol clusters: new stars mass (brown), SFR (purple), SFE (pink), cold-dense gas mass (blue), and cool gas mass (green). {\it Left-to-right} within each symbol cluster: entire galaxy (0$-$10 kpc, medium-tone filled circles), centre (0$-$1 kpc, dark filled triangles), and outskirts (1$-$10 kpc, light filled squares). \textcolor{black}{Error bars indicate 1$\sigma$ standard deviations.} The vertical gray lines separate symbol clusters. The horizontal black line denotes unity.
}
\label{fig:averages}
\end{figure}

\begin{table}
  \begin{center}
    \begin{tabular}{l|l|l|l} 
    \hline \hline
     Quantity & Entire galaxy & \,\, Centre & \, Outskirts  \\
              & \, (0$-$10 kpc)  & (0$-$1 kpc) & (1$-$10 kpc) \\
     \hline \hline
     {\bf Secondary galaxy}: & & & \\
     \hline
     New stars      & $30^{+16}_{-14}$ \% & $127^{+82}_{-50}$   \% & $-7^{+13}_{-11}$  \% \\ [1.5ex]
     SFR            & $26^{+36}_{-25}$ \% & $160^{+263}_{-131}$ \% & $-13^{+27}_{-21}$ \% \\ [1.5ex]
     SFE            & $24^{+34}_{-23}$ \% & $19^{+107}_{-56}$   \% & $-6^{+15}_{-13}$  \% \\ [1.5ex]
     Cold-dense gas & $0^{+18}_{-15}$  \% & $132^{+101}_{-70}$  \% & $-9^{+19}_{-16}$  \% \\ [1.5ex]
     Cool gas       & $-7^{+24}_{-12}$ \% & $72^{+50}_{-39}$    \% & $-10^{+14}_{-12}$ \% \\ [0.1ex]
     \hline \hline
     {\bf Primary galaxy}: & & & \\
     \hline
     New stars      & $14^{+10}_{-9}$  \% & $20^{+13}_{-12}$    \% & $9^{+10}_{-8}$   \% \\ [1.5ex]
     SFR            & $35^{+38}_{-29}$ \% & $71^{+161}_{-83}$   \% & $21^{+28}_{-20}$  \% \\ [1.5ex]
     SFE            & $11^{+26}_{-21}$ \% & $-43^{+54}_{-28}$   \% & $3^{+10}_{-13}$   \% \\ [1.5ex]
     Cold-dense gas & $21^{+19}_{-15}$ \% & $216^{+255}_{-141}$ \% & $18^{+15}_{-16}$ \% \\ [1.5ex]
     Cool gas       & $8^{+9}_{-8}$    \% & $147^{+115}_{-78}$  \% & $7^{+9}_{-8}$    \% \\
  [0.1ex]
     \hline \hline
    \end{tabular}
  \end{center}
\caption{Average sample-wide enhancements. {\it Top (bottom)}: Secondary (primary) galaxy. {\it Top-to-bottom}: new stellar mass, SFR, SFE, cold-dense gas mass, and cool gas mass. {\it Left-to-right}: entire galaxy (0-10 kpc), centre (0-1 kpc), and outskirts (1-10 kpc). \textcolor{black}{Errors refer to 1$\sigma$ standard deviations.} Figure~\ref{fig:averages} displays these data.
}
\label{table:averages}
\end{table}

Before studying interaction-induced changes in the radial structure of galaxies, this section reports sample-wide averages. Figure~\ref{fig:averages} displays average enhancements across our merger suite. We calculate these values by marginalising time in the quantities displayed in Figure~\ref{fig:time_evolution_suite}. \textcolor{black}{The top panel shows results for the secondary galaxy, and the bottom panel for the primary galaxy.} Medium-coloured circles, dark triangles, and light squares represent the entire galaxy, the centre, and the outskirts respectively. \textcolor{black}{Error bars indicate 1$\sigma$ standard deviations.} Table~\ref{table:averages} presents these values in percentage format. 

During the interaction, the secondary galaxy (Figure~\ref{fig:averages}, top panel) contains, on average, $\sim$30\% more mass in new stars relative to its isolated counterpart. When we focus on the central region, this level of enhancement increases to $\sim$130\%. Some orbital configurations (e.g., our fiducial run) exhibit suppression in the outskirts (defined here as the region spanning radii of 1$-$10 kpc) -- but on average, this effect is not statistically significant. Instantaneous SFR is mildly enhanced globally ($\sim$25\%), and strongly enhanced in the centre ($\sim$160\%). SFR in the outskirts is, on average, statistically consistent with unity. See equation~(\ref{eqn:mnew_vs_sfr}) for an explanation of why average enhancements in SFR and new-stellar mass are not necessarily identical. SFE is marginally enhanced across the entire galaxy ($\sim$25\%), but consistent with unity when the centre and outskirts are considered separately. For the cold-dense and cool gas mass budgets, there is also no statistically significant deviation from unity within the outskirts or the entire galaxy. On the other hand, the centre experiences intense levels of mass enhancement: $\sim$130\% in cold-dense gas and $\sim$70\% in cool gas. 

Mass in new stars in the primary galaxy (Figure~\ref{fig:averages}, bottom panel) is enhanced everywhere, not just in the centre. Specifically, the outskirts experience an enhancement of $\sim$10\%. The central and global enhancements are weaker than in the secondary: only $\sim$20\% and $\sim$15\%, respectively. \textcolor{black}{SFR is marginally enhanced in the outskirts ($\sim$20\%) and globally, and is statistically consistent with unity in the centre.} SFE is consistent with unity (within uncertainty) globally, and for both the centre and the outskirts. As in the secondary, the centre exhibits \textcolor{black}{boosts} of cold-dense and cool gas mass: $\sim$200\% and $\sim$150\%, respectively. Unlike the secondary, the elevation of cold-dense gas mass in the outskirts is statistically significant: $\sim$20\%. The same is not true for the cool gas phase.

This analysis adds a new layer of detail to the work by \cite{Moreno2019}, who report values for the {\it entire system} (i.e., the two galaxies combined). These authors report that interacting pairs experience a boost of $\sim$20\% in cold-dense gas and no statistically-significant change in cool gas content. Here we report mass enhancement in both of these gas phases in the centres of both galaxies, and no statistically-significant change in the outskirts -- {\it except} for the cold-dense gas budget in the primary galaxy, which explains the aforementioned trends for the entire system. We warn the reader to interpret average sample-wide results from large sets with care. It is tempting to draw definitive conclusions from simple averages, which often conceal subtle but important details. For example, the statement `new-stellar mass suppression in the outskirts is not statistically significant' should {\it not} be interpreted as `new-stellar mass suppression never occurs' -- it definitely occurs! We certainly do witness this effect in our fiducial run and in many other merging configurations, but just not in the majority of our runs. Furthermore, this result is also sensitive to our rather generous definition of `outskirts' (i.e., from 1$-10$kpc). Figure~\ref{fig:profile_definitions} demonstrates that mass suppression of new stars {\it is} indeed statistically-significant between $\sim$5$-$8 kpc in the secondary galaxy, and beyond $\sim$8 kpc in the primary -- across the entire merger suite! The rest of this paper is devoted to unveiling such details, by exploring how interaction-induced effects \textcolor{black}{depend} on properties such as radial location within the galaxy (in finer spatial detail, beyond the centre versus outskirts dichotomy), the time after the first encounter, the geometry of the encounter, and the level of global SFR enhancement.

\begin{figure*}
\centerline{\vbox{
\hbox{
\includegraphics[width=3.5in]{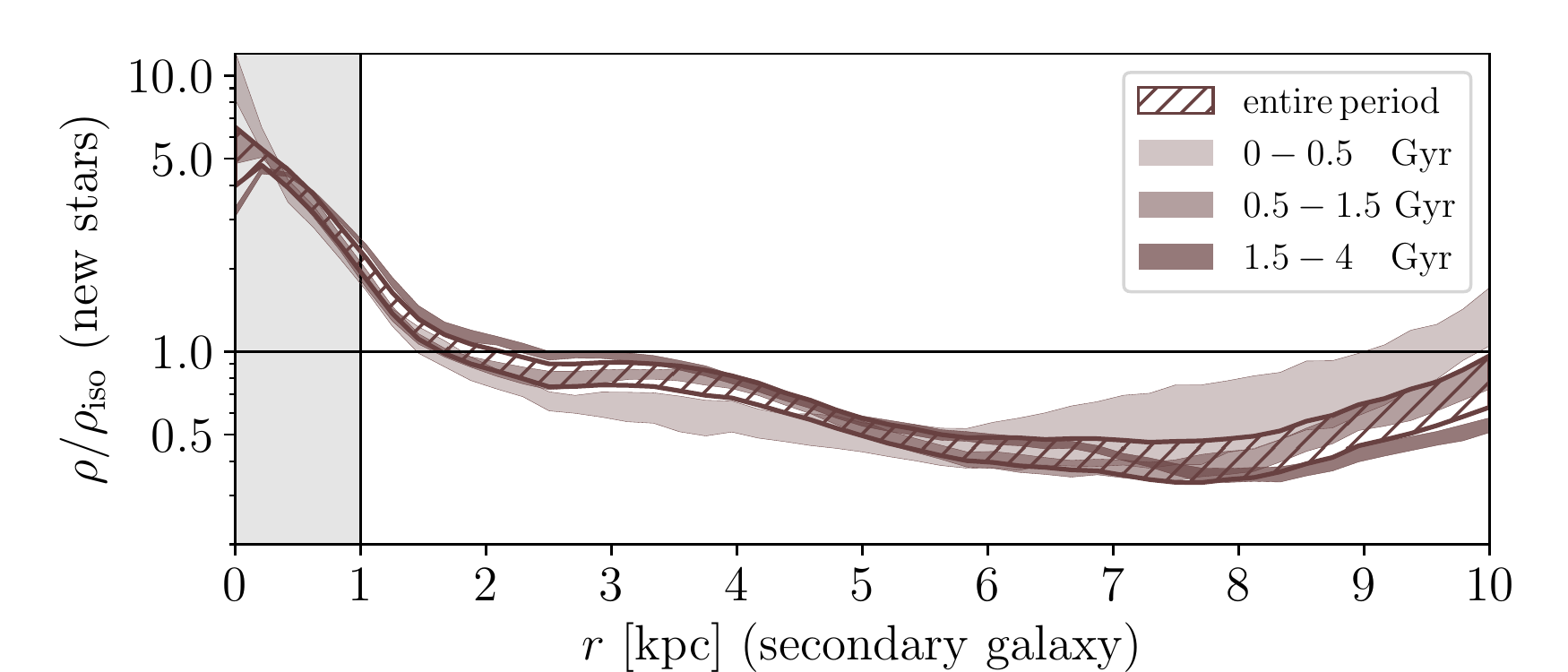}
\includegraphics[width=3.5in]{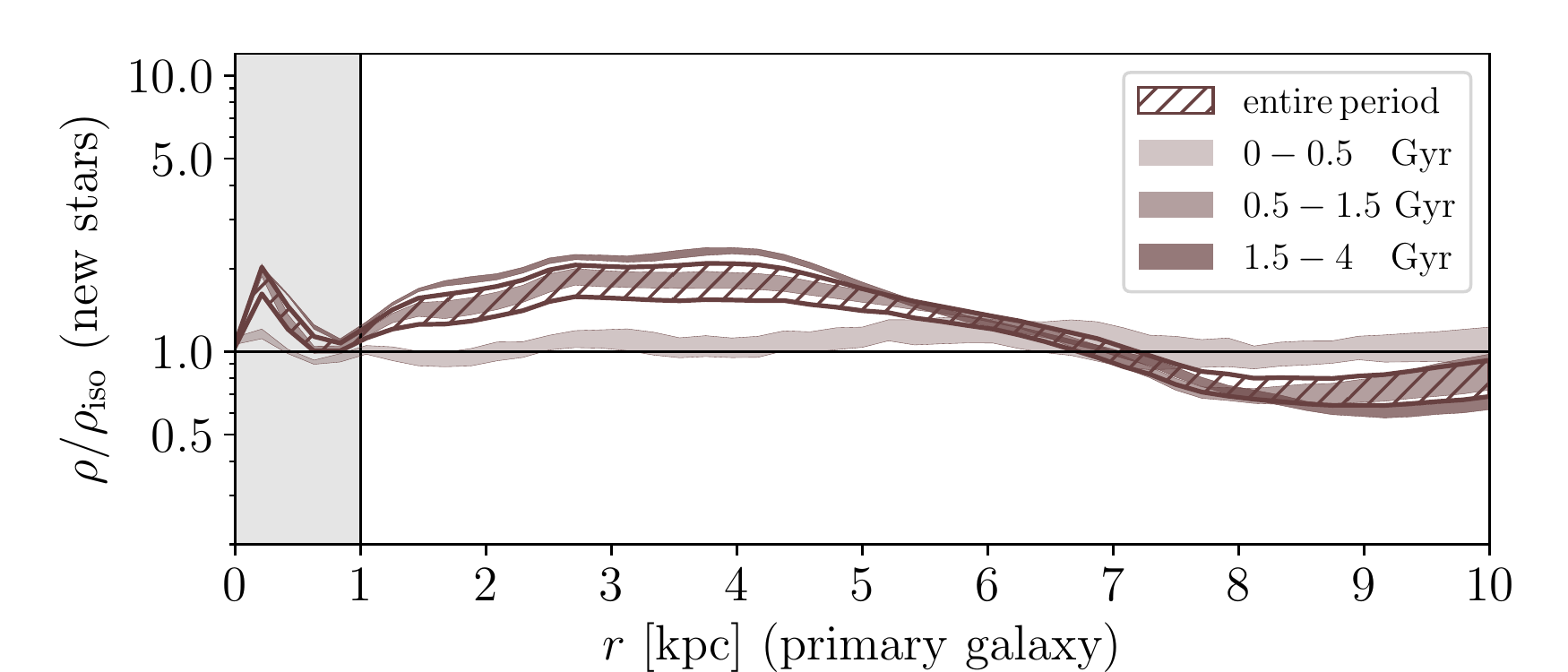}
}
\vspace{-.00in}
\hbox{
\includegraphics[width=3.5in]{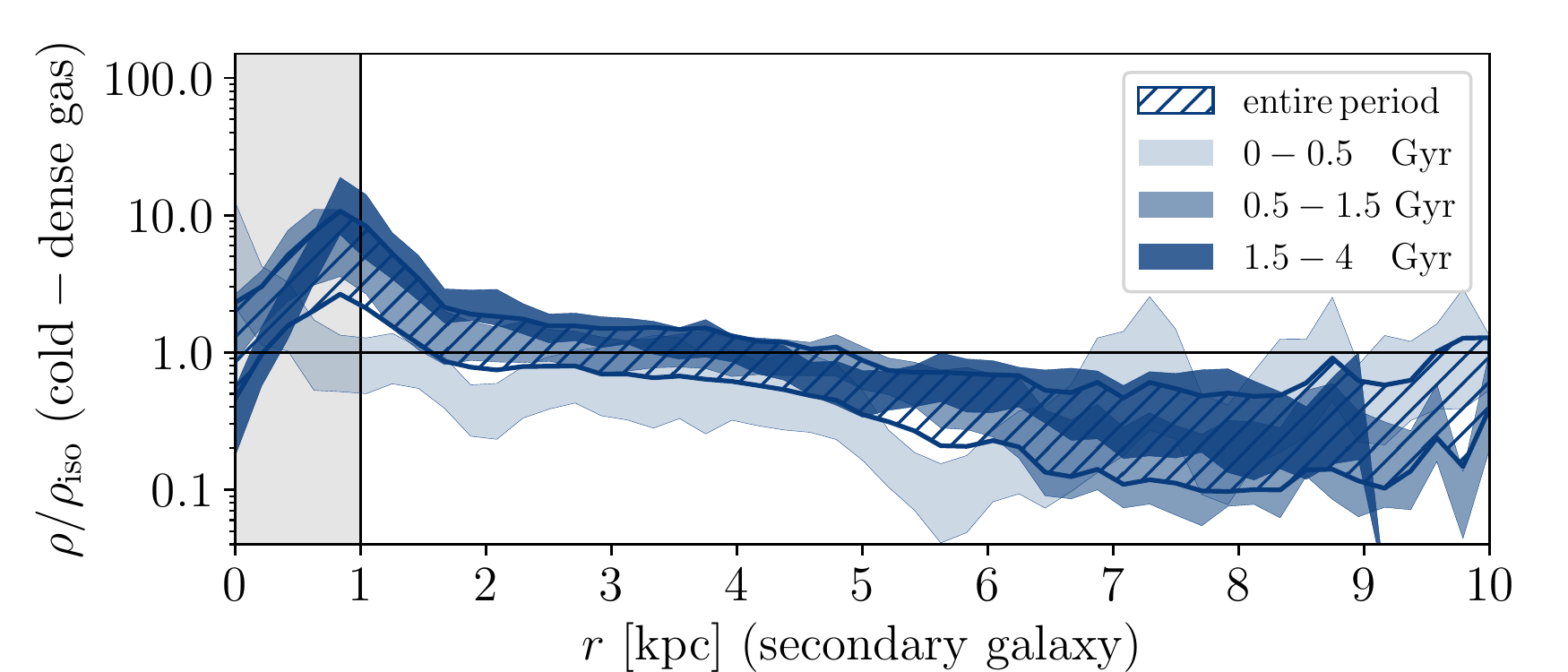}
\includegraphics[width=3.5in]{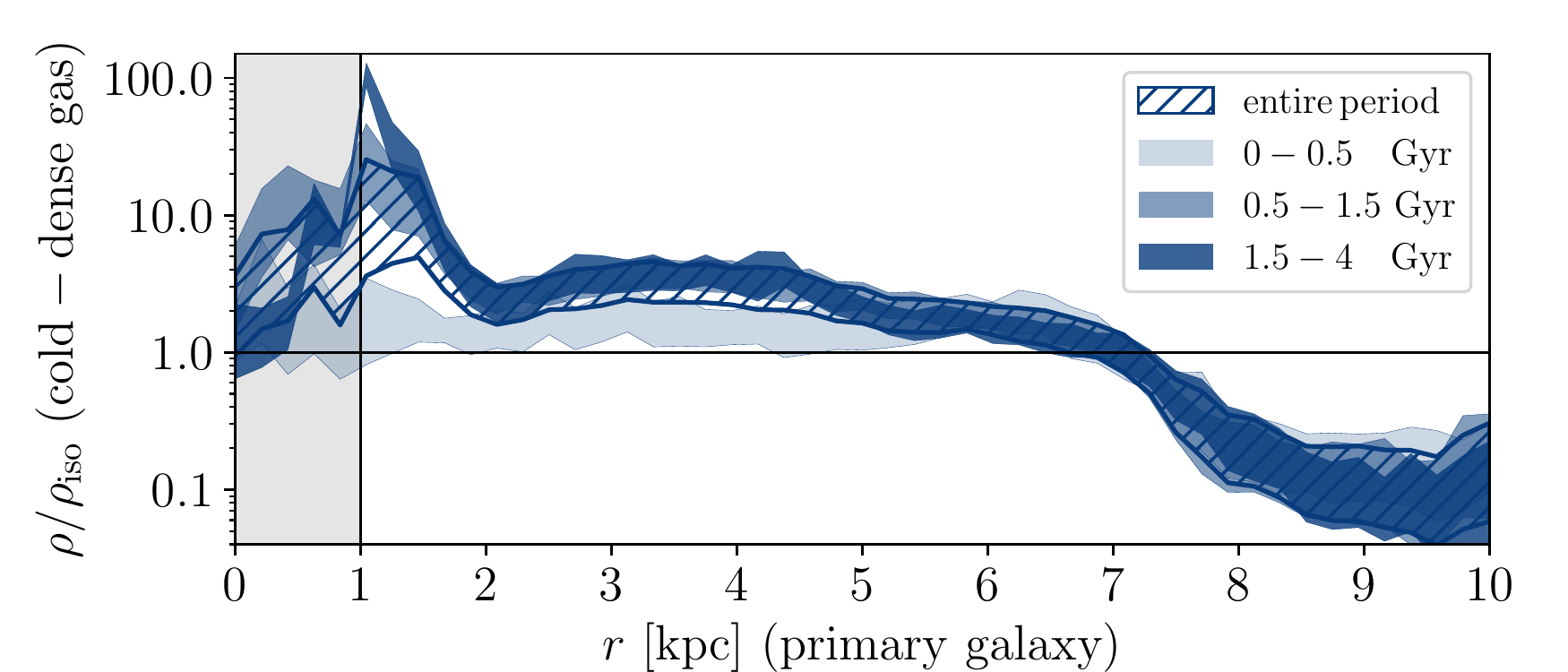}
}
\vspace{-.00in}
\hbox{
\includegraphics[width=3.5in]{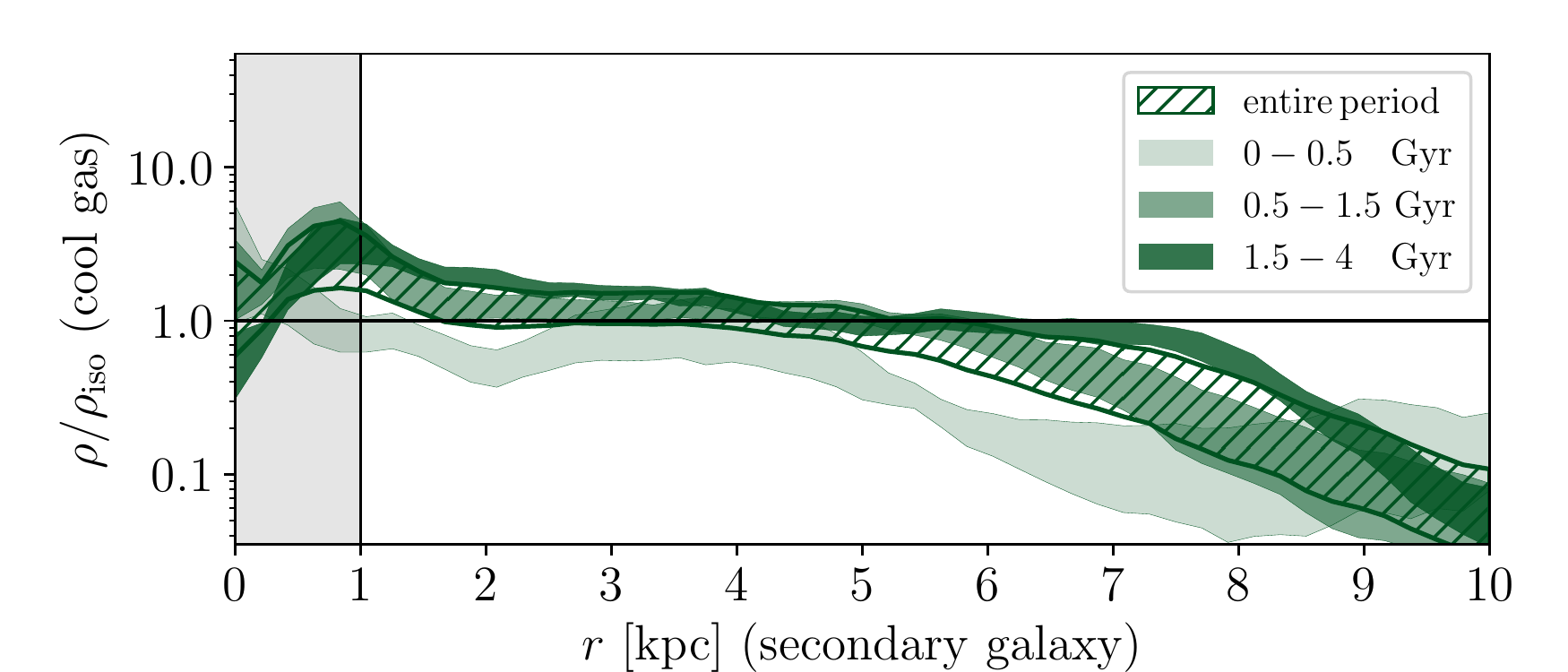}
\includegraphics[width=3.5in]{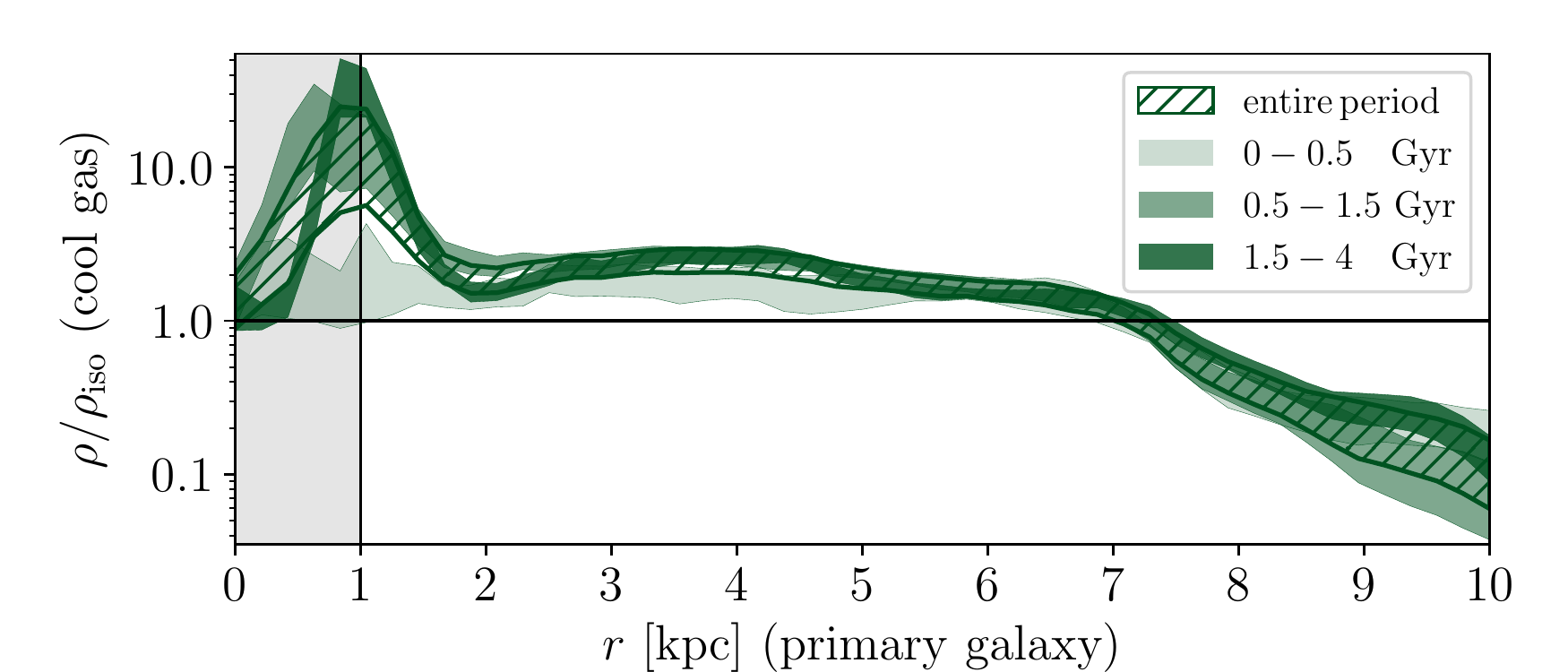}
}
\vspace{-.02in}
}}
\caption{Average profile ratios for the fiducial run. {\it Left (right) panels}: secondary (primary) galaxy. {\it Top-to-bottom panels}: new stars (brown), cold-dense gas (blue), and cool gas (green). The hatched band denotes entire galaxy-pair period, defined to be between first and second pericentric passage, with separation greater than 20 kpc. The light, medium, and dark coloured bands represent early (0$-$0.5 Gyr), intermediate (0.5$-$1.5 Gyr) and late ($>$1.5 Gyr) times (within the galaxy-pair period). Band thickness refers to one standard deviation. The vertical black line and gray box highlight central region ($r<1$ kpc). The horizontal line represents unity.
}
\label{fig:profiles_fiducial}
\end{figure*}

Observationally, it is now established that galaxy pairs in the local Universe exhibit moderate SFR enhancements, accompanied with strong central enhancements -- e.g., \cite{Ellison2013} and \cite{Patton2013}, who use the Sloan Digital Sky Survey \citep[SDSS,][]{Abazajian2009}. Using the CO Legacy Database for the {\it Galex}-Arecibo-SDSS (COLD GASS) Survey, \cite{Saintonge2012} reports that mergers and morphologically-disturbed galaxies tend to have shorter depletion timescales (equivalent to higher SFEs) than the general population, in line with our predictions (at least for the secondary galaxy). Using IRAM 30-m CO(1–0) observations of SDSS paired-galaxies and controls from the extended COLD GASS (xCOLDGASS) Survey \citep{Saintonge2017}, \cite{Violino2018} find that interactions elevate SFEs \citep[see also][]{Combes1994}. In contrast, \cite{Pan2018} find no interaction-induced deviations in SFE. Neither of these two works compare secondary versus primary galaxies (recall that we predict enhanced global SFE for the former, but not for the latter). Both papers measure interaction-induced $H_2$ mass enhancements in their galaxy pairs, in line with our simulations (at least for the primary galaxy). See also \cite{Lisenfeld2019}, who find enhancement in H$_2$ content, but not in SFE. Regarding HI content (equivalent to our cool gas component), observations by \cite{Knapen2009} and \cite{DiazGarcia2020} do not find enhancements in atomic gas mass in their galaxy pairs, in line with our global values. We mention, however, that \cite{Ellison2018} report HI enhancement in their post-merger sample.

Cosmological simulations also address interaction-induced star formation. Using Illustris \citep{Vogelsberger2013}, EAGLE \citep[the Evolution and Assembly of GaLaxies and their Environments project,][]{Schaye2015}, and IllustrisTNG \citep{Pillepich2018} --  \cite{Patton2020} demonstrate that SFR is enhanced in galaxies with relatively close companions relative to carefully-matched non-interacting controls \citep[mimicking the observational methodology of][]{Patton2013,Patton2016}, and that this is a generic feature across simulations with very distinct physics assumptions. This result \textcolor{black}{confirms} earlier findings with smaller simulation volumes \citep{Perez2006} -- but see \cite{TonnesenCen2012}, who do not identify enhanced specific SFR (sSFR) in their cosmologically-selected galaxy pairs with separations below the Roche limit. Using SIMBA \citep{Dave2019}, \cite{RodriguezMontero2019} investigate what drives elevated SFR in mergers. They find that for low-mass galaxies (stellar mass below $10^{10.5}M_{\odot}$), increases in SFR are attributed to increases in $H_2$ content, not SFE -- and that this trend reverses for more massive galaxies. Surprisingly, we find that for the secondary galaxy, global SFR enhancement is driven by enhanced SFE (Figure~\ref{fig:averages}, top panel, medium purple and pink circles) -- whilst, for the primary, it is driven by enhanced cold-dense gas content (Figure~\ref{fig:averages}, bottom panel, medium purple and green circles). Unfortunately, these authors identify mergers by selecting systems with sudden jumps in stellar mass (above that expected from in situ star formation), which may include both galaxy-pairs and mergers past the coalescing period. Interestingly, \cite{Sparre2016} use zoom-in simulations from Illustris to highlight the importance of making this distinction: during the galaxy-pair period, SFR enhancement is driven by increases in $H_2$ content -- but after coalescence, it is driven by increases in SFE.

\subsection{Radial structure and evolution}
\label{subsec:profiles}

The previous section studies time evolution in three radial regions: the centre (0$-$1 kpc), the outskirts (1$-$10 kpc), and the entire galaxy (0$-$10 kpc). In this section we dissect radial structure in finer detail, at the expense of having to use cruder time bins: the early (0$<t<$0.5 Gyr), intermediate (0.5$<t<$1.5 Gyr), and late ($t>$1.5 Gyr) period of interaction. See Section~\ref{sec:simulations_and_terminology} for a justification, and the thin vertical lines in Figures~\ref{fig:time_evolution_fiducial} and \ref{fig:time_evolution_suite} for reference. We focus on the fiducial run only, and explore other merging configurations in Section~\ref{subsec:profiles_subsuites}. We present three baryonic components: new stars, cold-dense gas and cool gas - and defer an analysis of the spatial extent of SFR and SFE to Section~\ref{subsec:driving_global_sfr}. Figure~\ref{fig:profiles_fiducial} displays these average profile ratios. The hatched bands represent the entire galaxy-pair periods, whilst the coloured bands (from light-to-dark) represent averages constraint to times in the early, intermediate, and late periods. 

For the secondary galaxy (left panels), enhancement in new stellar mass is centrally peaked (within $\sim$2 kpc), and suppressed at larger galactocentric radii. This behaviour is particularly accentuated during the early period, and becomes weaker at later times. The cold-dense and cool components behave similarly, except that enhancement does not peak at the centre, and suppression starts at different radii: $\sim$4.5 kpc and $\sim$5.5 kpc, respectively. Recall that suppression at the centre does not necessarily imply the presence of a `hole' in the gas distribution, but rather it is commonly attributed to a diminishment in mass relative to the isolated control galaxy (e.g., Figure~\ref{fig:terminology}, fourth-row$/$third-column panel). As a function of time, the peak shifts outwards (from $0$ kpc to $\sim$0.8 kpc) and suppression at larger radii becomes weaker. At early times, both new stars and cold-dense gas also exhibit an uptick at the very largest radii. By inspecting a video from which the images in Figure~\ref{fig:terminology} were drawn, we find that this effect is explained by material that was originally launched into tidal tails and \textcolor{black}{a bridge} after the first encounter \citep{Donghia2010,Blumenthal2018}, becomes compressed there, and is now settling on to the outer portions of the disc. This phenomenon is a modern version of the sequences of events described by the original \cite{BH96} paper -- but here we employ a model that resolves the structure of the ISM. In a future paper, we investigate how cold-dense gas and new stars are formed in tidal tails and bridges, and later migrate back onto the discs.

The radial structure of mass enhancement in new stars in the primary galaxy (Figure~\ref{fig:profiles_fiducial}, right panels) is strikingly different. On average, two peaks are formed: one spanning $\sim$0$-$0.7 kpc, and the other spanning $\sim$0.8$-$7 kpc. Initially, this distribution is weak and flat, but it gets stronger around these two radii at later times as new stars are born. The cold-dense and cool gas components exhibit strong enhancement between $\sim$0$-$2 kpc, accompanied by a milder plateau that extends out to $\sim$7 kpc, with suppression beyond that radius. Initially, the plateau is weak and extends all the way to the centre. At later times, the above secondary concentration at small radii becomes stronger and its peak shifts to slightly larger radii. This extended reservoir of cold-dense (and cool) gas explains the \textcolor{black}{build-up} of the corresponding plateau in new stars (Figure~\ref{fig:profiles_fiducial}, top-right panel). In contrast to the secondary galaxy, the primary has a {\it stronger} increase in cold-dense (and cool) gas mass at small radii. This does result in a increase in new stars in that region, but this boost is weaker than in the centre secondary galaxy. The presence of new stars is the cumulative effect of SFR, which in turn is governed by the amount of fuel available and SFE. Recall that SFE is suppressed in the centre, and remains close to unity in the outskirts (Figure~\ref{fig:time_evolution_fiducial}, third-row$/$second-column panel, dark versus light pink). We address the connection between SFE, available cold-dense gas fuel, and SFR in detail in Sections~\ref{subsec:driving_global_sfr} and \ref{subsec:fuel}.

There are very few statistical studies on the spatial extent of interaction-induced star formation and molecular-gas fuelling using observations in the local Universe. \cite{BarreraBallesteros2015sfr} use a sample of over 100 merging galaxies drawn from the CALIFA survey \citep{Sanchez2012}, and report sSFR enhancement in the central region, and moderate-to-null suppression in the outskirts, in line with our findings. These authors do not report results in terms of primary versus secondary galaxy. Similarly, using a MaNGA dataset containing over 200 merging galaxies, \cite{Pan2019sfr} split their sample into four evolutionary stages, two of which are directly relevant to our work: `Stage 2' (corresponding to our early period) and `Stage 3' (corresponding to our intermediate and late periods). Their observations suggest tantalising similarity with our results: steep centrally-concentrated star formation, with decaying excess profiles at large galactocentric radii; and with a decrease in central steepness at later times. Direct comparison with their work is difficult for two reasons: (1) they only probe out to 1.5 half-light radii, and (2) they do not split their sample into primary and secondary galaxies. See also \cite{Thorp2019}, who perform a similar analysis for the post-merger period, and \cite{Ellison2018sfr} for more general (non-merger specific) results. Interestingly, in the stellar-mass regime we cover here, \cite{Spindler2018} find that satellites exhibit enhanced star formation in the centre and suppression in the outskirts relative to centrals. If this result remains true for systems where the satellite is (1) the most massive satellite within the host dark matter halo, and (2) it is interacting with the central -- these observations would be consistent with our predictions (Figure~\ref{fig:profiles_fiducial}, top-left versus top-right panel).

To investigate the spatial extent of the cold molecular gas component (our `cold-dense' gas phase), \cite{Yamashita2017} use CO-observations with the Nobeyama Radio Observatory on a sample of 58 interacting galaxies drawn from the Great Observatories All-sky LIRGs (Luminous Infrared Galaxies) Survey \citep[GOALS, ][]{Armus2009}. These authors find that, on average, the CO radii of galaxies in widely-separated pairs is larger than their isolated counterparts, and become smaller as they approach second pericentre. This is contrary to our findings (at least for the secondary galaxy, Figure~\ref{fig:profiles_fiducial}, bottom-left panel, light-to-dark bands). However, these authors also normalise their CO-sizes relative to the size of the stellar component, which may also experience changes in physical size during the interaction (Moreno et al., in prep). Direct comparison with our work is also challenging because they focus on extreme systems (LIRGS), whilst our suite is designed to model more common galaxy interactions found in the SDSS. To empirically corroborate this behaviour, CO-observations of a larger set of interacting galaxies (beyond LIRGs) with more finely-sampled evolutionary stages is required.

\subsection{Dependence on orbital merging geometry}
\label{subsec:profiles_subsuites}

\begin{figure*}
\centerline{\vbox{
\hbox{
\includegraphics[width=3.5in]{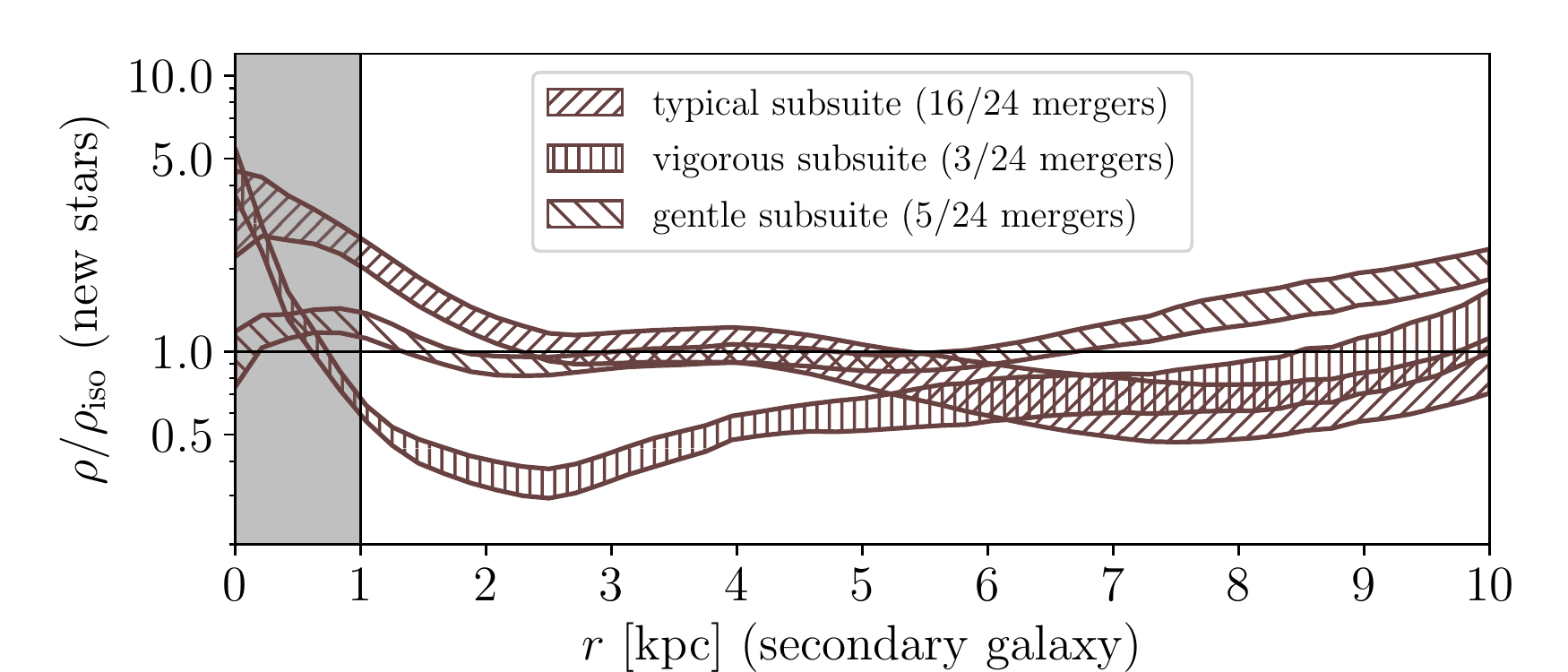}
\includegraphics[width=3.5in]{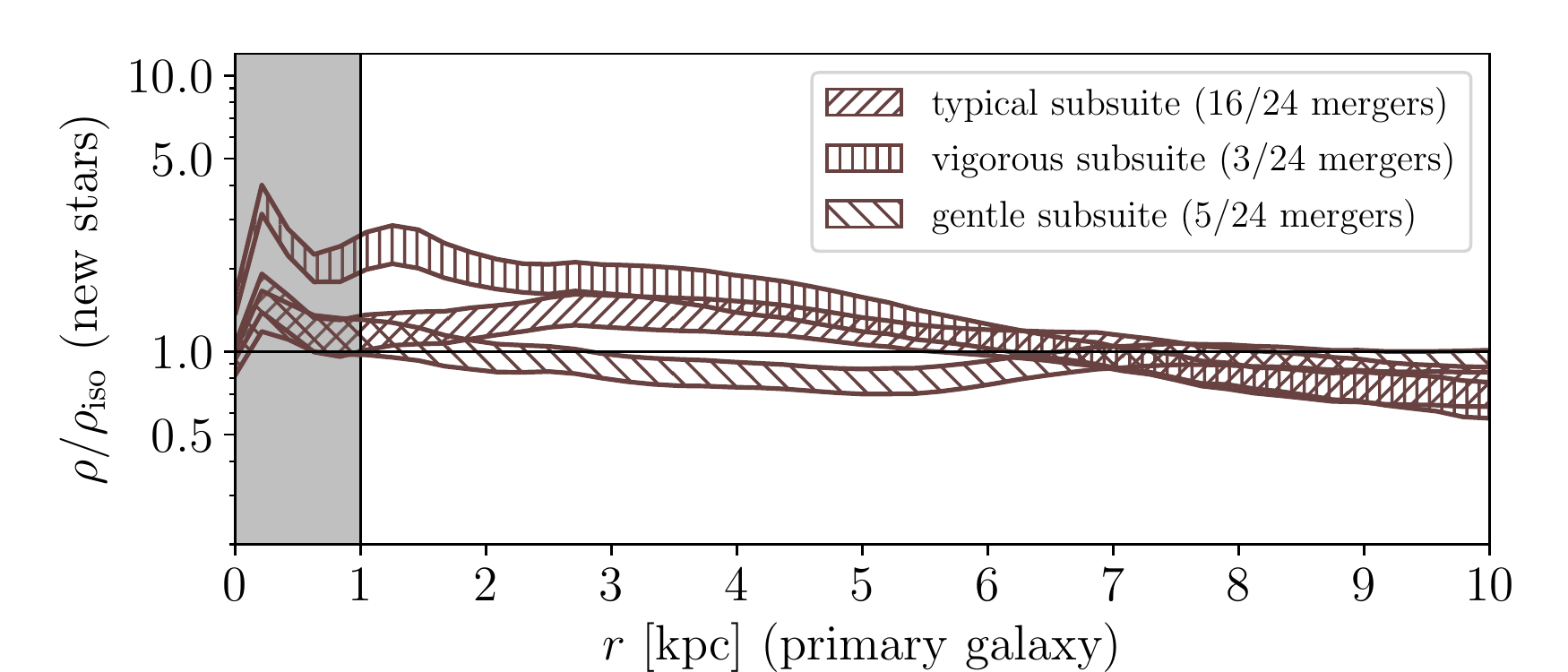}
}
\vspace{-.02in}
\hbox{
\includegraphics[width=3.5in]{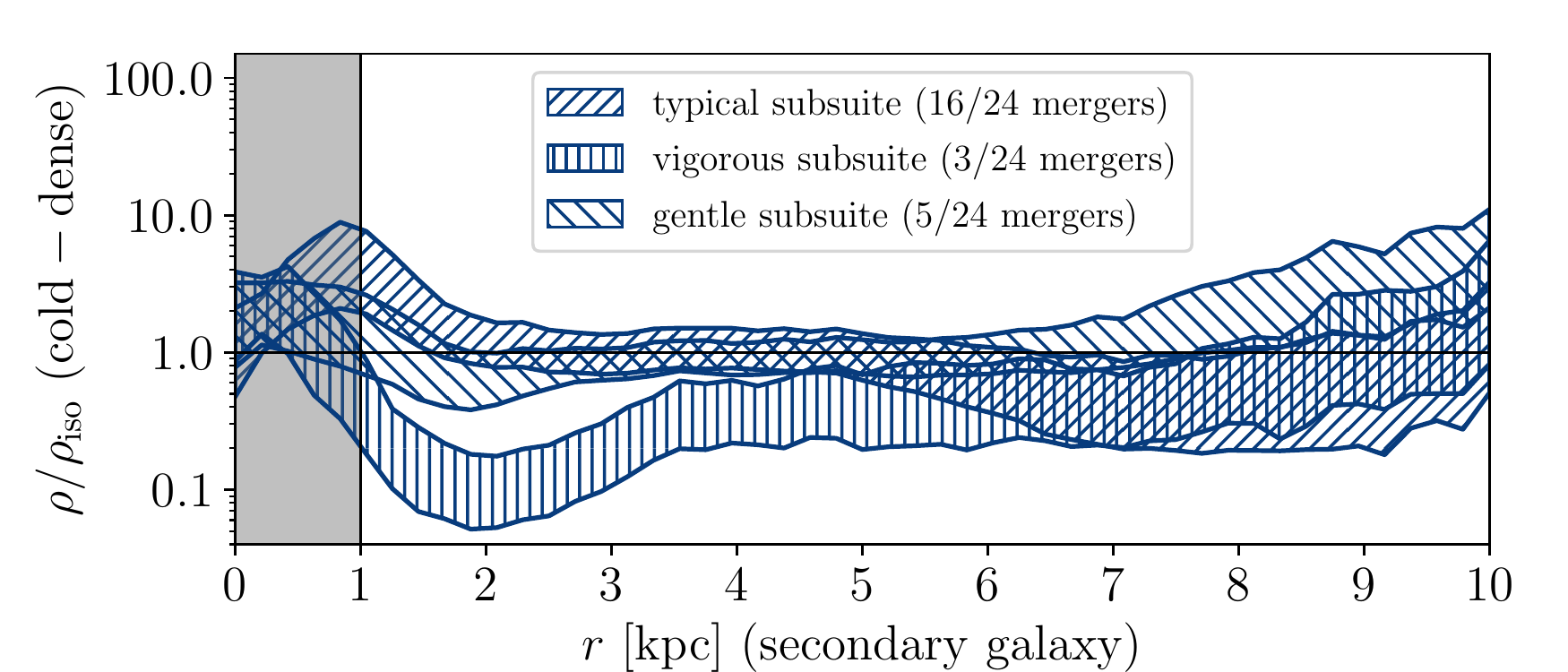}
\includegraphics[width=3.5in]{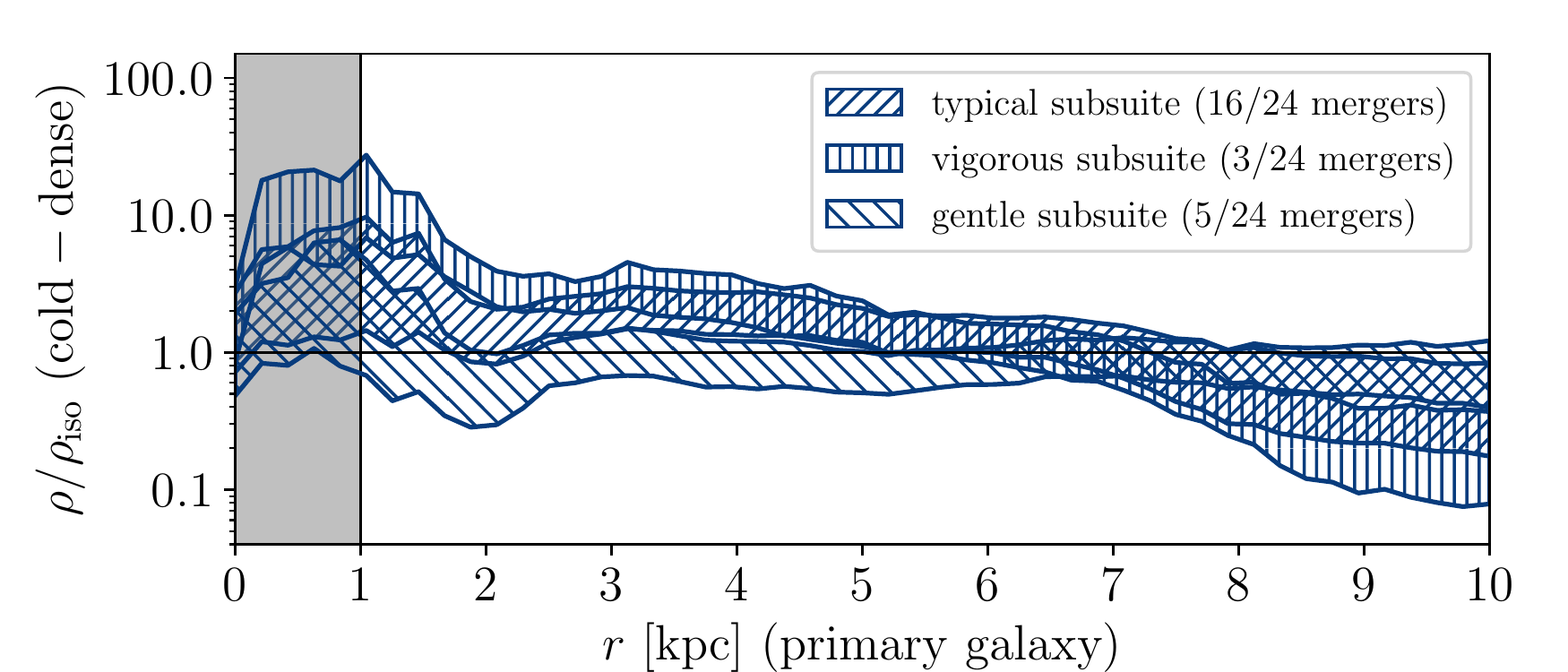}
}
\vspace{-.02in}
\hbox{
\includegraphics[width=3.5in]{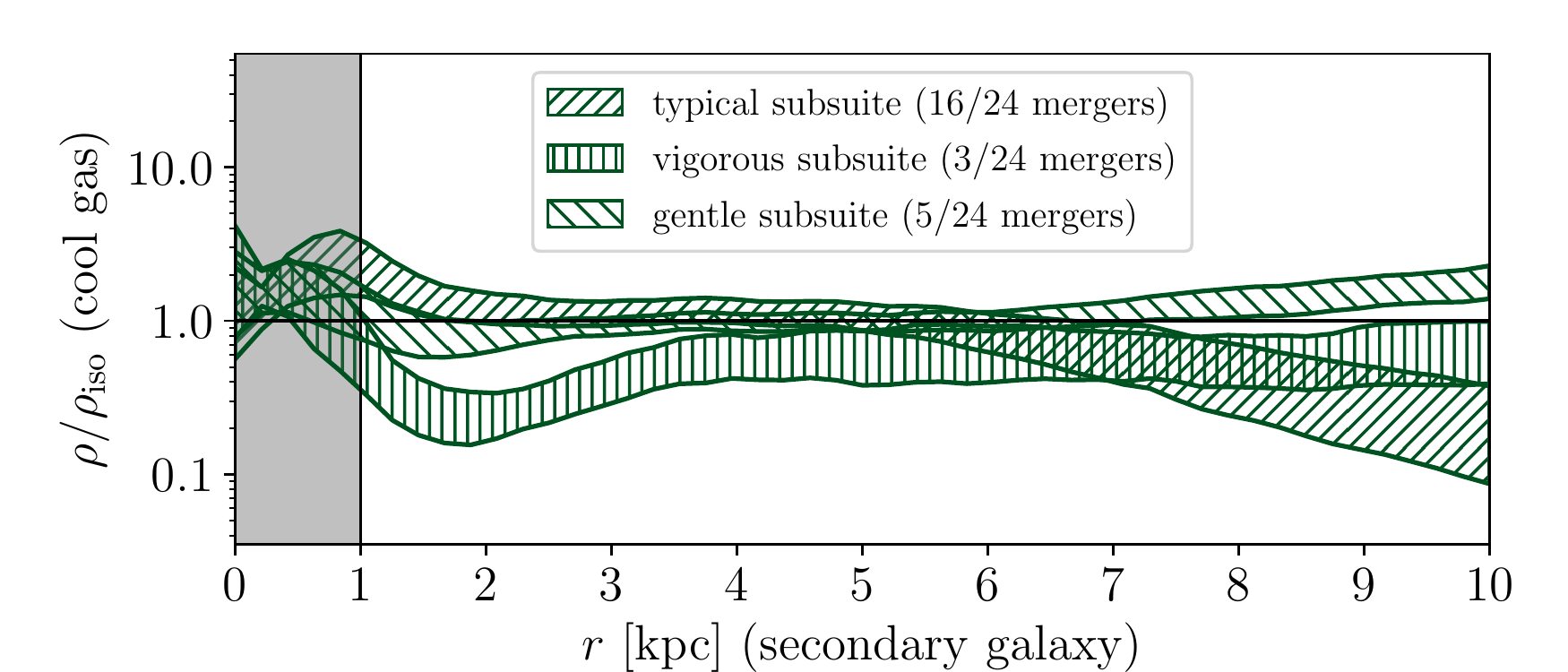}
\includegraphics[width=3.5in]{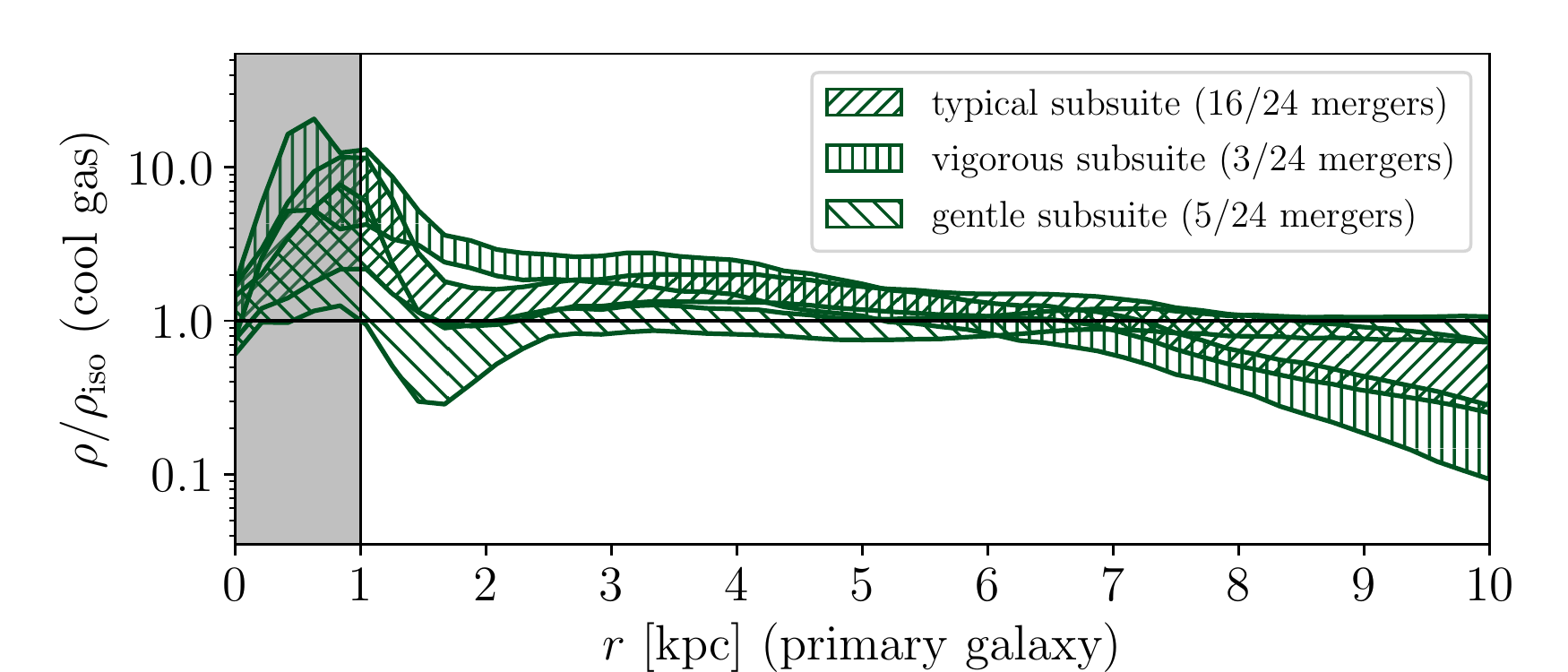}
}
\vspace{-.02in}
}}
\caption{Average profile ratios for three subsets of galaxy-merger configurations (subsuites). {\it Left (right) panels}: secondary (primary) galaxy. {\it Top-to-bottom panels}: new stars (brown), cold-dense gas (blue), and cool gas (green). The forward-slash, vertical, and backward-slash hatched bands represent the typical, vigorous, and gentle subsuites (keys indicate fraction of mergers in each subsuite -- see Section~\ref{sec:simulations_and_terminology} for definitions). Band thickness refers to one standard deviation. The vertical black line and gray box highlight the central region ($r<1$ kpc). The horizontal line represents unity. Galaxy-pair periods only.  
}
\label{fig:profiles_minisuites}
\end{figure*}

\begin{figure*}
\centerline{\vbox{
\hbox{
\includegraphics[width=3.5in]{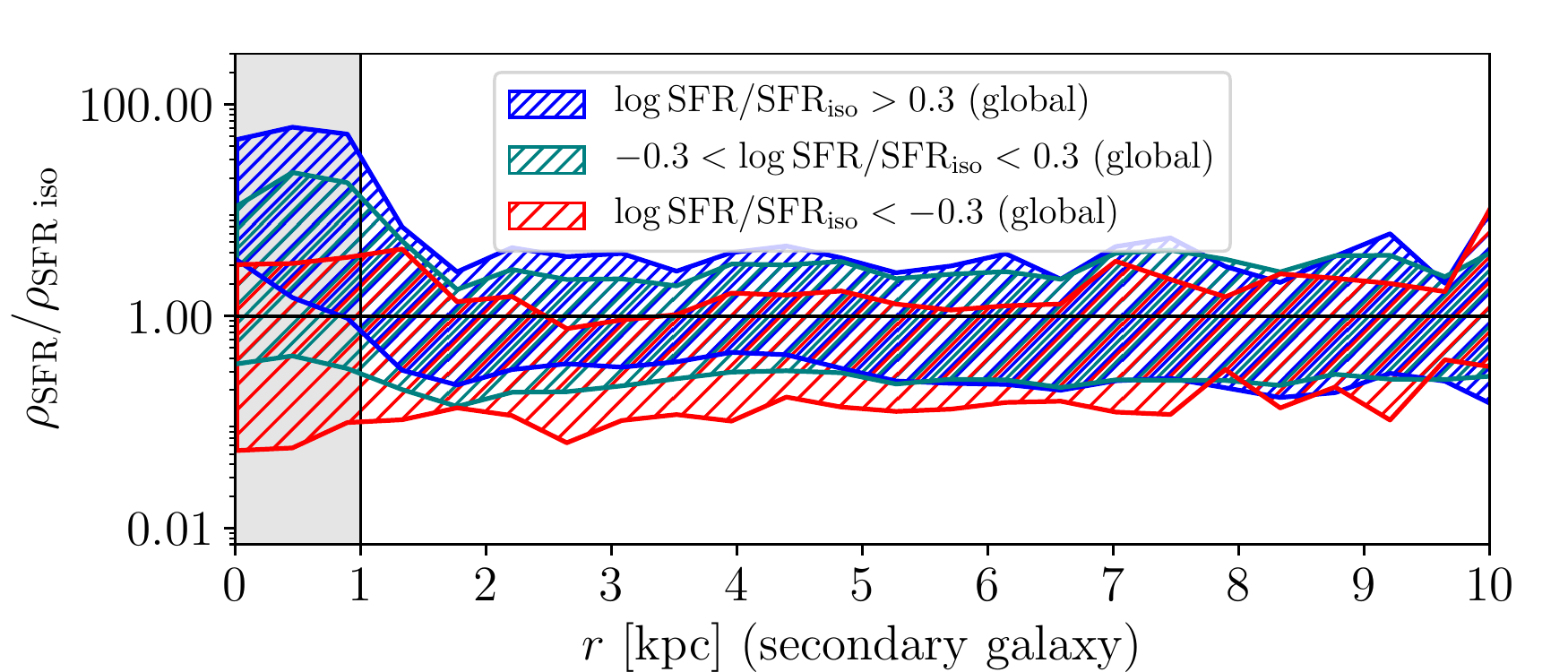}
\includegraphics[width=3.5in]{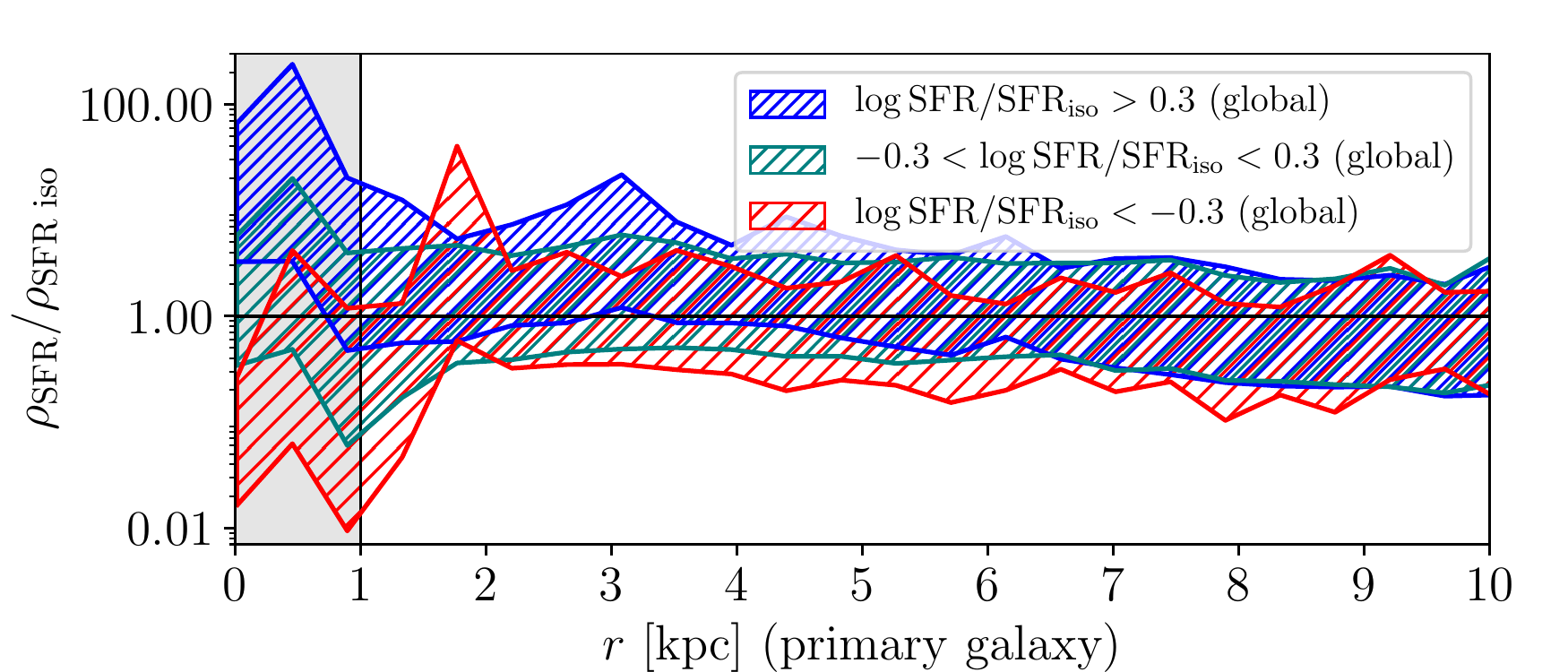}
}
\vspace{-.02in}
\hbox{
\includegraphics[width=3.5in]{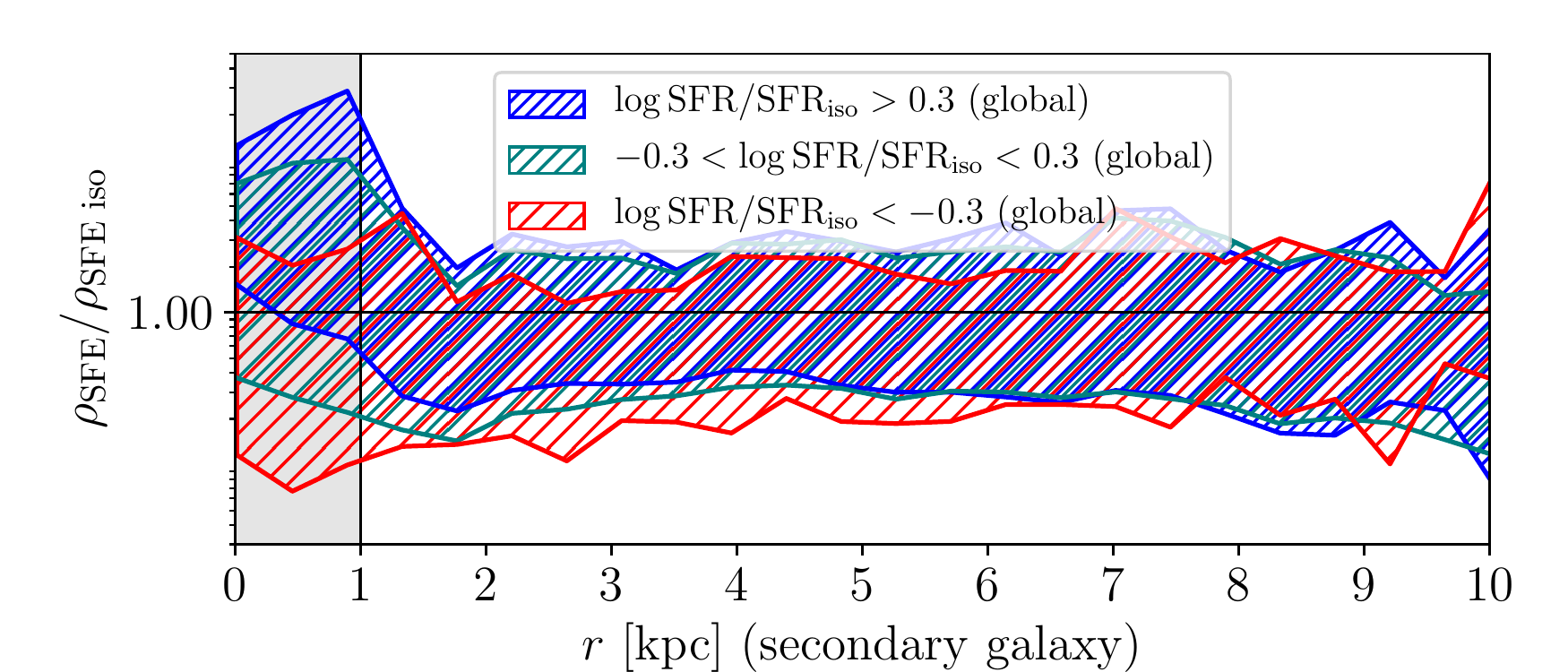}
\includegraphics[width=3.5in]{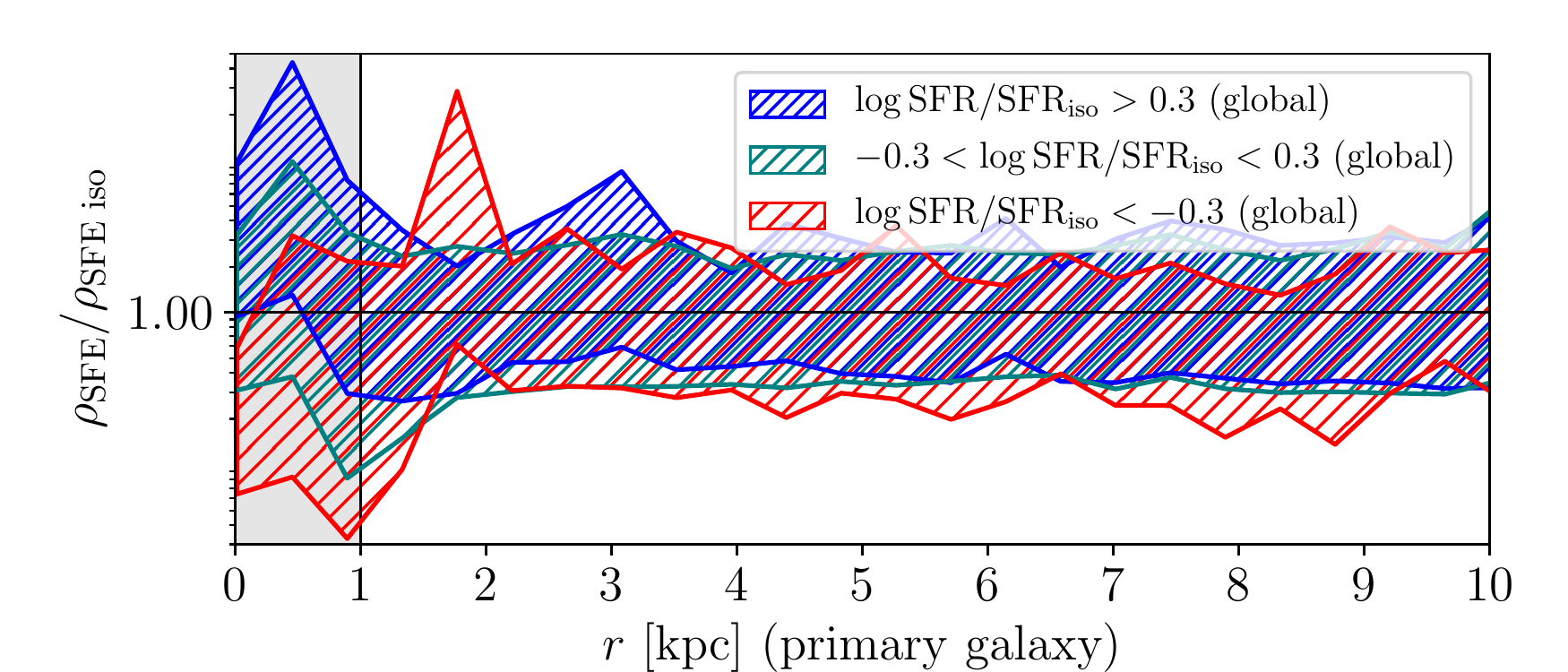}
}
\vspace{-.02in}
\hbox{
\includegraphics[width=3.5in]{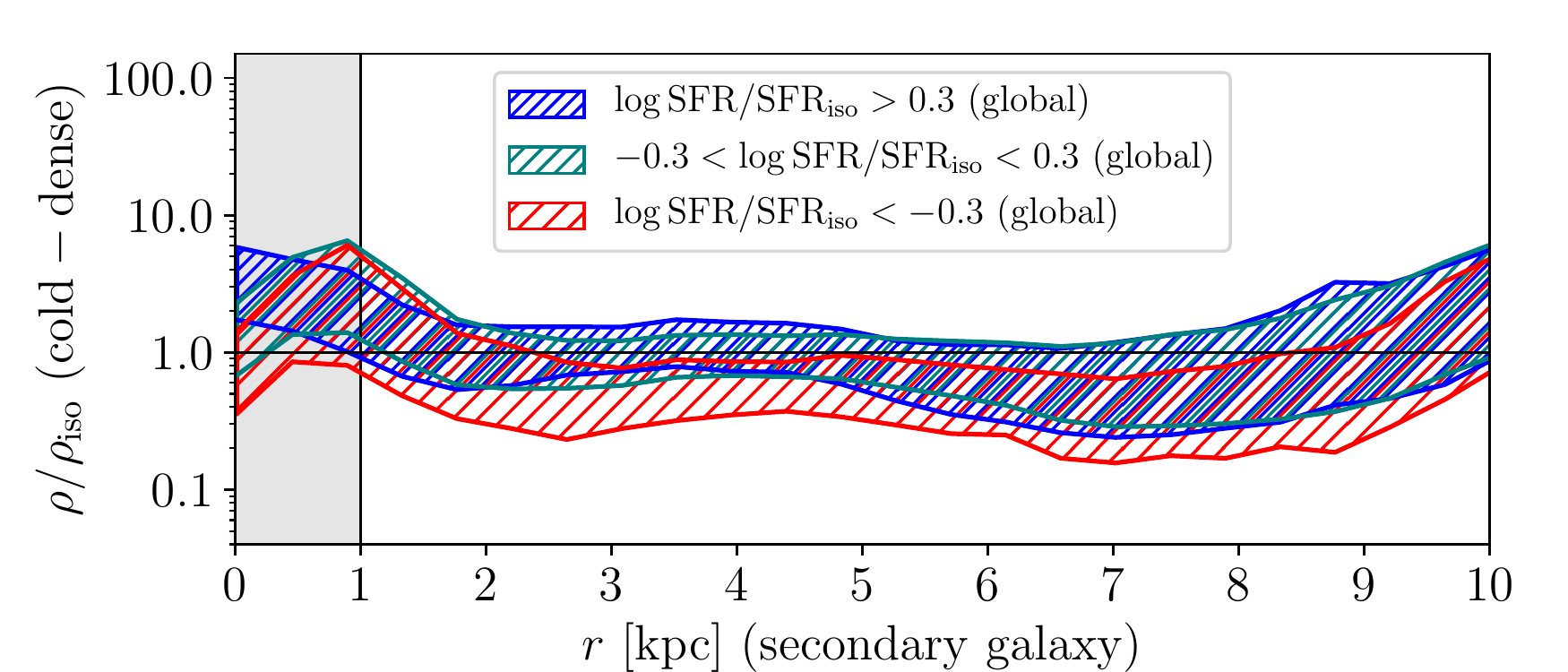}
\includegraphics[width=3.5in]{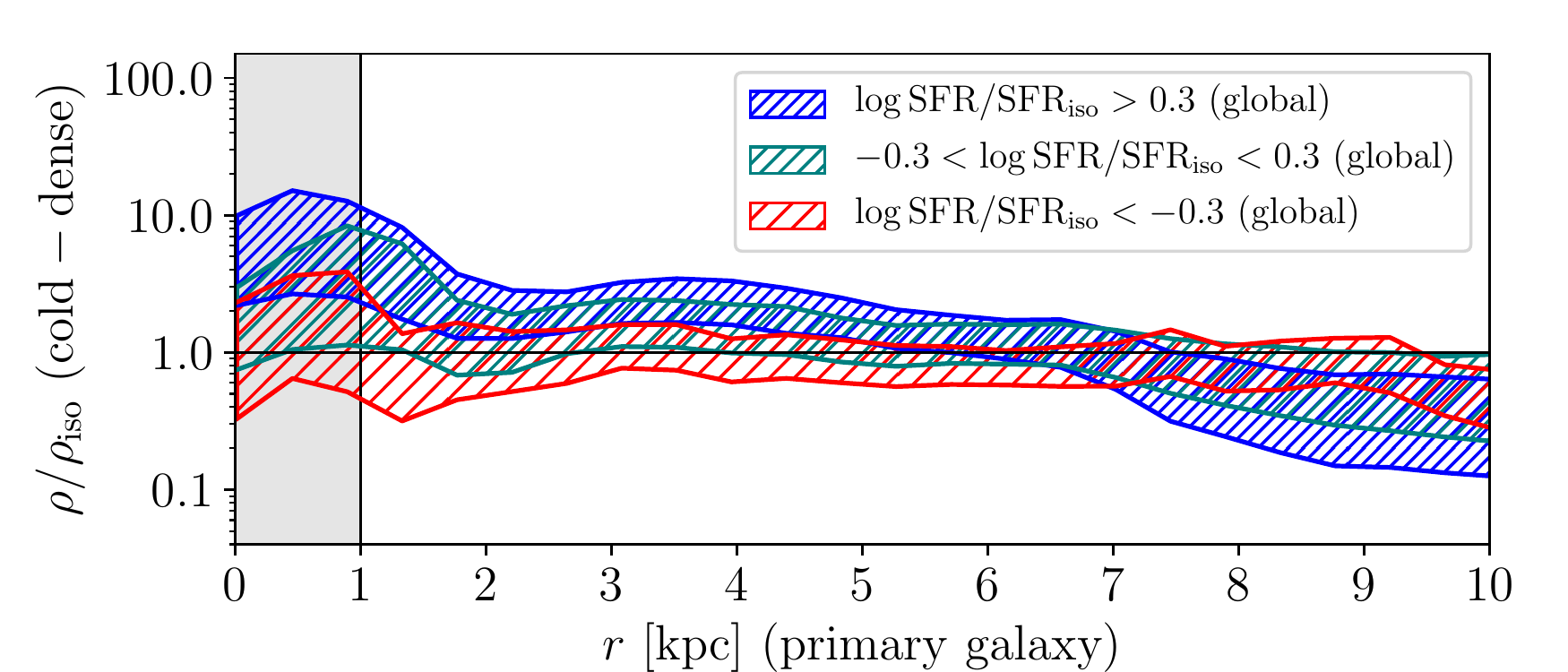}
}
\vspace{-.1in}
}}
\caption{Average profile ratios split into three global star formation rate enhancement bins. {\it Left (right) panels}: secondary (primary) galaxy. {\it Top-to-bottom panels}: instantaneous SFR, SFE, and cold-dense gas mass. The densely-packed (blue), regular (green), and sparsely-packed (red) hatched bands represent the result of averaging over galaxies with global (within 10 kpc) SFR enhancement greater that $+$0.5 dex, between $-$0.5 and $+$0.5 dex, and below $-$0.5 dex, respectively. Band thickness refers to one standard deviation. The vertical black line and gray box highlight the central region ($r<1$ kpc). The horizontal line represents unity. Galaxy-pair periods only. Table~\ref{table:sfr_bins} lists percentages per SF type and galaxy.
}
\label{fig:sfr_bins}
\end{figure*}

So far we have only reported average profile ratios for the fiducial run. Generalising these results to all mergers in the suite is challenging  because one needs to disentangle variations caused by time evolution within an individual merger and differences amongst various merger configurations. Simply averaging across the entire suite (e.g., as in Figure~\ref{fig:profile_definitions}) might mask subtle, but important, details. It is not practical to repeat the analysis presented in Figure~\ref{fig:profiles_fiducial} for all 24 mergers in the suite. Rather, we visually inspect these single-run average profile ratios for the three baryonic components displayed in Figure~\ref{fig:profiles_fiducial} (new stars, cool gas, and cold-dense gas), and group them together if they share common features. We find that we can split the 24 configurations in our suite into three main categories, or {\it subsuites}:

\begin{enumerate}
    \item The {\it typical subsuite} includes every near-prograde configuration (8 mergers), plus those near-polar configurations with first pericentric passages at intermediate ($\sim$16 kpc) and large ($\sim$27 kpc) separations (5 mergers), and those near-retrograde with small separation ($\sim$7 kpc) at first pericentric passage (3 mergers). This amounts to 16/24 mergers, or 66.7\% of the entire suite. The fiducial run belongs to this subsuite.
    \item The {\it vigorous subsuite} includes near-polar configurations with first pericentric passage at $\sim$7 kpc. This amounts to only 3/24 mergers, or 12.5\% of the suite.
    \item The {\it gentle subsuite} contains those near-retrograde configurations with intermediate ($\sim$16 kpc) and large ($\sim$27 kpc) separations at first pericentric passage. This corresponds to 5/24 mergers, or 20.1\% of the suite.
\end{enumerate}
We adopt the terms `typical', `vigorous', and `gentle' to informally describe how encounters in these subsuites alter the average profile ratios (for the three baryonic components) with respect to unity {\it during the galaxy-pair period} -- i.e., this naming scheme might {\it not} be appropriate to describe the impact of these orbital geometries on radial structure after coalescence. We emphasize that we choose to group our mergers in these three subsuites as an alternative to the cumbersome presentation of 24 individual configurations. There is no apriori rigourous physical reason why we should group our orbits in this particular fashion. \textcolor{black}{At the time of writing, we are not aware of any published work where merger simulations are segregated in this manner. Commonly, authors group orbits in terms of spin-orbit orientation only \citep[e.g.,][]{DiMatteo2007,Moreno2015}. Here, we note that the combination of orientation, and whether or not the two discs intersect, governs how radial structure evolves after first pericentric passage. Investigating dynamical processes in this context is the subject of future work.}

Figure~\ref{fig:profiles_minisuites} displays configurations in the typical, vigorous, and gentle subsuites using forward-slash, vertical, and back-slash hatched bands respectively. We calculate these average profile ratios by taking the average of individual profile ratios across every member configuration of each subsuite, and across their respective galaxy-pair periods. Note that the results corresponding to the typical subsuite resemble those corresponding to the fiducial run (hatched bands here versus hatched bands in Figure~\ref{fig:profiles_fiducial}). Below we describe features appearing in the vigorous and gentle subsuites in detail, and compare them to those in the (larger) typical subsuite (which contains the fiducial case -- discussed thoroughly already).

The secondary galaxy (left panels) displays the most dramatic effects in the vigorous subsuite. New stellar mass is suppressed severely between $\sim$0.5$-$8.5 kpc. The cold-dense and cool gas mass budgets are also strongly suppressed between $\sim$1$-$5 kpc and $\sim$1$-$9 kpc, respectively. Interacting galaxies in the gentle subsuite, on the other hand, experience the mildest effects. Galaxies in this category experience weak mass enhancement in new stars within $\sim$1.5 kpc and beyond $\sim$6 kpc. The cold-dense and cool gas mass budgets are elevated within the central $\sim$0.5 kpc, and beyond $\sim$8 kpc and $\sim$7 kpc, respectively. The primary galaxy (right panels) in the  vigorous subsuite acts as a more-intense version of the typical subsuite, with slightly larger central baryonic concentrations, and spatially-extended plateaus. Overall, average profile ratios belonging to the gentle subsuite tend to be flat and close to unity, with dips between $\sim$3$-6$ kpc for the new-stellar component, and near $\sim$1.5 kpc for the cold-dense and cool gas components. 

Very little numerical work exists on how the spatial distribution of new stars and cold-dense$/$cool gas depends on orbital merging geometry. \cite{DiMatteo2008} \textcolor{black}{present} surface density maps of star-forming gas for a handful of time frames for two mergers, one with prograde and one with retrograde spin-orbit orientation. Their prograde merger exhibits large concentrations of gas and star-forming \textcolor{black}{regions} in the centres, with secondary contributions from tidal tails, the bridge, and ring structures \citep[see also][who report the existence of star-forming rings in their EOS simulations]{Moreno2015}. Their retrograde merger only exhibits strong gas concentrations and intense star formation in the centres. These authors do not present a quantitative analysis like ours, making a direct comparison with their work unfeasable. \cite{Moreno2015} split their sample into three subsuites of identical size: the near-prograde, near-polar, and near-retrograde orientations. We use exactly the same spin-orbit orientations as in that paper, but with a new model (\textsc{fire-2}). In that paper, the authors find that star formation in the secondary galaxy is enhanced in the centre, and suppressed in the outskirts. This effect is particularly strong for near-prograde and near-polar orbits, and weak for near-retrograde configurations. We find similar trends when we compare the typical subsuite (containing all the near-prograde orbits) against the gentle subsuite (containing most of the near-retrograde mergers): enhancement in the centre and suppression at large galactocentric radii is more pronounced in the former category than in the latter one (if at all). Unlike those older simulations, we generally find that our levels of enhancement and suppression are weaker, and suppression appears at larger radii: beyond $\sim$6 kpc in this work, versus $\sim$1 kpc in \cite{Moreno2015}. In that older work, the primary galaxy exhibits strongly boosted star formation in the centre for the near-prograde interactions, and weak enhancement out to larger radii (out to $\sim$3 kpc for near-polar orbits, and everywhere for near-retrograde orbits). The primary galaxy in our \textsc{fire-2} simulations produces more new stars in the typical subsuite than in the gentle subsuite. The former is weakly enhanced within $\sim$6 kpc, whilst the latter is weakly enhanced within $\sim$1 kpc. Overall, our results suggest that incorporating resolved, feedback-regulated, physics -- and their effect on the turbulent structure of the ISM -- serves to mitigate the pronounced centre-versus-outskirts disparity prevalent in older models. 

At the time of writing, we do not find any observational work exploring the connection between orbital orientation and the spatial extent of new stars and cold-dense$/$cool gas. To address connections between orbital orientation and global properties, \cite{Mesa2014} use a sample of $\sim$1500 visually-classified galaxy pairs selected from the SDSS. They use spiral-arm direction to break their sample into subsamples of co-rotating and counter-rotating pairs. These authors report bluer colours and younger stellar populations in their counter-rotating systems. To compare directly with our work, it would be interesting to follow up these systems with spatially-resolved IFU observations, which would allow the measurement of the spatial structure of the star-forming component as a function of orbital orientation. Comparing the results from such an exercise with our work presents two caveats. (1) The \cite{Mesa2014} SDSS sample only selects galaxies displaying tidal tails. Inspection of videos \textcolor{black}{of} our simulations (analysis not included here) show that such selection is biased towards the early period (\textcolor{black}{second}-row$/$first-column image versus other \textcolor{black}{second}-row images in Figure~\ref{fig:terminology}). This claim is consistent with findings by \cite{Blumenthal2020}, who use a cosmological simulation of galaxy formation \citep[IllustrisTNG,][]{Pillepich2018} to infer that only $\sim$45\% of interacting pairs display visually-identified features. Lastly (2), our results suggest that it is not enough to break our suite by spin-orbit orientation alone \citep[as in][]{Moreno2015} -- but information on whether or not the two disc interpenetrate one another at first pericentric passage is just as pertinent. We explore these details in future work.

\subsection{Global SFR enhancement versus radial structure}
\label{subsec:driving_global_sfr}

It has been known for almost two decades that star forming galaxies follow a tight SFR-$M_{\star}$ relation, known as the star-forming `main sequence' \citep[SFMS,][]{Brinchmann2004,Noeske2007,Elbaz2007,Whitaker2012,Salim2014}. Following this discovery, it has been suggested that regions above this sequence (i.e., the starburst regime) tend to be populated by merging systems \citep{Jogee2009,Hung2013,Willett2015,Pearson2019}. However, the majority of these systems tend to be late-stage mergers, with galaxy pairs spending more time in regions closer to the main ridgeline of this relation \citep{Puech2014,Cibinel2019}. Conversely, observations by \cite{Silva2018} show that only 12$-$20\% of close galaxy pairs with separations between 3$-$15 kpc are starbursts, depending on which stellar-mass bin is considered. Likewise, using a cosmological simulation \citep[IllustrisTNG,][]{Pillepich2018}, \cite{Blumenthal2020} show that interacting galaxies in the pre-merger phase are more likely to lie well above the SFMS only at or soon after their most recent close pericentric passage. Similarly, \cite{Wilkinson2018} use the original Illustris \citep{Vogelsberger2013} to show that $\sim$55\% of starburst in the local Universe are activated by tidal interactions with their neighbours.

\begin{table}
  \begin{center}
    \setlength\tabcolsep{5.0pt} 
    \begin{tabular}{l|l|c|c} 
    \hline \hline
     SF Type & Definition & Secondary & Primary \\
     \hline
     Enhanced   & $\log$ SFR$/$SFR$_{\rm iso} > +$0.3 & 15.2\% & 16.9\%  \\ 
     Regular   & $-$0.3 $<\log$ SFR$/$SFR$_{\rm iso}< +$0.3 & 82.5\% & 81.5 \%\\ 
     Suppressed & $\log$ SFR$/$SFR$_{\rm iso} < -$0.3 & 2.3\% & 1.6 \% \\ 
     \hline
    \end{tabular}
  \end{center}
\caption{Types of star-formers (SFs): definitions and percentages per galaxy. SFR$/$SFR$_{\rm iso}$ denotes {\it global} SFR enhancement. Figure~\ref{fig:sfr_bins} shows radial structure per SF type in terms of SFR, SFE, and cold-dense gas mass. 
}
\label{table:sfr_bins}
\end{table}

The emergence of spatially-resolved integral-field unit (IFU) surveys has stimulated an interest in how the internal structure of a galaxy is connected to its location relative to the SFR-$M_{\star}$ relation \citep[e.g.,][]{Sanchez2012,CanoDiaz2016,GonzalezDelgado2016,Hsieh2017}. This motivates us to analyse the connection between global SFR enhancement and radial structure in interacting galaxies. To continue teasing out the effects caused by the encounter, here we elect to compare against the SFR$/$SFR$_{\rm iso}=$1 line, not the SFMS. Figure~\ref{fig:sfr_bins} splits our sample into three types of star-formers (SFs): enhanced SFs (at least 0.3 dex above unity, blue closely-hatched), regular SFs (between $-$0.3 and $+$0.3 dex from unity, green hatched), and suppressed SFs (at least 0.3 dex below unity, red loosely-hatched). We choose 0.3 dex as a compromise between the 0.2 dex half-scatter of the SFMS \citep{Belfiore2018} and the 0.5$-$0.6 dex thresholds commonly adopted to identify starbursts \citep[e.g.,][]{Wilkinson2018,Ellison2020aqii}. We note that our galaxies are predominantly regular SFs at the great majority of times ($\sim$80\%). See Table~\ref{table:sfr_bins} for a list of percentages per SF type and galaxy. Note that individual galaxies are not permanent members of any specific SF type -- rather, they may `visit' all three regimes throughout the duration of their interaction \citep{MartinezGalarza2016}. Also, we emphasize that we quote enhancements relative to their respective isolated galaxies, not to the global SFMS.

For the secondary galaxy (Figure~\ref{fig:sfr_bins}, left panels), the average SFR profile ratios (top panels) for the three SF-type bins are similar at large radii, and diverge towards the centre. The large scatter displayed by the thickness of all three bands (corresponding to one standard deviation) is driven by the bursty nature of star formation in our \textsc{fire-2} physics model \citep[e.g.,][]{Orr2017bursty}. Globally enhanced star-formers experience elevated SFR within $\sim$0.9 kpc. This is caused by the combination of enhanced SFE (middle panels) and elevated cold-dense gas content (bottom panels) in that region. The regular star-formers also experience a central boost in cold-dense gas mass, but not in SFE, causing SFR to be consistent with unity in that region. The globally suppressed star-formers, on the other hand, exhibit a flat average SFR profile ratio at most radii, except near $\sim$2.5-3 kpc, where it dips below unity. This is driven by a deficit of cold-dense gas at those radii, which prevails over the lack of SFE suppression in that region.

The primary galaxy (Figure~\ref{fig:sfr_bins}, right panels) also experiences a bifurcation of average SFR profile ratios near the centre: globally-enhanced SFs bend upwards, whilst their suppressed counterparts bend downwards with decreasing radius (in the case of the secondary, the globally suppressed SFs bend downward relative to the enhanced star-formers, but not relative to unity). \textcolor{black}{The primary} experiences a boost in cold-dense gas mass within $\sim$5 kpc for the globally-enhanced SF population. However, SFE is only enhanced within $\sim$0.5 kpc, which explains why SFR enhancement occurs only within that smaller region. SFR enhancement near $\sim$3 kpc, on the other hand, is explained by excess in cold-dense gas (SFE is consistent with unity there). For the globally-suppressed SF sample, the deficit in SFR within $\sim$0.3 kpc is explained by suppressed SFE alone -- cold-dense gas content is consistent with null enhancement in that region. 

The upward versus downward bifurcation discussed above is observed by IFU surveys for galaxies with global SFR above and below the SFMS. Using SAMI, \cite{Medling2018} find that SFR surface density ($\Sigma_{\rm SFR}$) profiles bend downward in the inner regions of galaxies inhabiting the regime located 1-3$\sigma$ below the SFMS. \cite{Belfiore2018} find similar results with MaNGA: the $\Sigma_{\rm SFR}$ profiles of galaxies in the Green Valley -- i.e., those located 1$\sigma$ (0.39 dex) below the SFMS, experience a downward turn near the centre \citep[see also][]{Brownson2020}. Using the SIMBA cosmological simulation \citep{Dave2019}, \cite{Appleby2020} explain this trend as the combination of two effects: lower H$_2$-gas mass presence towards smaller radii, and a decrease in SFE in that region. In contrast, our simulations predict a downturn in SFE (primary galaxy only), and no statistically-significant change in cold-dense gas content (Figure~\ref{fig:sfr_bins}, middle and bottom panels, loosely-hatched red bands). For a more direct comparison with SIMBA, it would be interesting if those authors performed an analysis of paired-versus-isolated galaxies similar to ours. 

\begin{figure*}
\centerline{\vbox{
\hbox{
\includegraphics[width=3.35in]{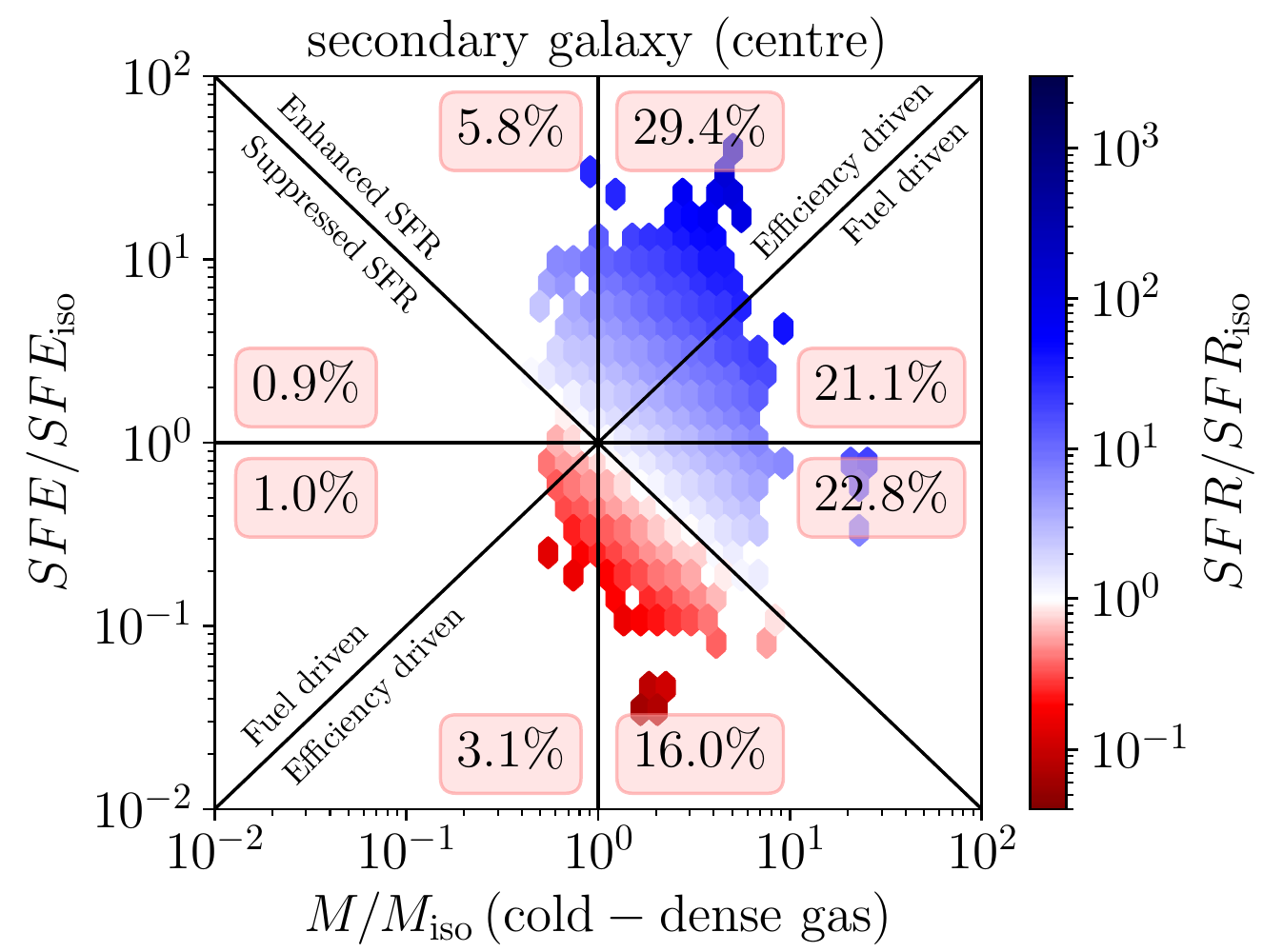}
\hspace{.25in}
\includegraphics[width=3.35in]{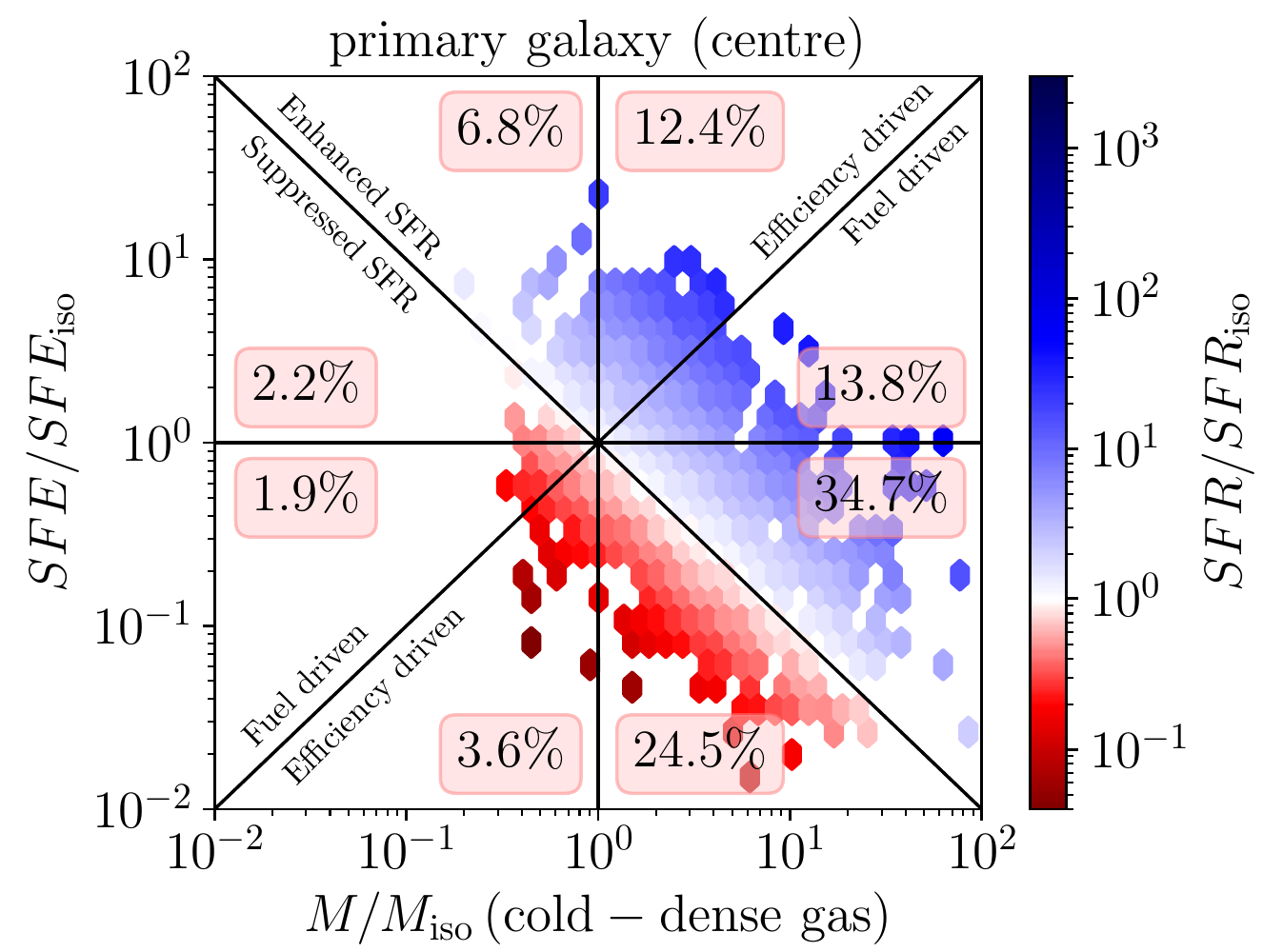}
}
\vspace{-.1in}
}}
\caption{Dependence of central SFR enhancement on central SFE and cold-gas mass enhancements. {\it Left (right) panels}: secondary (primary) galaxy. Galaxy-pair periods only. Central kiloparsec only. The 2D histogram is colour-coded by central SFR enhancement, split into enhanced (red hexagons) and suppressed values (red hexagons) -- also indicated by the diagonal line with negative slope. The vertical (horizontal) line splits central cold-dense gas mass (central SFE) enhancement into enhanced and suppressed sub-samples. \textcolor{black}{The diagonal line with positive slope splits the sample into situations where the dominant factor driving central SFR$/$SFR$_{\rm iso}$ {\it away from unity} is either efficiency or fuel mass content. Note that these designations depend on whether or not SFR is enhanced or suppressed.} These four lines split each panel into eight triangular regions. Sample percentage in each triangular region is indicated.}
\label{fig:fuelling}
\end{figure*}

Using EDGE-CALIFA \citep{Bolatto2017}, \cite{Chown2020} find that galaxies with H$_{\alpha}$ inner upturns tend to be barred galaxies, mergers or galaxy in close pairs. However, not all mergers$/$pairs have such upturns. Also, they find that galaxies with upturns tend to have higher molecular-gas concentrations, but the presence of such fuel reservoirs are not always required for galaxies with upturns. This observational study does not separate galaxies in terms of their location to the global SFMS -- nor relative to non-interacting controls (in the case of pairs). \cite{Ellison2018sfr} report both downturns {\it and upturns} towards the smaller radii in their MaNGA-selected $\Sigma_{\rm SFR}$ profiles at low galactocentric radii for galaxies with SFR above and below their resolved (spaxel-by-spaxel) \textcolor{black}{SFR-$M_{\star}$ relation} \citep[the `rSFMS' -- see e.g.,][]{Ellison2018sfr,Hani2020rsfms}. \cite{Wang2019} find similar results with the same survey, but using the median $\Sigma_{\rm SFR}$ in stellar-mass bins (rather than the rSFMS) as reference. \cite{Morselli2019} report similar bifurcations out to $z=1.2$ using multi-wavelength {\it Hubble} Space Telescope (HST) data selected from the GOODS$+$CANDELS campaign\footnote{GOODS stands for Great Observatories Origins Deep Survey, whilst CANDELS stands for Cosmic Assembly Near-infrared Deep Extragalactic Legacy Survey.}. Using a sample of 12 starbursts selected from ALMaQUEST \citep{Lin2019}, \cite{Ellison2020aqii} and \cite{Ellison2020aqiii} find that galaxies well above the global SFMS exhibit upturns in their SFE profiles, in agreement with our results (Figure~\ref{fig:sfr_bins}, middle panels, densely-packed blue versus loosely-hatched red bands). We also predict these galaxies to have an upturn in cold-dense gas mass at lower galactocentric radii (Figure~\ref{fig:sfr_bins}, middle panels). Instead, ALMaQUEST starbursts exhibit a mix of $f_{\rm H_2}$ profile shapes, many of which bend downwards, with suppressed ${\rm H_2}$ mass near the centre. 

Whilst encouraging, comparisons between the above observations and our simulations must be interpreted with care. Firstly, we do not compare profiles relative to systems on the SFMS (or the rSFMS), but rather relative to profiles associated to identical galaxies simulated in isolation. Although we do not expect this correction to cause substantial effects (our isolated galaxies remain close to the global SFR-$M_{\star}$ relation), discrepancies may be amplified when our isolated control galaxies momentarily deviate from the SFMS due to secular processes. Secondly, samples drawn from spatially-resolved surveys -- especially those with follow up observations using interferometric arrays -- tend to be small. \textcolor{black}{The above sections demonstrate} that the shapes of density profiles depend not only on global SFR, but also on the time of observation (\textcolor{black}{Figure~\ref{fig:profiles_fiducial}}) and the orbital geometry (Figure~\ref{fig:profiles_minisuites}). Thus, it is possible that recently-published observed samples are not necessarily representative. For these reasons, it is critical that the next generation of spatially-resolved galaxy surveys (1) expand campaigns similar to that conducted by \cite{Pan2019sfr} -- who compare spatially-resolved interacting systems against properly-matched controls -- but with substantially larger and more varied galaxy samples; and (2) that these programmes are coupled with follow-up interferometric observations to infer the role of SFE, molecular gas content, and their spatial structure. 

\textcolor{black}{Our results suggest that, in a few ways, the inner regions of our interacting systems in the local Universe resemble high-redshift galaxies experiencing `{\it compaction}': i.e., disc-contraction episodes believed to be driven by cold streams and mergers, producing star-bursting `{\it blue nuggets}' as a result \citep{Dekel2014,Zolotov2015}. Concretely, when we segregate interacting galaxies in terms of their global SFR, we find a bifurcation in SFR, SFE and available fuel at small galactocentric radii. At high redshift, galaxies in the upper envelope of the SFMS exhibit high gas fractions, high SFEs, and cuspy gas profiles -- whilst those in the lower envelope are endowed with low gas-fractions, low SFEs, and gas-depleted cores \citep{TacchellaMS2016,TacchellaProfiles2016}. Investigating the importance of the relative contribution of galaxy-galaxy interactions to compaction and quenching in the local Universe \citep{Woo2019} is the subject of future work.}

\subsection{What drives star formation in the central kiloparsec?}
\label{subsec:fuel}

The previous section demonstrates that the strongest interaction-induced modifications to radial SFR structure occur at small galactocentric radii (Figure~\ref{fig:sfr_bins}). Figures~\ref{fig:time_evolution_fiducial}, \ref{fig:time_evolution_suite}, and \ref{fig:averages} confirm this for the central kpc. We find that such variations are driven by changes in SFE and the cold-dense gas reservoir - in line with prior observational work focused on how SFE and available fuel in galaxies regulate \textcolor{black}{their location within (and departure from) the global SFMS} \citep[e.g.,][]{Saintonge2017,Ellison2020aqiii,Piotrowska2020}. In this section, we probe deeper into the following question: {\it what drives star formation in the central kiloparsec, efficiency or available fuel?} Expressing SFR as in equation~(\ref{eqn:sfe}) facilitates this. Specifically, it allows us to write SFR enhancement as follows:
\begin{equation}
\label{eqn:fuel}
\frac{\rm SFR}{{\rm SFR}_{\rm iso}} = \frac{\rm SFE}{{\rm SFE}_{\rm iso}} \times \frac{M_{\rm cold-dense}}{M_{\rm cold-dense, \, iso}}.
\end{equation}
Namely, in order to enhance SFR, either both SFE and cold-dense gas mass are enhanced simultaneously, or the enhancement of one supersedes the suppression of the other. Similarly, in order to suppress SFR, either both SFE and cold-dense gas mass are suppressed simultaneously, or the suppression of one supersedes the enhancement of the other. \textcolor{black}{To describe which factor on the right-hand side of equation~(\ref{eqn:fuel}) dominates in driving their product {\it away from unity}, we adopt the following terminology:} 

\noindent \textcolor{black}{In the enhanced-SFR regime:}

\noindent \textcolor{black}{{\small $\bullet$} {\it Efficiency-driven:} ${\rm SFE}/{{\rm SFE}_{\rm iso}} > {M_{\rm cold-dense}}/{M_{\rm cold-dense, \, iso}}.$}

\noindent \textcolor{black}{{\small $\bullet$} {\it Fuel-driven:} \,\,\,\,\,\,\,\, \,\, ${\rm SFE}/{{\rm SFE}_{\rm iso}} < {M_{\rm cold-dense}}/{M_{\rm cold-dense, \, iso}}.$}

\noindent \textcolor{black}{In the suppressed-SFR regime:}

\noindent \textcolor{black}{{\small $\bullet$} {\it Fuel-driven:} \,\,\,\,\,\,\,\, \,\, ${\rm SFE}/{{\rm SFE}_{\rm iso}} > {M_{\rm cold-dense}}/{M_{\rm cold-dense, \, iso}}.$}

\noindent \textcolor{black}{{\small $\bullet$} {\it Efficiency-driven:} ${\rm SFE}/{{\rm SFE}_{\rm iso}} < {M_{\rm cold-dense}}/{M_{\rm cold-dense, \, iso}}.$}

Figure~\ref{fig:fuelling} displays a 2D histogram of central SFE enhancement versus central cold-dense gas mass enhancement, colour-coded by central SFR enhancement. I.e., the vertical and horizontal axes correspond to the two factors in equation~(\ref{eqn:fuel}), and the colour bar displays their product -- hence the utility of the format adopted in equations~(\ref{eqn:sfe}) and (\ref{eqn:fuel}). We include data from the entire merger suite (galaxy-pair periods only), {\it focusing exclusively on the central kiloparsec}.  The left panel represents the secondary galaxy, whilst the right panel shows the primary. The \textcolor{black}{horizontal} and \textcolor{black}{vertical} lines at unity split the sample into objects with enhanced versus suppressed central SFE, and enhanced versus suppressed central cold-dense gas mass, respectively. The diagonal line with negative slope separates values with enhanced central SFR (blue hexagons) from those with suppressed central SFR (red hexagons). The diagonal line with positive slope splits the sample into efficiency-driven versus fuel-driven categories. \textcolor{black}{This definition differs for SFR-enhanced and SFR-suppressed cases -- i.e., it flips across the accompanying negatively-sloped diagonal line (see above definitions).} These four lines segregate our data into eight triangular regions. Percent contribution per triangular region is indicated.

Overall, the great majority of data points in our sample (i.e., the majority of times during the galaxy-pair period across our entire merger suite) experience enhanced central SFR: 79\% for the secondary galaxy and 68\% for the primary. Namely, not only is the average magnitude of central SFR enhancement larger in the \textcolor{black}{secondary-galaxy population} (by factors of 160\% and 70\% for the secondary and primary respectively -- see Table~\ref{table:averages} and Figure~\ref{fig:averages}), but the {\it frequency} of systems with enhanced central SFR is higher as well. Similarly, enhanced cold-dense gas content in the centre is a generic feature in our simulations: 90\% of cases for the secondary and 85\% for the primary (with average enhancement values of 
130\% and 220\%). However, despite this almost ubiquitous enhancement in available cold-dense fuel within the innermost kiloparsec, central SFR is not always enhanced (e.g., red hexagons to the right of the vertical line -- corresponding to 16\% and 24.5\% of cases for the secondary and primary respectively). 

For the secondary galaxy (Figure~\ref{fig:fuelling}, left panel), 79\% of our sample exhibits enhanced central SFR. Within this sub-sample, 71\% have enhanced central SFE, and 92\% have enhanced central cold-dense gas mass. This sub-sample is almost evenly split between \textcolor{black}{fuel driven and efficiency driven} systems, with 55\% in the former category. We note that systems with the most extreme central SFR enhancements (darkest-blue hexagons) are efficiency-driven (and accompanied by enhanced central cold-dense gas content). In the central SFR-suppressed sub-sample (the remaining 21\% of the entire sample), 96\% have suppressed central SFE, and 91\% occupy the \textcolor{black}{efficiency}-driven regime. The most extreme central SFR deficits (darkest-red hexagons) are caused by strongly-suppressed central SFE, despite the presence of an abundant cold-dense gas reservoir in the centre.

For the primary galaxy (Figure~\ref{fig:fuelling}, right panel), a smaller fraction (68\%) exhibit central SFR enhancement. Within this sub-sample, only 49\% have enhanced central SFE, and 90\% have enhanced cold-dense gas mass in the centre. This sub-sample has a larger fraction of fuel driven systems (71\%) than their secondary galaxy counterparts (only 55\%). Also, unlike their secondary companions, the most extreme central-SFR enhancements here (darkest-blue hexagons) are fuel-driven (with central-SFE enhancement near unity). In the central SFR-suppressed regime (32\% of the full sample), 93\% of our galaxies have suppressed central SFE, and 87\% are in the \textcolor{black}{efficiency}-driven regime. As in the secondary galaxy case, the most extreme central-SFR deficits (darkest-red hexagons) occur in the \textcolor{black}{efficiency}-driven regime -- but unlike the secondary case, these systems also contain low levels of available cold-dense gas within the inner kiloparsec.

For both galaxies, the majority of central SFR-enhanced systems are fuel driven (55\% and 71\%), and the majority of central SFR-suppressed systems are efficiency driven (91\% and 97\%) -- i.e., central SFE is more strongly suppressed than their central cold-dense gas content. Also note that central SFE enhancement is more common in the secondary galaxy (57\% versus 35\% in the primary), and reaches higher levels than the primary (126\% versus 11\% at the upper 1$\sigma$ level -- Table~\ref{table:averages} and Figure~\ref{fig:averages}). Although central cold-dense gas mass enhancement is slightly less frequent in the primary (85\% versus 89\% in the secondary), it reaches higher levels than the secondary (470\% versus 230\% at the upper 1$\sigma$ level). This is manifested by an overall shift downwards and to-the-right (lower efficiencies and higher cold-dense gas masses) in the 2D distribution (left versus right panels in Figure~\ref{fig:fuelling}). In sum, this population shift explains why, even though the primary galaxy has a healthier central cold-dense gas reservoir, the secondary galaxy is more efficient at making stars in the centre (Figures~\ref{fig:time_evolution_fiducial},  \ref{fig:time_evolution_suite}, and \ref{fig:averages}).

Observations using the ALMaQUEST survey suggests that central starbursts are primarily driven by enhancements in central SFE \citep{Ellison2020aqii}. Here we find that, for the secondary galaxy, systems with the highest central SFR enhancements are efficiency driven -- but this is not necessarily true for the primary galaxy. Note also that the central region of the secondary galaxy achieves higher central SFR enhancements relative to its primary counterpart, and has higher central SFE levels (darkest-blue hexagons in the triangular region with the `29.4\%' label). In other words, according to our simulations, the centres of secondary galaxies are more likely to achieve starburst status. It would be interesting to check if the three starbursts with signs of merger features in the \cite{Ellison2020aqii} sample are secondary companions in a galaxy pair. For galaxies {\it not} classified as starbursts, \cite{Ellison2020aqiii} find that enhanced central SFR is more likely to be driven by high levels of H$_{2}$ gas in the inner regions \citep[see also][]{Piotrowska2020,Bluck2020}. The large fraction of fuel-driven cases with mild SFR enhancement in the centre (lighest-blue hexagons in both panels) are in line with these observations. Lastly, Thorp et al. (in prep) report that the majority of their SFR-enhanced post-mergers are SFE-driven, whilst the majority of the SR-suppressed post-mergers are fuel-driven. Testing this result against our simulations in the post-coalescence regime is the subject of future work.

Beyond the near universal trends we describe above, there exist a few cases with enhanced central SFR and {\it suppressed} cold-dense gas mass in the centre (blue-hexagons to the left of the vertical line, corresponding to 7\% and 10\% for the secondary and primary samples, respectively) -- in agreement with merging galaxies existing in nature  \citep[e.g., Arp 240,][]{He2020}. At the other extreme, a substantial fraction of our sample experiences enhancements in central SFR and cold-dense gas mass, but with {\it suppressed} central SFE (29\% and 51\% for the secondary and primary cases). Examples of systems following this trend also exist in nature, including the two nuclei of the famous Antennae galaxies \citep[NGC 4038/39,][]{Bemis2019}. \textcolor{black}{It would be interesting to verify if simulations tailored to model this specific system \citep[e.g.,][]{Renaud2014,Renaud2015} reproduce this central behaviour.}

The fact that the central kiloparsec of the primary galaxy frequently experiences low SFE levels, despite experiencing huge boosts in cold-dense gas fuel, is intriguing. This is particularly true between 0.5 and 1.5 Gyr after first pericentric passage (Figures~\ref{fig:time_evolution_fiducial} and \ref{fig:time_evolution_suite}). Concretely, in those situations, stellar feedback (and possibly other dynamical processes) prevents our cold-dense gas (at $n>10 \, {\rm cm}^{-3}$ and $T<300$K) from reaching (or maintaining) the thresholds required for star formation (in our model, self-gravitating with $n>1000 \, {\rm cm}^{-3}$). Indeed, \cite{Moreno2019} shows that, in our suite of galaxy merger simulations, under $\sim$0.15\% of our cold-dense gas budget achieves star-forming status at any given time. Such a small fraction is a reflection of the dynamic and turbulent nature of the ISM at small scales, wherein gas undergoes a cycle of collapse, star formation, and cloud dispersal -- governed by feedback and possibly other dynamical processes \citep[see e.g.,][]{Torrey2017,Semenov2017,Semenov2018,Orr2019}. Detailed understanding of which processes govern low SFE levels in the presence of abundant cold-dense gas in our merger simulations is beyond the scope of this paper. \textcolor{black}{Nevertheless, we highlight a well-known (non-merging) system that exhibits this behaviour: our own Milky Way. Namely, the Central Molecular Zone \citep[CMZ,][]{Ferriere2007,Ginsburg2016} experiences low levels of star-formation activity, despite the copious presence of molecular-gas fuel \citep[e.g.,][]{Barnes2017}. Such low SFE levels are likely driven by turbulent pressure \citep{Krumholz2015}. \cite{Jeffreson2018} suggest that between $\sim$120-500 pc, galactic shear dissipates clouds -- whilst epicyclic perturbations incite tidally-driven collapse along a stream at 100-pc from the centre. Using \textsc{fire-2} physics, Orr et. al. (in prep.) find that their simulated galaxies reproduce the properties of the CMZ at some point in their evolution, but only in situations where asymmetric and bursty galactic cores are produced.}

\section{Summary}
\label{sec:summary}

We use an extensive suite of parsec-scale galaxy merger simulations \citep{Moreno2019} to track the spatial structure and evolution of star formation and the interstellar medium (ISM) in interacting galaxies. These simulations employ the `Feedback in Realistic Environments-2' physics model (\textsc{fire-2}), which is capable of capturing the multi-phase structure of the ISM \citep{FIRE2}. In this paper we focus on major mergers (stellar mass ratio $=$ 2.5:1) in the galaxy-pair period, between first and second pericentric passage, with distance greater 20 kpc. We point the reader to Section~\ref{sec:simulations_and_terminology} for relevant terminology (\textbf{\textit{boldface italics}}) and summarise our main results below.
\begin{enumerate}
    \item[1.] {\it Evolution in the central kiloparsec and the outskirts.}
    \begin{itemize}
        \item Both galaxies experience strong central mass boosts in cool$/$cold-dense gas and new stars during the interaction.
        \item Despite the presence of a healthy reservoir of cold-dense gas, nuclear star formation in the primary galaxy is weak. This is caused by low star formation efficiency (SFE) levels.
        \item Suppression of star formation is not statistically significant in the 1$-$10 kpc region -- although suppression exists for a small range of radii within that region, and its strength depends on orbital geometry and time of observation after the first encounter. 
    \end{itemize}
    \item[2.] {\it Radial structure and evolution.}
    \begin{itemize}
        \item The radial extent of mass enhancement in new stars and cool$/$cold-dense gas is more centrally concentrated in the secondary galaxy than in the primary (within $\sim$2 versus $\sim$7 kpc).
        \item The primary galaxy builds a healthier cold-dense gas reservoir, but this does not translate into more mass in new stars due to low SFE levels.
        \item As a function of time, cold-dense and cool gas mass enhancement at large radii grows, and the central peak shifts outwards and becomes more intense.
    \end{itemize}
    \item[3.] {\it Dependence on orbital geometry.}
    \begin{itemize}
        \item The great majority of orbits in our galaxy merger suite (two-thirds) display the behaviour described in item (2) above.
        \item A small sub-sample (one-fifth -- mostly near-retrograde orbits with large separation at first pericentre) exhibit gentler changes to  radial structure during the interaction.
        \item A much smaller subset (one-eight -- mostly near-polar orbits with small separation at first pericentre) experience vigorous changes in radial structure, including deep baryonic suppression at intermediate radii for the secondary galaxy, and more intense enhancement in that region for the primary.
    \end{itemize}
    \item[4.] {\it Connections between global SFR and radial structure.}
    \begin{itemize}
        \item Globally-enhanced star-formers experience strong SFR enhancement in the inner region, driven by similar enhancements in nuclear SFE and available cold-dense ISM fuel.
        \item When the primary belongs to the globally-suppressed star-former population, SFR at small radii is suppressed due to decrements in nuclear SFE, without significant changes in available cold-dense ISM fuel.
    \end{itemize}
    \item[5.] {\it Star formation in the central kiloparsec.}
    \begin{itemize}
        \item Both secondary and primary galaxies experience enhancements in central SFR (79\% and 68\%), and available cold-dense gas mass in the centre (89\% and 85\%).
        \item Central SFE is enhanced in the majority of cases for the secondary galaxy (57\%), and suppressed in the primary (67\%).
        \item In most cases, central SFR-enhancement in both galaxies is fuel-driven (55\% and 71\%), whilst central SFR-suppression is efficiency-driven (91\% and 97\%).
    \end{itemize}
\end{enumerate}

Our results advocate for a new class of IFU surveys-- i.e., either immense efforts with substantially larger galaxy samples, such as HECTOR \citep{hector}; and$/$or programmes focused chiefly on galaxy mergers, such as the Snapshot Optical Spectroscopic Imaging of Mergers and Pairs for Legacy Exploration (SOSIMPLE, PI: B. Husemann). Such initiatives must also be coupled with multi-wavelength follow-up campaigns (e.g., EDGE-CALIFA and ALMaQUEST) capable of capturing large samples of interacting galaxies at various stages, orbital geometries, locations relative to the global SFR-M$_{\star}$ relation; with representative ISM content. For direct comparison with our numerical predictions, it is imperative to measure the radial structure of interacting galaxies in relation to carefully-matched non-interacting controls, and to ascribe secondary versus primary galaxy status in these targets.

Our work focuses exclusively on major galaxy interactions, meant to represent typical galaxy pairs in the local Universe. In future work, we plan to explore the effect of varying the mass ratio \citep{Cox2008,Lotz2010}, and will extend our analysis into the post-coalescence period \citep{Thorp2019,Hani2020,Peschken2020}. \textcolor{black}{We also intend to explore the region outside our two 10-kpc spheres, and evaluate the role of tidal compression in driving extended star formation \citep{Renaud2014,Renaud2015}.} These investigations can also be expanded into (1) the dwarf regime, where gas content and SFE might be substantially different \citep{Stierwalt2015,Pearson2016,KadoFong2020,Martin2020}; (2) the high-redshift regime, where modifications to ISM content and structure \citep{Bournaud2011,Fensch2017,Calabro2019} are accompanied by an increased frequency of merging \citep{Bluck2009,LopezSanjuan2009,Bluck2012,LopezSanjuan2012,LopezSanJuan2013,Vicente2015,Mantha2018,Duncan2019}; (3) minor mergers, which are expected to be more frequent \citep{Villalobos2008,Villalobos2009,Qu2011,Kaviraj2014,Martin2018}; and (4) the massive regime, where mixed and dry encounters tend to appear \citep{Lin2008,Stewart2009,Lin2010} -- and where AGN feedback and environmental quenching processes are likely to collaborate in concert with tidal interactions, and assist galaxies in achieving their retirement into the passive sequence \citep{Bluck2014,Bluck2020}.


\section*{Acknowledgements} 

\textcolor{black}{We thank the anonymous reviewer for their thorough and insightful comments, which were shared with us on a heroically reasonable timescale in the middle of the Covid-19 pandemic. Their suggestions certainly improved the quality of this manuscript.} The computations in this paper were run on the Odyssey cluster supported by the FAS Division of Science, Research Computing Group at Harvard University. Support for JM is provided by the NSF (AST Award Number 1516374), and by the Harvard Institute for Theory and Computation, through their Visiting Scholars Program. DRP and SLE gratefully acknowledge NSERC for Discovery Grants which helped to fund this research. CB acknowledges the support of a National Sciences and Engineering Research Council of Canada (NSERC) Graduate Scholarship. AB acknowledges ERC Advanced Grant 695671 `Quench' and support from the STFC. MHH acknowledges the receipt of a Vanier Canada Graduate Scholarship. The Flatiron Institute is supported by the Simons Foundation. Support for PFH is provided by NSF Collaborative Research Grants 1715847 \& 1911233, NSF CAREER grant 1455342, NASA grants 80NSSC18K0562, JPL 1589742. We honour the invaluable labour of the maintenance and clerical staff at our institutions -- whose contributions make our scientific discoveries a reality -- and urge institutions everywhere to protect these essential workers from the worldwide economic contraction caused by the pandemic. This research was conducted on Tongva-Gabrielino Indigenous land.

\section*{Data Availability} 

The data underlying this article will be shared on reasonable request to the corresponding author.

\bibliographystyle{mnras}
\bibliography{bibliography}

\bsp
\label{lastpage}
\end{document}